\documentclass[11pt]{article}

\usepackage{graphicx,epsf, epsfig, amssymb, amsmath, amsfonts, mathrsfs}
\usepackage{longtable}
\usepackage{color}
\usepackage{enumitem}
\usepackage{float}
\usepackage{caption}
\usepackage{subcaption}
\usepackage{upgreek}
\usepackage{cite}
\usepackage{xcolor}
\usepackage[section]{placeins}
\usepackage[colorlinks=true,linktocpage=true,linkcolor=blue,citecolor=blue]{hyperref}
\usepackage{tensor}
\usepackage{multirow}
\usepackage[utf8]{inputenc}
\usepackage{authblk}
\usepackage{breqn}

\newcommand{\be}{\begin{equation}}
\newcommand{\ee}{\end{equation}}
\newcommand{\bea}{\begin{eqnarray}}
\newcommand{\eea}{\end{eqnarray}}

\newcommand{\dmoff}[1]{}

\newtheorem{Prop}{Proposition}
\usepackage
[
a4paper,
left=1.8cm,
right=1.8cm,
top=3cm,
bottom=4cm,
]{geometry}
\newcommand{\beqa}{\begin{eqnarray}}
\newcommand{\eeqa}{\end{eqnarray}}

\numberwithin{equation}{section} 

\author[1]{Carlos Herdeiro}
\author[1]{Eugen Radu}
\author[1]{Etevaldo dos Santos Costa Filho}

\affil[1]{Departamento de Matemática da Universidade de Aveiro and
	Centre for Research and Development in Mathematics and Applications (CIDMA),
	Campus de Santiago, 3810-193 Aveiro, Portugal}

\newcommand\blfootnote[1]{%
  \begingroup
  \renewcommand\thefootnote{}\footnote{#1}%
  \addtocounter{footnote}{-1}%
  \endgroup
}

\begin{document}
	\title{\bf Spinning Proca-Higgs balls, stars and hairy black holes} 
	
	\date{June 2024}
	
	\maketitle

	\begin{abstract}
		
		Recently, spherical and static flat space solitons (balls) and self-gravitating, everywhere regular, asymptotically flat solitons (stars) were constructed in an Einstein-Proca-Higgs model~\cite{herdeiro2023procahiggs}, where a complex vector field gains mass by coupling to a real scalar field with a Higgs-type potential. The Proca-Higgs model serves as a UV completion of a complex Proca model with self-interactions. Here, we construct and examine the mathematical and physical properties of rotating configurations. In particular,  rotation allows horizon-bearing solutions, including stationary clouds surrounding Kerr black holes and their non-linear continuation into black holes with Proca-Higgs \textit{hair}.
	\end{abstract}

\vfill
\blfootnote{E-mail: {\tt herdeiro@ua.pt}}
\blfootnote{E-mail: {\tt eugen.radu@ua.pt}}
\blfootnote{E-mail: {\tt etevaldo.s.costa@ua.pt}}

	\newpage
	
	\tableofcontents
	
	\newpage 
	
	\section{Introduction}

Inaugurated by the landmark detection of gravitational waves in 2015~\cite{LIGOScientific:2016aoc}, the era of precision observations of compact objects is upon us. Since then, over one hundred gravitational wave events have been observed, which can provide empirical validation of theoretical astrophysical models. So far, the most successful theory of gravity is General Relativity. Within its framework, these data make it possible to test not only the theory itself but also some of its key predictions, such as the Kerr hypothesis \cite{10.1063/1.3022513,Herdeiro:2022yle}. 

While observations are consistent with a substantial presence of black holes in the universe, there is a growing interest in theoretical models proposing horizonless compact objects and non-Kerr black holes. This interest is driven both by the theoretical challenges posed by General Relativity and the Kerr solution, acknowledged to be incomplete descriptions of the physical nature, and by the relevant question of degeneracy: whether other models can replicate key astrophysical phenomena attributed to Kerr black holes, including the EHT imaging observations~\cite{EventHorizonTelescope:2019dse,EventHorizonTelescope:2022wkp} (see e.g.~the discussions in \cite{Herdeiro:2021lwl,Olivares:2018abq,Sengo:2022jif,Rosa:2022tfv}) and gravitational wave events, see e.g.~the discussions in \cite{CalderonBustillo:2020fyi,Sanchis-Gual:2018oui,CalderonBustillo:2022cja,Palenzuela:2006wp,Atteneder:2023pge,Tsukada:2020lgt}. An additional motivation is that some non-Kerr models for compact objects could make contact with other open problems in strong gravity, such as dark matter and dark energy \cite{Lee:1995af,Suarez:2013iw,Eby:2015hsq,Chen:2020cef,Jones:2023fzz}.

In (possible) connection to dark matter, scalar boson stars have been studied for over 50 years (see Refs. \cite{Jetzer:1991jr,Schunck:2003kk,Liebling:2012fv,Shnir:2022lba} for reviews). These nontopological solitons have been studied, in particular as black holes mimickers~\cite{Schunck:1998cdq,Mielke:2000mh,Yuan:2004sv,Guzman:2005bs,Berti:2006qt,Guzman:2009zz,Vincent:2015xta,Sennett:2017etc,Grould:2017rzz,Olivares:2018abq,Herdeiro:2021lwl,Rosa:2022tfv}. Although they remain speculative, their study pioneered some techniques and concepts to deal with observational degeneracies between putative exotic compact objects and black holes~\cite{Ryan:1996nk,ryan1995gravitational,Siemonsen:2023age}. From a high energy physics viewpoint, most of the studied models accommodating boson stars are regarded as effective field theories (EFTs) rather than fundamental ones, and require beyond the Standard Model constructions.

Scalar boson stars have a cousin model composed of spin 1 particles. The corresponding self-gravitating configurations are known as Proca stars, which have been considered much more recently for a massive complex free field \cite{Brito:2015pxa} (see also \cite{Herdeiro:2023wqf}). The Proca model, even in its most basic form, is yielding intriguing phenomenological results in the astrophysical context: stable configurations of these stars are able to mimic the shadow of a Kerr black hole \cite{Herdeiro:2021lwl,Sengo:2022jif}; and the gravitational wave signal GW190521 that is presumed to be the result of a binary merger of two Kerr black holes was shown to have an equally valid description (or even with a slight statistical preference) as the collision of two Proca stars \cite{CalderonBustillo:2020fyi} - therefore, as a proof of concept, as a collision between two dark matter stars. More recently~\cite{CalderonBustillo:2022cja}, other peculiar gravitational-wave signals from Advanced LIGO and Virgo, were compared with Proca star merger simulations, which in some cases corroborates the issue of degeneracy. Notably, this study not only explored the viability of Proca stars in explaining some astrophysical events but also yielded an interesting finding: a consistent boson mass across multiple gravitational-wave events, reinforcing the potential astrophysical relevance of Proca stars.

For further studies on self-gravitating solitonic solutions involving bosonic fields, see \cite{Mourelle:2024qgo,Gutierrez-Luna:2021tmq,Jaramillo:2023lgk,Dzhunushaliev:2021vwn}, which explore various aspects and models of bosonic field configurations and their astrophysical implications. Investigations focusing on the non-relativistic regime are detailed in \cite{Zhang:2021xxa,Wang:2023tly,Jain:2022kwq,Annulli:2020ilw}.

From the perspective of high-energy physics, it is important to view even the simplest Proca model as an EFT. In fact, the model's lack of gauge invariance, a consequence of the explicit mass term, raises issues like non-renormalizability. In a more fundamental description, however, one could expect this mass term to be generated through a Higgs-mechanism, as for the massive vector bosons observed in the electroweak sector of the Standard Model.

Additionally, as another token of the effective character of simple Proca models, introducing self-interactions to the Proca field \cite{Minamitsuji:2018kof,Herdeiro:2020jzx} significantly impacts its consistency, primarily due to hyperbolicity problems and various instabilities \cite{Coates:2022nif,Mou:2022hqb,PhysRevLett.129.151103,Barausse:2022rvg,PhysRevLett.129.151102}. This is in contrast to the scalar case \cite{baer2008wave}. A proof that a massive free Proca field is hyperbolic can be found in \cite{Baer2015}. Even so, self-interacting Proca models and their generalized versions have been proposed in different scenarios, e.g., \cite{Esposito-Farese:2009wbc,Annulli:2019fzq,Barton:2021wfj,Loginov:2015rya,Brihaye:2017inn}. To develop these results within a coherent field theory framework, a UV completion for the self-interacting Proca model is necessary. It has been shown that coupling the Proca field with an additional scalar field can resolve these issues \cite{PhysRevD.106.084022,herdeiro2023procahiggs}, at least at the linear level. Moreover, depending on the nature of this coupling, it is possible to revert to a self-interacting Proca model under certain limits.

In our previous work \cite{herdeiro2023procahiggs}, we introduced the Proca-Higgs model, a vector analog of the Friedberg-Lee-Sirlin model \cite{Friedberg:1976me}, enabling a complex vector field to dynamically acquire mass, through a Higgs-like mechanism. This model serves as a UV completion for a quartic self-interacting Proca model, effectively resolving its hyperbolicity issues. Additionally, albeit the free Proca theory is not a consistent truncation of the Proca-Higgs model, the latter reduces to the former in certain limits.

In Ref. \cite{herdeiro2023procahiggs}, we have studied solitonic solutions within the Proca-Higgs model, both as non-gravitating solitons (balls) and as self-gravitating solitons (stars) subject to spherical symmetry. Proca-Higgs balls arise from effective self-interactions due to scalar-vector coupling. It is important to note that, within our chosen ansatz, pure Proca and scalar field nontopological solitons do not exist. The existence of these solutions relies on the coupling term $\phi^2\mathcal{A}_\alpha\bar{\mathcal{A}}^\alpha$ in the Lagrangian. 

The solution space of spherical Proca-Higgs balls and stars exhibits distinct properties, such as multiple solutions with the same mass and frequency. Additionally, Proca-Higgs stars' physical properties, like compactness and circular geodesic structures, might also differ significantly from mini-Proca stars. On the other hand, no spherical black hole horizon can coexist in equilibrium with Proca-Higgs stars, similarly to spherical Proca and scalar boson stars. Under strong gravitational or self-interaction forces, Proca-Higgs stars tend towards mini-Proca stars (Proca stars without self-interactions), which defines an upper mass and Noether charge limits, connecting the two models.  

In the present study, we build upon our previous research by incorporating rotation into the Proca-Higgs model. Our focus is to explore how rotation influences both the flat spacetime and self-gravitating solitons within this model, extending the analysis beyond the static, spherically symmetric configurations previously examined. One first difference arising from rotation is that the pure Proca configuration does not define an upper mass limit for the Proca-Higgs configurations. Also, rotation enables the existence of solutions for the Proca-Higgs configurations at smaller values of the frequency parameter $\omega$ as compared to the static case.

Furthermore, our investigation shows that the inclusion of rotation allows the existence of solutions with horizons. Employing the synchronization condition~\cite{Herdeiro:2014goa,Herdeiro:2014ima}, we have constructed stationary Proca-Higgs clouds~\cite{Hod:2012px,Hod:2013zza,Benone:2014ssa} around a Kerr black hole. Unlike the pure Proca scenario, which is confined to a one-dimensional parameter space (for fixed $m$ and given $\omega$, solutions exist for a single mass), the Proca-Higgs model accommodates a more extensive range of parameters. Similar behavior can be found in non-linear Q-clouds around Kerr black holes \cite{HERDEIRO2014302}, when the scalar field has self-interactions. However, without self-interactions, these scalar Q-clouds also only exist along a 1-dimensional subspace. Additionally, we delve into the fully non-linear realization of these clouds, allowing for backreaction between the matter field and the spacetime, forming stationary black hole solutions: Kerr black holes with Proca-Higgs hair. As in the Proca and scalar cases \cite{Herdeiro_2015,Herdeiro:2014goa,HERDEIRO2014302,Herdeiro:2016tmi,Santos:2020pmh}, the solutions naturally transition into rotating Proca-Higgs stars as the black hole size vanishes.

\textbf{Conventions.} Our analysis is conducted on a 4-dimensional manifold equipped with a pseudo-Riemannian metric that is symmetric, denoted as $g_{\mu\nu}=g_{\nu\mu}$ and possesses a signature of $(- + + +)$. We use Greek alphabet letters to label spacetime coordinates, which range from 0 to 3. Additionally, we employ Einstein's summation convention throughout our calculations. Throughout this paper we shall use units with $c=1$.

	\section{The Proca-Higgs model}
	\label{sec2}

	\subsection{Action, field equations and limits}

	We consider a model with a real scalar, $\phi$, and a complex vector field, $\mathcal{A}_\alpha$, both minimally coupled to Einstein's gravity. Unlike previous works dealing with Proca stars \footnote{These works consider the following action, that we shall refer to as the standard (complex) Proca model (where $\mu$ is a constant mass term)}
	\begin{equation}
		\nonumber
		\mathcal{S}=\int d^4x \sqrt{-g}\left[
		\frac{R}{16 \pi  G}
		-\frac{1}{4}\mathcal{F}_{\alpha\beta}\bar{\mathcal{F}}^{\alpha\beta}
		-\frac{\mu^2}{2}\mathcal{A}_\alpha\bar{\mathcal{A}}^\alpha
		\right]~. 
	\end{equation}
	-- $e.g.$~\cite{Brito:2015pxa,Herdeiro:2016tmi} --, in this model, the mass of the vector field results from a scalar-vector coupling. A non-zero vector mass results from the Higgs-like scalar potential, dynamically imposing a non-trivial scalar vacuum expectation value ($v.e.v.$) at infinity, $v=$constant.
	Explicitly, the model is  described by the action:
	\begin{equation}
		\label{action}
		\mathcal{S}=\int d^4x \sqrt{-g}\left[
		\frac{R}{16 \pi  G}
		+\mathcal{L}_{m}
		\right]~, 
	\end{equation}
	where $R$ is the Ricci scalar of the spacetime metric $g_{\alpha\beta}$, with determinant $g$. To understand better the contribution of each field, we can split the matter Lagrangian density, $\mathcal{L}_{m}$, into three parts: one pure scalar, pure vector, and the interaction term
 \begin{equation}
		\label{lagrangianmatter}
		\mathcal{L}_{m}=	\mathcal{L}^{(s)} + \mathcal{L}^{(v)}+ \mathcal{L}^{(i)}\,,
	\end{equation}
 \begin{equation}
     \mathcal{L}^{(s)} =
		-\frac{1}{2} \partial_\alpha \phi \partial^\alpha \phi	-U(\phi)	~, \qquad \mathcal{L}^{(v)}= -\frac{1}{4}\mathcal{F}_{\alpha\beta}\bar{\mathcal{F}}^{\alpha\beta}~, \qquad \mathcal{L}^{(i)} =-\frac{1}{2}\phi^2\mathcal{A}_\alpha\bar{\mathcal{A}}^\alpha\,.
 \end{equation}

 Here the vector field strength is $\mathcal{F}_{\alpha\beta}=\nabla_{\alpha}\mathcal{A}_{\beta}-\nabla_{\beta}\mathcal{A}_{\alpha}$, with overbar denoting complex conjugation. We shall focus on the `Mexican-hat' potential for the scalar field
	\begin{equation}
		~U(\phi)=\frac{\lambda}{4} (\phi^2-v^2)^2 \,, 
	\end{equation}
	where $\lambda>0$ is a constant. 
	
	The vector and scalar equations of the model are, respectively,
	\begin{eqnarray}
		&&
		\label{procafe}
		\nabla_\alpha\mathcal{F}^{\alpha\beta}=\phi^2\mathcal{A}^\beta \ ,
		\\
		&&
		\label{scalarfe}
		\Box\phi
		=\frac{dU}{d\phi}
		+\phi ~\mathcal{A}_\alpha\bar{\mathcal{A}}^\alpha\ ,
	\end{eqnarray}
	whereas the Einstein equations read
	\begin{equation}
		R_{\alpha \beta}-\frac{1}{2}Rg_{\alpha \beta}=8 \pi G\,T_{\alpha \beta}\,,
		\label{Einstein-eqs}
	\end{equation}
	where $T_{\alpha \beta}$ is the total energy-momentum tensor of the theory, related to the matter Lagrangian \eqref{lagrangianmatter}. The scalar, vector and interaction components of the energy-momentum tensor are
	\begin{eqnarray}
 		&&
		T_{\alpha\beta}^{(s)}=\partial_\alpha \phi \partial_\beta \phi-g_{\alpha \beta}
		\left[\frac{1}{2} (\partial \phi)^2 +U(\phi) \right]~\,,\label{scalar_EMT}\\
		&&
		T_{\alpha\beta}^{(v)}=\frac{1}{2}
		( \mathcal{F}_{\alpha \sigma }\bar{\mathcal{F}}_{\beta \gamma}
		+\bar{\mathcal{F}}_{\alpha \sigma } \mathcal{F}_{\beta \gamma}
		)g^{\sigma \gamma}
		-\frac{1}{4}g_{\alpha\beta}\mathcal{F}_{\sigma\tau}\bar{\mathcal{F}}^{\sigma\tau} \label{vector_EMT}\ ,\\
  		&&
		T_{\alpha\beta}^{(i)}=\phi^2\left[ \frac{1}{2}
		(
		\mathcal{A}_{\alpha}\bar{\mathcal{A}}_{\beta}
		+\bar{\mathcal{A}}_{\alpha}\mathcal{A}_{\beta}
		)
		-\frac{1}{2}g_{\alpha\beta}\mathcal{A}_\sigma\bar{\mathcal{A}}^\sigma \right] \label{inter_EMT}\, .
		\end{eqnarray}

	The Proca-like vector field equations  (\ref{procafe})	imply the Lorenz-like condition, which is not a gauge choice, but rather a dynamical requirement:
	\begin{equation}
		\nabla_\alpha (\phi^2 \mathcal{A}^\alpha)= 0 \ .
		\label{lorentz}
	\end{equation}

 	Due to the global $U(1)$ invariance of the complex vector field,  $\mathcal{A}_\mu \rightarrow e^{i \chi}\mathcal{A}_\mu$, where $\chi$ is a constant, the Proca-Higgs model \eqref{action} is associated with a conserved 4-current, $j^{\alpha}$, and conserved charge, $Q$, respectively given by
	\begin{equation}
		j^\alpha=\frac{i}{2}\left[\overline{\mathcal{F}}^{\alpha \beta} \mathcal{A}_\beta-\mathcal{F}^{\alpha \beta} \overline{\mathcal{A}}_\beta\right],
	\end{equation}
	\begin{equation}\label{noether}
		Q=\int_{\Sigma}  j^\alpha d\Sigma_\alpha\,.
	\end{equation}

 Compact stationary axially symmetric solutions to the system defined by equations \eqref{procafe}, \eqref{scalarfe}, \eqref{Einstein-eqs}  are not viable for a real vector field (as can be shown by a similar procedure as in \cite{Herdeiro:2016tmi,Ayon-Beato:2002kxy}). However, as in the scalar/Proca case, a way to circumvent those no-go theorems is to consider a harmonic time dependence for the (complex) field. Although the field oscillates with a frequency $\omega$, the energy-momentum tensor remains stationary. Moreover, for flat spacetime solutions, this is not enough to ensure the existence of solutions. In order to have soliton solutions, there must be nonlinear couplings (and for nontopological solitons, an additional conservation law). Scalar Q-balls exist for quartic-order self-interactions \cite{Coleman:1985ki}, while known solutions of Proca balls were constructed for sixth-order self-interactions \cite{Loginov:2015rya}. In contrast, given its nonlinear nature, Proca-Higgs balls exist even in the limit $\lambda=0$.

 \medskip
	
Our prior analysis of the model represented by~\eqref{action} interprets it as an ultraviolet (UV) completion of a self-interacting Proca model, which can be understood as follows. A closer examination of the field equations~\eqref{procafe}-\eqref{Einstein-eqs} indicates that while $\phi=0$ simplifies the model to a complex vector minimally coupled to Einstein's gravity (essentially an Einstein-(double)Maxwell theory). On the other hand, the scenario with $\phi=v\neq 0$ does not offer such a straightforward truncation. In this latter case, the Einstein-vector equations lead to an Einstein-complex-Proca system. However, the scalar equation introduces an additional constraint, $\mathcal{A}_\alpha\bar{\mathcal{A}}^\alpha=0$.

Although the model does not precisely reduce to the conventional Proca model, certain limits approach it. By expanding $\phi$ around its v.e.v.,  it acquires an effective mass $M_\rho\equiv\sqrt{2\lambda}v$. At energies significantly lower than $M_\rho$, the scalar field effectively becomes constant, approximated by its vacuum expectation value, leading to the low energy EFT given by:
\begin{equation}
\mathcal{S}_{\text {eff }}=\int \mathrm{d}^4 x\left[\frac{R}{16 \pi G}-\frac{1}{4} \mathcal{F}_{\alpha \beta} \overline{\mathcal{F}}^{\alpha \beta}-\frac{v^2}{2}\left(\mathcal{A}_\alpha \overline{\mathcal{A}}^\alpha-\frac{\left(\mathcal{A}_\alpha \overline{\mathcal{A}}^\alpha\right)^2}{M_\rho^2}\right)\right]\,.
\end{equation}
This results in a complex Proca model with quartic self-interactions. The effective Proca field theory obtained has a specific sign for the self-interactions, and the mass term $\mu^2=v^2$ is determined by the scalar v.e.v. The self-interactions are given by:
\begin{equation}
-\frac{v^2}{2M_\rho^2}=-\frac{1}{4\lambda}<0 \ ,
\end{equation}
which is influenced by the scalar self-interaction parameter. Therefore, for large values of $v^2 \lambda$ and not so large values of $\lambda$, the Proca-Higgs can be well described by a quartic self-interacting Proca. But as we increase $\lambda$ the model tends to a Proca theory without self-interactions.

Once more, the Proca-Higgs model can be seen as a UV completion for a quartic self-interacting Proca model. However, a clear advantage of such model is that free of hyperbolicity issues \cite{Coates:2022nif,PhysRevLett.129.151102,Mou:2022hqb,PhysRevLett.129.151103,Barausse:2022rvg}, which plague self-interacting vector fields. These issues primarily stem from the dynamical evolution towards a singular effective metric, which depends on the background and on the vector field. The Proca-Higgs model circumvents hyperbolicity issues, at least at the linear level, by ensuring that \textit{only} the spacetime metric governs the principal part of the differential operator describing the dynamics of the vector field. Indeed, in the Proca-Higgs case, this conclusion is confirmed by putting the Proca equation in the form
\begin{equation}
	\nabla_{\mu }\nabla^{\mu }\mathcal{A}_{\nu } - \mathcal{A}^{\mu } R_{\nu \mu } -  \mathcal{A}_{\nu } \phi^2  + 2  \nabla_{\nu }\mathcal{A}^{\mu } \nabla_{\mu }\ln\phi+ 2 \mathcal{A}^{\mu } \nabla_{\nu }\nabla_{\mu }\ln\phi=0 \ .
\end{equation}

	\subsection{Horizon Properties and Mass Formula}\label{komar_quantities}

 The close relationship between gravitational theory and thermodynamics is exemplified by the establishment of black hole mechanics laws and the discovery of Hawking radiation \cite{Bardeen:1973gs,Wald2001}. In General Relativity, the entropy $S$ of a black hole follows the Bekenstein-Hawking area law, expressed as $A/4$, which equals a quarter of the black hole's event horizon area \cite{PhysRevD.9.3292}. Over the years, numerous methodologies have been developed to derive mass formulas\cite{PhysRevD.46.1453,PhysRevD.47.R5209,Wald:1993ki,Wald:1993nt}, and these have been applied across various models \cite{PhysRevD.47.R5209,Heusler:1993cj,ballesteros2023hairy}. In this section, we present a derivation of a generalized Smarr mass formula for black holes with Proca-Higgs hair. We construct the global charges associated with these solutions and their symmetries constructing a mass formula that can be easily generalized to cases with other matter fields, and we delineate the characteristics of such generalizations. It turns out that the scalar field does not directly contribute to the angular momentum. Instead, the total angular momentum is composed of the horizon angular momentum and the Noether charge, which is linked to the global $U(1)$ symmetry of the Proca field (see Eq. \eqref{first_law} below). Therefore, the influence of the scalar field on angular momentum is implicit, emerging as the fields satisfy the equations of motion.

	On the manifold $\mathcal{M}$ let $\pi_u$ be a one-parameter group of
	diffeomorphisms (where $\pi_0$ is the identity) generated by the smooth vector field $\kappa^\mu$, which is tangent to the orbits of the diffeomorphisms. The diffeomorphisms carry a tensor field $T$, with components $T^{\lambda_1 \lambda_2\cdots}\,_{\sigma_1\sigma_2\cdots}$, along their orbits 
	\begin{equation}
		T\rightarrow(\pi_u)^* T\,.
	\end{equation}
	
	Specifically, we are interested in the fields described in \eqref{lagrangianmatter}, which transform as
	\begin{equation}
		\mathcal{A}\rightarrow(\pi_u)^* \mathcal{A}=e^{i k u} \mathcal{A}\ ,\qquad \phi\rightarrow(\pi_u)^* \phi=\phi\,,
	\end{equation}
	such that the Lie derivative along $\kappa$ of these fields is
	\begin{equation}\label{liefields}
		L_{\kappa}\mathcal{A}= i k \mathcal{A}\ , \qquad 	L_{\kappa}\phi= 0\ .
	\end{equation}

	Let us now specialize it for our case of interest. The assumptions of axisymmetry and stationarity imply the existence of two Killing vector fields, $\eta$ and $\xi$. These vector fields commute, without loss of generality, under the condition that the spacetime is asymptotically flat. With this in mind, we can choose coordinates adapted simultaneously to both these vector fields. Consequently, we introduce a temporal coordinate, $t$, and an angular coordinate, $\varphi$, along the paths of the Killing vector fields. Hence, following condition \eqref{liefields}, we can specialize the Ansatz of the matter field to be \cite{martini1985geometric}
	\begin{equation}
		\mathcal{A}=e^{-i\omega t+ i m \varphi} A(x^1,x^2)\ , \qquad \phi=\phi(x^1,x^2)\ ,
	\end{equation}
 where $x^1,x^2$ are the two non-Killing coordinates, $\omega$ is a real frequency parameter, assumed to be non-negative without loss of generality and $m\in \mathbb{Z}^{+}$ is the azimuthal harmonic index.
	
Although $\xi$ and $\eta$ generate symmetries of the spacetime, they do not necessarily give rise to symmetries of the matter field in the sense that their Lie derivative does not vanish (see Eq.\eqref{liefields}). In contrast, the sole symmetry that applies to the entire solution (geometry plus matter) is derived from the helicoidal vector field \cite{Herdeiro_2015,Herdeiro:2016tmi}
 \begin{equation}\label{helicoidal}
	\chi=\xi+\frac{w}{m} \eta,
\end{equation}
which ensures that
\begin{align}
	L_{\chi}\mathcal{A} &= 0, \\
	L_{\chi}\phi &= 0.
\end{align}
Consequently, the field strength of the Proca potential, $\mathcal{F}=d\mathcal{A}$, also satisfies 
\begin{equation}\label{lieF}
	L_{\chi}\mathcal{F}=0.
\end{equation}

\subsubsection{Horizon Properties }

In the study of electrically charged black holes, we often consider three key quantities: $l$, $\Omega_H$, and $\Phi_H$. $l$ being the Killing vector defined as $l=\xi+\Omega_H\,\eta$, and $\Omega_H$ the black hole horizon velocity. The last of these, $\Phi_H$, represents the electric potential on the black hole's horizon, which remains constant across it.  It is well known that this quantity enters the Smarr relation, playing the role of a chemical potential \cite{carter2010,Dyson:2023ujk,Compere:2006my}. 

The chemical potential associated with the electric charge also has an intrinsic relation with the conservation of electric charge and with the existence of a Gauss law. In the context of the Proca-Higgs model, there is no such Gauss law, but still, some parallelism can be made with horizon quantities. In addition, after tracing this parallelism, we present a mass formula for black holes with Proca-Higgs hair.

Analogously to the ``electric potential", we can define a potential $\Phi$ (here, we imply a similar methodology as presented in \cite{Frolov:1998wf})
\begin{equation}
	\Phi_{(\kappa)}=-\mathcal{A}_\alpha \kappa^\alpha\,,
\end{equation}
with $\kappa$ an arbitrary Killing vector. Differentiating it, we obtain
\begin{equation}
	\Phi_{(\kappa)\,; \mu}=-L_{\kappa}\mathcal{A}_{\mu}-\mathcal{F}_{\mu \nu} \kappa^\nu\,.
\end{equation}

Now introduce the vector $E_{(\kappa)\,\mu}=-\mathcal{F}_{\mu \nu} \kappa^\nu$, analogous to the electric field.
Considering that the horizon of a stationary black hole is a Killing horizon and that the Proca-Higgs model obeys the null energy condition (see Appendix \ref{sec_ec}), we can establish the following condition on the horizon
\begin{align}
	R_{\mu \nu}l^{\mu}l^{\nu} &= 0 \quad \text{where} \quad l^\mu=\xi^\mu+\Omega_H \eta^\mu\,.
\end{align}
Consequently, by means of Einstein's equation, we have
\begin{equation}
	0=T_{\mu \nu}l^{\mu}l^{\nu}=E_{(l)\,\mu} \bar{E}_{(l)}^\mu+ \phi^2|\Phi_{(l)}|^2.
\end{equation}
Therefore, the boundary conditions should be chosen so that this equality holds. Notice, however, that the vector $E_{(l)}^\mu$ is perpendicular to the null generator $l$, so it must be either spacelike or null \cite{Ayon-Beato:2002kxy}. Thus, each term on the right-hand side must vanish identically, which implies that
\begin{equation}\label{Ahorizon}
	\left.l^\alpha \mathcal{A}_\alpha\right|_{H}=0\,,\qquad\qquad E_{(l)\,\mu} \bar{E}_{(l)}^\mu=0.
\end{equation}
Hence, the ``electric potential" $\Phi$ is zero along the event horizon. This leads to the following equality holding for any vector $s^\mu$ tangent to the event horizon
\begin{equation}
	\left.s^\mu \Phi_{;\mu}\right|_{H}=0\,,
\end{equation}
	which implies the synchronization condition
\begin{equation}\label{sync}
	\omega=m \Omega_H\,,
\end{equation}

Due to $0=\left.s^\mu \Phi_{;\mu}\right|_{H}=s^\mu L_l \mathcal{A}_\mu$, implying that $l$ must be proportional to $\chi$. When the synchronization is met, we have
\begin{equation}
	E_{[\mu; \nu]}=0.
\end{equation}
Here, we did not use the equations of motion, but solely that $\mathcal{F}$ obeys \eqref{lieF} and a Bianchi identity. To conclude, when the synchronization is met, the flux over the horizon, $T_{\mu\nu}l^\mu\xi^\nu$, is zero.

\subsubsection{Mass Formula}
	
	Symmetries in spacetime have a direct connection to conservation laws for fields and particles. This connection can be established by using Killing vectors, which are vector fields that generate the isometries of the metric of spacetime. Being $\kappa$ a Killing vector, we can then introduce the Komar current, $J^\mu$, of the form:
	\begin{equation}
		J^\mu(\xi)=R^{\mu\nu}\kappa_{\nu}\ .
	\end{equation}
	
	The Komar integral can be expressed in terms of the Killing field $\kappa$
	\begin{equation}
		Q(\kappa) = -\int_{S_{\infty}^2}\kappa^{\mu ; \nu} d \Sigma_{\mu \nu} \ .
	\end{equation}	
	Here, we denote $Q(\kappa)$, the global conserved quantity associated with the symmetry of the spacetime (a Komar charge). For an arbitrary Killing vector field, we have 
	\begin{equation}
		\kappa^{\mu ; \nu}{ }_\nu=-R_\nu^\mu \kappa^\nu,
	\end{equation}
	hence
	\begin{equation}
		Q(\kappa)= \int_{\Sigma}R_{\mu  \nu} \kappa^\nu d \Sigma^{\mu}+Q_{\text{int}}(\kappa),
	\end{equation}
where the charge evaluated in a possible interior boundary is denoted by $Q_{\text{int}}(\kappa)$. For solitons, this charge is zero, while for black holes, the interior boundary is represented by the event horizon,
	\begin{equation}
		Q_{\text{int}}(\kappa)= -\int_{\partial B}\kappa^{\mu ; \nu} d \Sigma_{\mu \nu} \ .
	\end{equation}

	Using the Einstein equations (and therefore assuming that the equations of motion are satisfied), we can rewrite the Komar charge as
	\begin{equation}\label{komarcharge}
		Q(\kappa)=8\pi G\int_{\Sigma}\left( T^{\mu\nu}+\dfrac{g^{\mu\nu}R}{16\pi G}\right) \kappa_{\nu} d \Sigma_{\mu} +Q_{\text{int}}(\kappa)\ .
	\end{equation}	
	Moreover, since $\xi$ is a time-like Killing vector, we have $Q(\xi)=-4\pi G M$, where $M$ is the Komar mass. For the rotation Killing vector, $\eta$, we have $Q(\eta)=8 \pi G J$, where $J$ is the Komar angular momentum. It is important to note that the Komar expressions for the mass, $M$, and angular momentum, $J$, are not identical, differing by a factor of 2 and a sign. Such that the Komar mass and angular momentum,
	\begin{equation}
		M= \dfrac{1}{4\pi G}\int_{S_{\infty}^2}\xi^{\mu ; \nu} d \Sigma_{\mu \nu}\,, \qquad 	J= \dfrac{-1}{8\pi G}\int_{S_{\infty}^2}\eta^{\mu ; \nu} d \Sigma_{\mu \nu}\,,
	\end{equation}
	
	can be cast as \eqref{komarcharge}
	\begin{equation}
		M=-2\int_{\Sigma}\left( T^{\mu\nu}+\dfrac{g^{\mu\nu}R}{16\pi G}\right) \xi_{\nu} d \Sigma_{\mu} +M_H\, , \qquad
		J=\int_{\Sigma}\left( T^{\mu\nu}+\dfrac{g^{\mu\nu}R}{16\pi G}\right) \eta_{\nu} d \Sigma_{\mu} +J_H\ , \label{j_komar}
	\end{equation}

 	where
	\begin{equation}\label{mass_hor}
		M_H= \dfrac{1}{4\pi G}\int_{\partial B}\xi^{\mu ; \nu} d \Sigma_{\mu \nu}\,, \qquad		J_H=\dfrac{-1}{8\pi G} \int_{\partial B}\eta^{\mu ; \nu} d \Sigma_{\mu \nu}\,.
	\end{equation}
	
	Given the spacelike nature of the axisymmetry generator $\eta$, it is possible to choose the spacelike hypersurface $\Sigma$ such that $\eta$ is tangent to it, $\eta^\mu\,d \Sigma_{\mu}=0$, such that the second term in the integrand of equation \eqref{j_komar} vanishes,
	\begin{equation}
		J=\int_{\Sigma} T^{\mu\nu} \eta_{\nu} d \Sigma_{\mu} +J_H\,.
	\end{equation}

	From the definition of the energy-momentum tensor
	\begin{equation}
		T_{\mu\nu}=\dfrac{-2}{\sqrt{g}}\dfrac{\delta(\sqrt{g}\mathcal{L})}{\delta g^{\mu\nu}}=-2\dfrac{\delta\mathcal{L}_m}{\delta g^{\mu\nu}}+g_{\mu\nu}\mathcal{L}_m\, ,
	\end{equation}
	we can write	
	\begin{equation}
		M=4 \int_{\Sigma}\left(\dfrac{\delta\mathcal{L}_m}{\delta g^{\mu\nu}}\right) \xi^{\nu} d \Sigma^{\mu}-2 L + M_H\, ,\qquad
		J=-2 \int_{\Sigma}\left(\dfrac{\delta\mathcal{L}_m}{\delta g^{\mu\nu}}\right) \eta^{\nu} d \Sigma^{\mu}+ J_H\, ,
	\end{equation}
	where  the Lagrangian $L$ is given by
	\begin{equation}
		L=\int\mathcal{L}\ \xi^{\mu} d \Sigma_{\mu} \ .
	\end{equation}
	
	For the Lagrangian density \eqref{lagrangianmatter}, we have
	\begin{equation}
		\int_{\Sigma}\left(\dfrac{\delta\mathcal{L}_m}{\delta g^{\mu\nu}}\right) \xi^{\nu} d \Sigma^{\mu} =-	\frac{1}{4}\int_{\Sigma}\left(  \mathcal{F}_{\nu\sigma}\bar{\mathcal{F}}_{\mu}{}^{\sigma} +  \mathcal{F}_{\mu}{}^{\sigma} \bar{\mathcal{F}}_{\nu\sigma} +   \phi^2\left(\mathcal{A}_{\nu} \bar{\mathcal{A}}_{\mu}  +   \mathcal{A}_{\mu} \bar{\mathcal{A}}_{\nu} \right) +  2 \nabla_{\mu}\phi \nabla_{\nu}\phi \right) \xi^{\nu} d \Sigma^{\mu}\ .
	\end{equation}

 Moreover, the Proca and scalar fields are subject to the Lie derivative  as in \eqref{liefields}; and we can write
	\begin{equation}
		L_{\kappa}\mathcal{A}_\mu=i k \mathcal{A}_\mu =\kappa^{\nu}\mathcal{F}_{\nu\mu}+\nabla_{\mu}(\kappa^\nu \mathcal{A}_\nu )\ , \qquad L_{\kappa}\phi =0= \kappa^{\mu}\nabla_{\mu}\phi \ .
	\end{equation}
	
	Consequently, assuming that the equations of motion are satisfied	
	\begin{equation}
		\int_{\Sigma}\left(\dfrac{\delta\mathcal{L}_m}{\delta g^{\mu\nu}}\right) \kappa^{\nu} d \Sigma^{\mu} =-\dfrac{1}{4}\int_{\partial B}\left(\kappa^\nu \mathcal{A}_\nu \bar{\mathcal{F}}_{\mu\sigma}+\kappa^\nu \bar{\mathcal{A}}_\nu \mathcal{F}_{\mu\sigma}\right) d\Sigma^{\mu\sigma}-\dfrac{k}{2}Q \ .
	\end{equation}	
	Here, $Q$ stands for the Noether charge as given by \eqref{noether}. Following \cite{Bardeen:1973gs} and using the helicoidal Killing vector \eqref{helicoidal} and the synchronization condition \eqref{sync}, the mass of the horizon can be express as
	\begin{equation}
		M_H=2\Omega_H J_H+\frac{\kappa}{4 \pi} A_{H}\,,
	\end{equation}
	where $\kappa$ is the surface gravity defined by 
 \begin{equation}
 \kappa^2 = -\frac{1}{2} ( \nabla_a \chi_b ) ( \nabla^a \chi^b ) \bigg|_{H}\,,
  \end{equation}
 and $A_{H}$ is the horizon area of the black hole. The surface gravity relates to the black hole temperature through   $T_H=\dfrac{\kappa}{2\pi}$. Therefore, the mass of the black hole with Proca-Higgs hair becomes	
	\begin{equation}\label{first_law}
		M=2 \Omega_H J - 2L+\frac{\kappa}{4 \pi} A_{H}\,.
	\end{equation}

 Moreover, when we assume that the equations of motion are satisfied, the lagrangian can be written as
 \begin{equation}
     L=\int \left(U + \dfrac{1}{2}\phi^2\mathcal{A}_{\mu}\bar{\mathcal{A}}^{\nu}\right)\xi^\nu d \Sigma_\mu\,.
 \end{equation}
 Therefore, Equation \eqref{first_law} is the desired integral mass formula, or in an equivalent way
	\begin{equation}\label{first_law2}
		M=M_{\text{matter}}+2 \Omega_H J_H+\frac{\kappa}{4 \pi} A_{H}\,,
	\end{equation}
 with
 \begin{equation}
     M_{\text{matter}}=-2\int_{\Sigma}\left( T^{\mu\nu}-\dfrac{g^{\mu\nu}}{2}T\right) \xi_{\nu} d \Sigma_{\mu}\,.
 \end{equation}

     The first law of black hole mechanics, as formulated in \cite{Bardeen:1973gs}, establishes a relationship between the infinitesimal variations in the mass of a stationary, axisymmetric black hole and corresponding small changes in its horizon area, angular momentum, and the characteristics of any coupled matter. The variation of the mass formula \eqref{first_law}, can be  expressed as
     \begin{equation}\label{differential}
         d M=T_{H}d S+\Omega_{H}d J\ .
     \end{equation}

We have presented the first law in integral and differential forms, which are theoretically relevant and practically useful for verifying numerical solutions. The integral form of the first law plays an essential role in testing individual computed solutions. On the other hand, the differential form provides a broader perspective by evaluating the quality of the entire parameter space. Overall, using both forms together creates a strong framework for interpreting physical systems and confirming their accuracy.

\medskip

	To conclude this section, let us briefly compare with the electrovacuum case, in which case a quantity $Q_H$ appears, the electric charge of the Killing horizon. The electric potential of the black hole, $\Phi=\xi^\nu \mathcal{A}_\nu$, is constant at the horizon \cite{Compere:2006my,carter2010}, which can be seen as merely a boundary condition over the horizon or as a generalized zeroth law. Also, the surface integral over $\mathcal{F}$ gives the electric charge \cite{heusler_1996}. 
 However, in our case, the $\mathcal{F}$-contribution at infinity  vanishes
 (due to the exponential decay of the Proca field),
 and is zero also at the horizon due to boundary conditions.
 Another way to see or justify the vanishing of this term is to notice that our problem does not admit a charge given by a Gauss law. Even more, the electric potential of the black hole, $\Phi_H$, is also zero due to boundary conditions. Therefore,  the Proca field does not directly contribute to the angular velocity of the horizon either.

	\subsection{Ansatz, units and numerical approach}  \label{sec_ansatz} 

In the computations that follow, our focus will be on axially symmetric solutions. Utilizing the standard coordinates $(t,r,\theta,\varphi)$, the ansatz for the ``matter" fields  is 
	\begin{equation}
		\mathcal{A}=\left(i V d t+\frac{H_1}{r} d r+H_2 d \theta+i H_3 \sin \theta d \varphi\right) e^{i(m \varphi-w t)}\,,\qquad \phi=\phi(r,\theta)\,.
		\label{matteransatz}
	\end{equation}
	The four functions $(H_i,V)$ only depend on ($r$,$\theta$) and we shall only address the case with $m = 1$ in this work.

 For hairy black holes and clouds, the metric Ansatz is constructed to accommodate the presence of a horizon. The line element is:
\begin{equation}\label{line_BH}
d s^2=-e^{2 F_0} N d t^2+e^{2 F_1}\left(\frac{d r^2}{N}+r^2 d \theta^2\right)+e^{2 F_2} r^2 \sin ^2 \theta\left(d \varphi-\dfrac{W}{r^2} d t\right)^2\,,
\end{equation}
where
\begin{equation}
N \equiv 1-\dfrac{r_H}{r}\,.
\end{equation}
	and $(F_i, W)$ are functions of the spheroidal coordinates $(r, \theta)$. We are interested in the spacetime outside the event horizon, so we consider $r=[r_H,\infty)$.  For completeness, the full equations of motion are given in Appendix~\ref{eomap}.

Within the used matter Ansatz, the circularity condition is met \cite{carter2010}. Consequently, the vector $\chi$ tangent to the event horizon coincides in direction with the null vector that lies in the two-dimensional plane generated by the vectors $\xi$ and $\eta$. Thus, we can choose:
\begin{equation}
\chi=\xi+\Omega_H \eta,
\end{equation}
where $\Omega_H$ is the horizon velocity.
	
 The metric Ansatz is slightly different in the solitonic case. In addition to the non-existence of a horizon ($r_H=0$, implying $N=1$), for stars, we also work with $W\rightarrow r W$, which provides a better accuracy of the solutions).
 
 We shall now discuss some convenient rescalings of the model's variables and parameters, which shall be used in the remainder of the paper. Unlike the standard (free) Proca model, the theory~\eqref{action} allows for flat spacetime solitons. To simplify comparisons between such Proca-Higgs balls and the self-gravitating stars, we introduce the dimensionless quantities
	\begin{equation}\label{scale}
		r\rightarrow \dfrac{r}{v}\,, \qquad \omega\rightarrow v \omega \,,\qquad\phi\rightarrow v\phi\, , \qquad  \mathcal{A}_{\alpha}\rightarrow v \mathcal{A}_{\alpha}\,,
	\end{equation}
with $v$ the $v.e.v.$ of the Higgs field,
	such that the field equations~\eqref{procafe}-\eqref{Einstein-eqs} become
	\begin{eqnarray}
		&&
		\nabla_\alpha\mathcal{F}^{\alpha\beta}=\phi^2\mathcal{A}^\beta \ ,
		\label{v2}
		\\
		&&
		\Box\phi
		=\lambda(\phi^2-1)\phi
		+\phi ~\mathcal{A}_\alpha\bar{\mathcal{A}}^\alpha \ ,
		\label{s2}
		\\
		&&
		R_{ \beta \gamma}-\frac{1}{2}Rg_{\beta \gamma}=2\alpha^2
		\left[
		T_{\beta \gamma}^{(v)}
		+
		T_{\beta \gamma}^{(s)}
		\right] \ ,
		\label{e2}
	\end{eqnarray}
	where we have defined the  dimensionless coupling constant 
	\begin{equation}
		\alpha^2\equiv 4\pi G v^2 \ .
	\end{equation}
	Such scaling reduces the number of free couplings/parameters in the action from 3 $(G,\lambda,v)$ to 2 $(\alpha,\lambda)$. This leads to the notion of the gravitational decoupling limit, which is particularly relevant when studying solitonic configurations on a fixed background. Using the limit $\alpha=0$, we shall both study solitons on Minkowski spacetime and clouds surrounding a Kerr black hole.

	We are looking for symmetric solutions under reflections along the equatorial plane, $\theta=\pi/2$. Effectively, we limit our equation-solving efforts to the range of $0 \leq \theta \leq \pi/2$, thereby imposing symmetry through reflection across the equatorial plane. Therefore, all solutions generated using this approach exhibit such symmetry, and we only need to apply the subsequent boundary conditions at the equatorial plane, $\theta=\pi/2$, 
	\begin{equation}\label{equat_plane}
		\partial_\theta F_i=\partial_\theta W=\partial_\theta H_1=H_2=\partial_\theta H_3=\partial_\theta V=\partial_\theta \phi=0 \ .
	\end{equation}
	
	The
	metric's asymptotic behavior, in terms of the ADM quantities, is given by
	\begin{equation}\label{ADM_metric}
		g_{t t}=-1+\frac{2 M_{\text{ADM}}}{r}+\ldots, \quad g_{\varphi t}=-\frac{2 J_{\text{ADM}}}{r} \sin ^2 \theta+\ldots \ .
	\end{equation}
	
	Under the scaling defined in \eqref{scale}, the Noether charge remains invariant
	\begin{equation}
		Q=\int_\Sigma d^3x \sqrt{-g} j^0 \ .
		\label{qint}
	\end{equation}
	On the other hand, the energy/mass of (non) gravitating configurations scales as $M\rightarrow Mv$. In the case of stars, $M$ and $J$ can be computed in two ways: $(1)$ the Komar mass and Komar angular momentum
	\begin{equation}\label{mint}
		M_{\text{matter}}\equiv - \int_\Sigma d^3x \sqrt{-g} \left(2T^{0}\,_{0}-T\right)\, , \qquad J_{\text{matter}}\equiv\int_\Sigma d^3x \sqrt{-g} T^{0}\,_{\phi}\, ,
	\end{equation}
	where $T$ is the trace of the \textit{total}  energy-momentum tensor. We use the subscript ``matter" which stands for the mass and momentum stored in the matter field (Proca + Higgs). This equation extends to the case of balls, where it simplifies to the standard integral over the energy density
\begin{equation}\label{mint2}
M_{\text{matter}}^{\rm balls}\equiv \int_\Sigma d^3x \sqrt{-g} \, T_{00} \ .
\end{equation}
This simplification is a consequence of the Deser/Virial identity for flat spacetime solitons, which dictates that the integral of the spatial trace on a spacelike slice is zero~\cite{Herdeiro:2022ids}; 
	and 
	$(2)$ the ADM mass and ADM angular momentum given that can be read off from the metric behavior at infinity \eqref{ADM_metric}.
	
	In the black hole case, we have additional contributions from the horizon. Hence, we need to compute the horizon mass and angular momentum by using the appropriate Komar integrals associated to the corresponding Killing vector fields $\xi$ and $\eta$, through Eqs. \eqref{mass_hor}.
	The angular momentum/mass formula is defined in \eqref{first_law} or \eqref{first_law2} and provides a way to physically estimate the numerical errors of the solutions. Due to the scaling, they are connected through
	\begin{equation}
		M_{\text{ADM}}=\dfrac{\alpha^2}{4\pi} M_{\text{matter}}+M_H\ , \qquad J_{\text{ADM}}=\dfrac{\alpha^2}{4\pi} J_{\text{matter}}+ J_H \ .
	\end{equation}
	
	Moreover, to judge the quality of the whole parameter space, we can use the differential form given by \eqref{differential}. For horizonless objects, it becomes
	\begin{equation}
		\dfrac{d\, M}{d\, Q}=\omega \ .
	\end{equation}
	
	The numerical solution of the equations is a crucial aspect of the study, and we have taken a rigorous approach to ensure the accuracy and reliability of our results. To obtain the solutions for Proca-Higgs balls, stars and black holes, we have solved a set of coupled non-linear partial differential equations for the functions $ \mathcal{F}= (F_i, W; \phi, H_i, V)$ (with $F_i = W = 0$ for balls). The boundary conditions required for these equations have also been carefully considered and applied as described in the aforementioned sections.
	
	To perform the numerical calculations, we have utilized \textsc{fidisol/cadsol} \cite{SCHONAUER1989279,SCHONAUER1990279,SCHONAUER2001473}, a professional software package that employs a finite difference method with an arbitrary grid and arbitrary consistency order. This solver has been proven to be reliable and efficient for solving complex non-linear problems, and has been extensively used in the scientific community for a variety of applications. The Newton-Raphson method is utilized in the solver, which requires a good first guess to start a successful iteration procedure. In the present work, we have utilized the pure Proca stars and hairy black holes as initial guesses. Further details about the solver can be found in the references~\cite{Herdeiro_2015,Delgado:2022pwo}, providing a more comprehensive explanation of the methodology used.

    After eliminating the time and azimuthal variables, the solutions are governed by two spatial coordinates. The equations are then discretized in a two-dimensional grid with $N = N_r \times N_\theta$ points. 	To simplify the numerical calculations and reduce computational costs, we introduce a compactified radial variable $x = r/(c + r)$ (with c being an input parameter typically set to one, but increased of decreased for better accuracy in the extreme parts of the domain) which maps the semi-infinite region $[0, \infty)$ to the finite region $[0, 1]$, which avoids the use of a cutoff radius\footnote{This approach remains valid for black holes after introducing the new radial coordinate defined in \eqref{newr}.}.  We then make the necessary substitutions.
	\begin{equation}
		\mathcal{F}_{, r} \longrightarrow \frac{1}{c}(1-x)^2 \mathcal{F}_{, x}, \quad \mathcal{F}_{, r r} \longrightarrow \frac{1}{c^2}(1-x)^4 \mathcal{F}_{, x x}-\frac{2}{c^2}(1-x)^3 \mathcal{F}_{, x} \ .
	\end{equation}
	
	This substitution has been applied inside the solver, allowing for more efficient calculations of the solutions. We have utilized an equidistant grid with a varying number of points, in $x$, ranging from (around) 360 to 400 (balls and stars/black holes, respectively), covering the integration region $0 \leq x \leq 1$, and 50 to 80 points in $\theta$ direction (stars/black holes and balls, respectively). The choice of grid spacing has been made carefully, taking into consideration the trade-off between accuracy and computational cost. Furthermore, we apply a sixth-order finite difference scheme. We have verified the convergence of the solutions with respect to the grid spacing, ensuring that our results are reliable and accurate.

	In addition to the careful choice of numerical methods and grid spacing, we have also employed several techniques to assess the quality and reliability of the computed solutions. The solver  provides error estimates for each unknown function, allowing us to adjust the accuracy of the numerical solutions. To further ensure the trustworthiness of our results, we have employed physical constraints to test the validity of the numerical solutions. These constraints include the equivalence between the ADM mass and the Komar mass, the virial identity, the first-law identity and the constraint equations, $i.e.$, the gauge condition and the unsolved Einstein's equations, which should be valid for numerical scheme consistency. By testing these constraints, we have confirmed the accuracy and reliability of our numerical solutions for Proca-Higgs balls and stars. We have found that the numerical error for the solutions reported in this work is estimated to be typically less than $10^{-5}$.

 In the preparation of this work, we have systematically constructed approximately 600 flat spacetime solutions (numerical points), 5\,000 stars solutions, 9\,000 black holes solutions, and over 10\,000 clouds solutions. These numerical points were subsequently interpolated to form a continuum for the plots described herein.

	\subsection{Overview of the constructed solutions} 

 Spinning Proca-Higgs balls are constructed in Sec. \ref{balls}, where we study some of their properties. Treating $\omega$, $m$ and $\lambda$ as input parameters, balls exist only in a certain frequency range,  $\omega<\omega_{\text{max}}=v$. For all computed solutions, there is also a minimal allowed frequency, $\omega_{\text{min}}$. For the Proca-Higgs balls with $m=0$ (spherical solutions) and small values of $\lambda$, we could not find this minimal frequency even in the fifth branch.

 In the complex free Proca model, nontopological solitons are precluded. As the parameter $\lambda$ increases, the Proca-Higgs model tends towards a quartic self-interacting Proca model, and for sufficiently large values of $\lambda$, it resembles a free Proca field. However, the introduction of self-interactions in the pure Proca model can lead to the formation of nontopological solitons \cite{Loginov:2015rya} (to our knowledge, rotating Proca balls were not yet presented in the literature). At a critical frequency within the interval $]\omega_{\text{min}}, \omega_{\text{max}}[$, both scalar and vector nontopological solitons reach their minimum mass and angular momentum. From this point, these values monotonically increase as the frequency approaches both the upper and lower limits. Proca-Higgs balls, on the other hand, seem to have a diverging mass/angular momentum only when $\omega\rightarrow\omega_{\text{max}}$. For rotating Proca-Higgs balls, as we decrease $\omega$, the numerics become too cumbersome and the fields start to have large amplitudes. Our findings indicate that all solutions for $m=0$ ball configurations can be effectively extended to a corresponding one with $m=1$, but the parameter space for the rotating balls has smaller values of $\omega$.
	
Differently then the free Proca/scalar model, the Proca-Higgs model is, on its most basic form, nonlinear, due to the coupling $\phi^2\mathcal{A}_\alpha\bar{\mathcal{A}}^\alpha$ on the Lagrangian, which spoils the possibility of separating variables \cite{Cayuso:2019ieu,Frolov:2018ezx}.
In the case without gravity ($\alpha=0$), we let the background be the Kerr black hole and solve the Proca-Higgs equation on this geometry, getting nonlinear Proca-Higgs clouds. Similarly to the scalar Q-clouds with self-interactions, the Proca-Higgs clouds parameter space is bounded by the ball and linear cloud limit. As the black hole vanishes, the cloud solution will tend to a corresponding ball solution. On the other hand, as the black hole becomes more massive, it reaches a point in which it cannot accommodate clouds anymore, only at the linear level (we call this upper bound existence line). Beyond such a point, there are only hairy black holes. 

Interestingly, both scalar \cite{Hod:2012px} and Proca \cite{Santos:2020pmh} clouds for fields without self-interactions only exist at the probe limit (at the existence line). Those models, at the same time, also do not admit flat spacetime balls.
In this sense, the Kerr geometry is not attractive enough to form bound states. As one can expect, for large enough $\lambda$, Proca-Higgs clouds also cease to exist at the nonlinear level. Our findings indicate that all solutions for $m=1$ ball configurations can be effectively extended or generalized to corresponding cloud configurations.

At the existence line, the Proca-Higgs fields exist only as test fields. In this way, for reasonable values of $\lambda$ (which should not be to small at the perturbation order), the existence line for the Proca-Higgs model coincides with the Proca one. Beyond the existence line, we have the fully nonlinear system $\alpha\neq 0$. Hairy black holes within this model are limited by two curves, the existence line, and by the corresponding stars.


	\section{Proca-Higgs Balls} \label{balls}
	
	We shall now present our numerical results on the solitonic solutions of \eqref{action} with the ansatz \eqref{matteransatz} and boundary conditions previously discussed. In this section, we considered only flat spacetime solutions in the decoupling limit with $\alpha = 0 = F_i = W$; we now include the effects of rotation and compare with the results in \cite{herdeiro2023procahiggs}. The equations are solved with the following set of boundary conditions

 	\begin{enumerate} [label=(\roman*)]
		\item at the origin, $r=0$,
		\begin{equation}
			H_i=V=	\partial_r\phi=0 \ ;
		\end{equation}
		
		\item 
  at infinity, $r=\infty$,
		\begin{equation}
			H_i=V=0\ , \qquad \phi=1 \ ;
		\end{equation}
		
		\item on the symmetry axis, $\theta=0, \pi$,
		\begin{equation}
			H_1=\partial_\theta H_2=\partial_\theta H_3=V=0\  , \qquad \partial_\theta\phi=0 \ .
		\end{equation}
	\end{enumerate}
	
	For the Proca-Higgs balls, the globally conserved quantities, Noether charge and mass/energy, become
	\begin{equation}
		Q=2 \pi \int_0^{\infty}\int_0^{\pi} \left[\omega\dfrac{H_i H_i}{r^2}+\frac{m \csc \theta  H_3 V}{r^2}+\dfrac{H_i\nabla_i V}{r} \right]r^2\sin\theta  d r\ d \theta \ ,
	\end{equation}
	\begin{equation}
		M=2 \pi \int_0^{\infty}\int_0^{\pi}\left\{\dfrac{1}{2} \left[|\vec{E}|^2+|\vec{B}|^2+\phi^2\left(|\mathcal{A}_0|^2+|\mathcal{\vec{A}}|^2\right)\right]+|\vec{\nabla}\phi|^2+U\right\}r^2\sin\theta  d r\ d \theta \ ,
	\end{equation}	
	where $H_i$ are the components of the vector potential as given in \eqref{matteransatz}, the summing index $i=1,2,3$ and $\nabla=(\partial_r,r^{-1}\partial_{\theta})$. Also, $|\vec{E}|^2=-\mathcal{F}_{0i}\mathcal{\bar{F}}^{0i}$, $|\vec{B}|^2=\dfrac{1}{2}\mathcal{F}_{ij}\mathcal{\bar{F}}^{ij}$ and $\mathcal{A}=\{\mathcal{A}_0,\mathcal{\vec{A}}\}$.
	
	These solutions obey the virial identity
	\begin{equation}
		\int_0^{\infty}\int_0^{\pi}\left\{\dfrac{1}{2}|\vec{\nabla}\phi|^2+ 3 U + \dfrac{1}{2} \left[-|\vec{E}|^2-|\vec{B}|^2+\phi^2\left(-3|\mathcal{A}_0|^2+|\mathcal{\vec{A}}|^2\right)\right]\right\}r^2\sin\theta  d r\ d \theta = 0 \ ,
	\end{equation}
	which can be used to test the numerical accuracy of the solutions. All solutions reported in this work obey the
	virial identity up to errors of typical order $10^{-5}\, -\, 10^{-7}$.

\begin{figure}[!ht]
		\centering
		\includegraphics[width=0.47\textwidth]{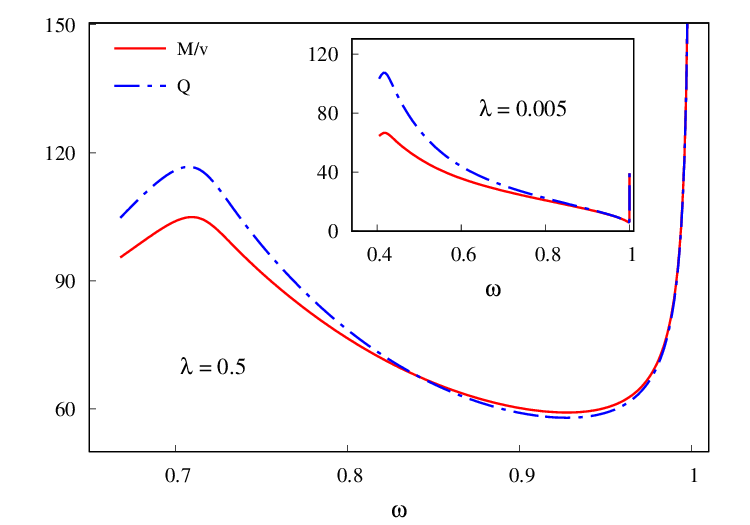}
		\includegraphics[width=0.47\textwidth]{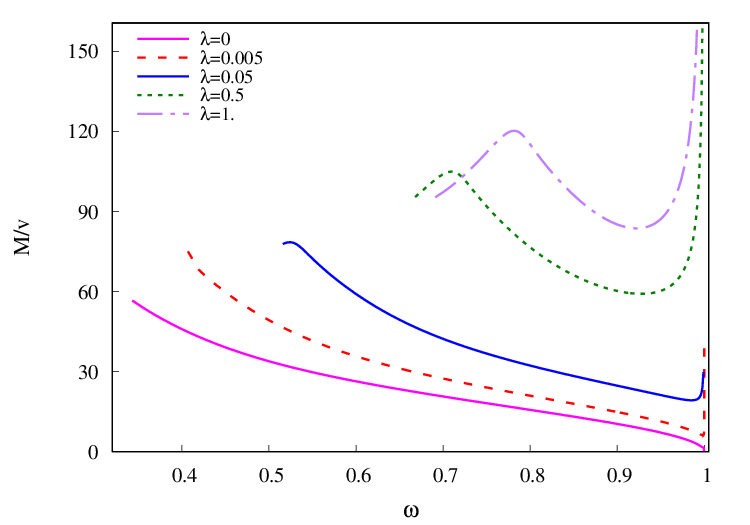}
		\includegraphics[width=0.47\textwidth]{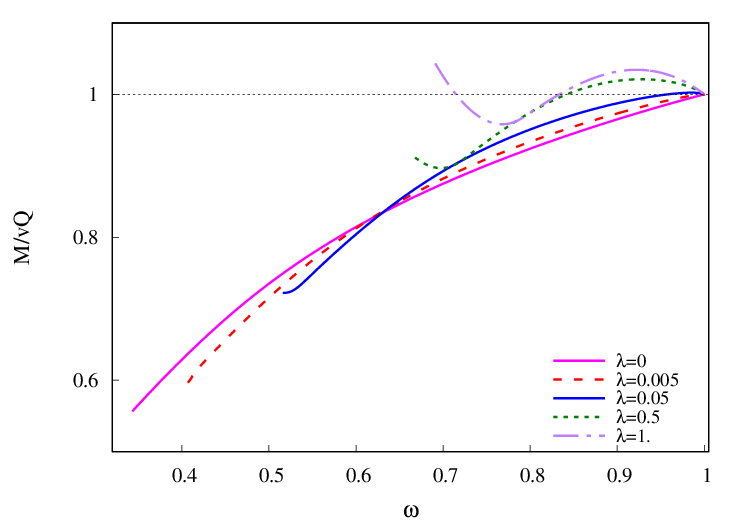}
         \includegraphics[width=0.47\textwidth]{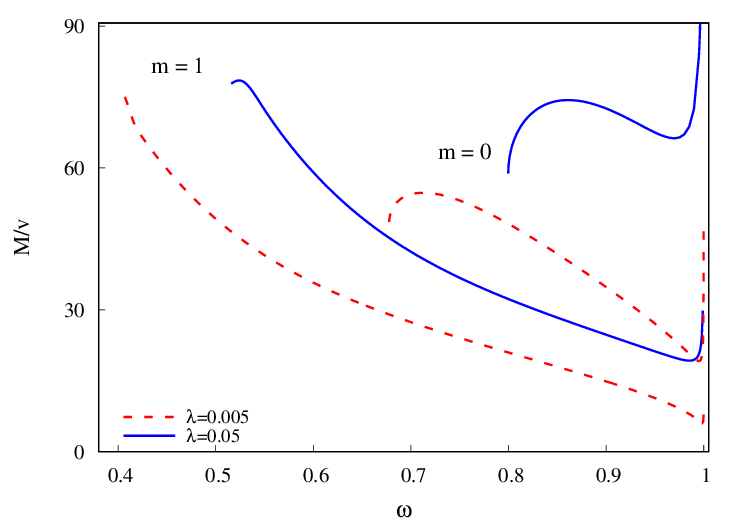}
		\caption{(Top left panel)  The mass/energy (red solid line) and Noether charge (blue dashed line) of the Proca-Higgs balls are plotted as a function of their frequency $\omega$ for two representative values of $\lambda$.
  (Top right panel) Variation of the Proca-Higgs balls mass vs. $\omega$ curve for different $\lambda$. (Bottom left panel) $M/vQ$ plotted against  $\omega$. For the first branch and small values of the self-interacting parameter, the Noether charge is always greater than the mass. 
   However, as we increase $\lambda$, the ratio $M/vQ$ might be greater than one. This suggests instability against fission. (Bottom right panel) A comparison of the M vs. $\omega$ diagram for balls with winding number $m=1$ and $m=0$ (spherical solutions), for two different values of $\lambda$.}
		\label{fig:massfreqcharge_flat}
	\end{figure}

 We study the frequency dependence of rotating Proca-Higgs balls with even parity, showing that their mass and Noether charge increases as $\omega$ goes to 1, tending to diverge. Moreover, these quantities attain a local maximum which, differently from rotating Q-balls, does not occur at the minimum allowed frequency \cite{Kleihaus:2005me,PhysRevD.66.085003}.  Additionally, we conjecture the existence of a lower frequency limit $\omega_{\text{min}}$, as observed in Fig.~\ref{fig:massfreqcharge_flat}, but numerical precision limitations prevent us from approaching $\omega_{\text{min}}$. In such regions, the Proca components become larger and larger. We display solutions with $\lambda=0, 0.005, 0.05, 0.5, 1$ in Fig. \ref{fig:massfreqcharge_flat}
	
	As we increase the self-interacting parameter, $\lambda$, we observe that the minimum frequency in the curve $\omega$ vs. M grows larger, resulting in a shorter existence line of Proca-Higgs balls. This behavior is expected as the model approaches the pure Proca model, which in previous literature has shown to not allow for flat spacetime solutions for the considered Ansatz. Our numerical findings corroborate this, indicating that the solutions cease to exist past some particular value of $\lambda$ that depends on the frequency $\omega$. Also, we have found that the spinning  solutions exist for larger values of $\lambda$ as in the spherical case.
 
\begin{figure}[h!]
		\centering
		\includegraphics[width=0.47\textwidth]{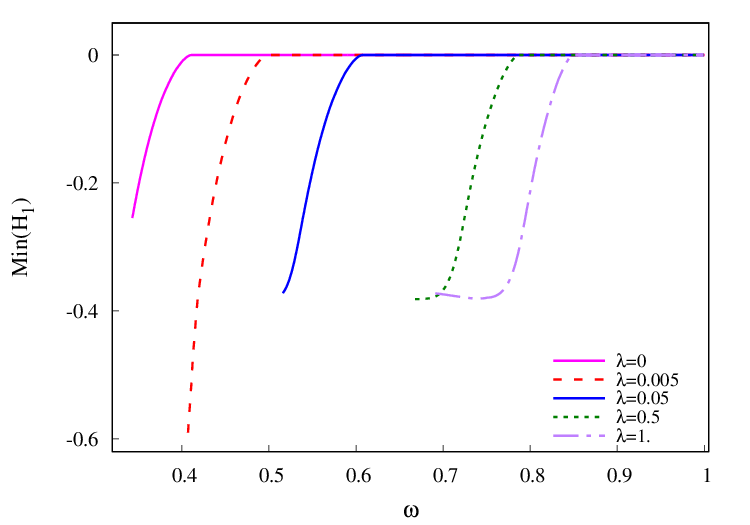}
		\includegraphics[width=0.47\textwidth]{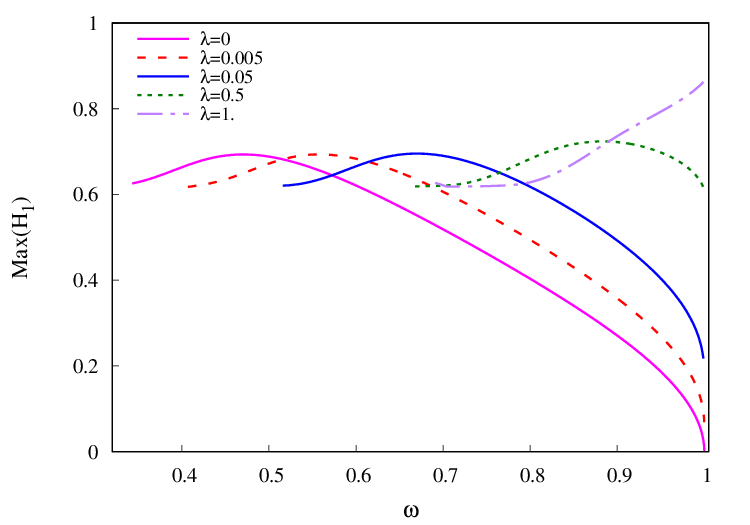}
		\includegraphics[width=0.47\textwidth]{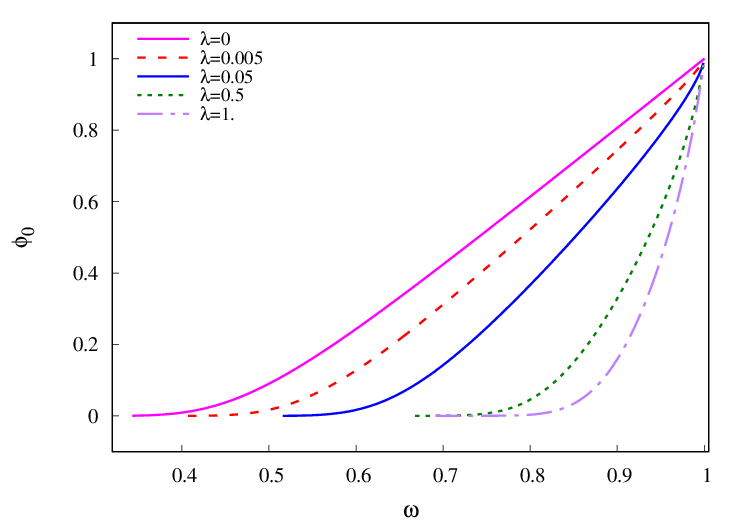}
        \includegraphics[width=0.47\textwidth]{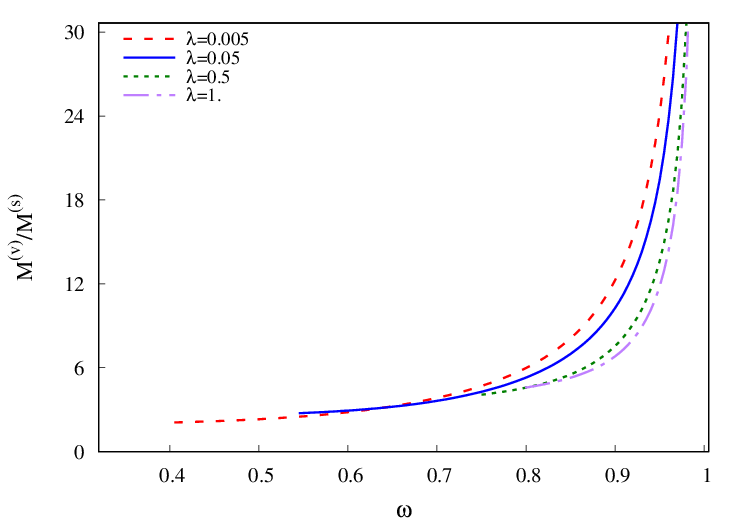}
		\caption{(Top panels) The minimum and maximum value of the function $H_1$. (Bottom panels) The scalar field at the origin, $\phi_0$, and the ratio between the mass in the Proca field and in the scalar field.}
		\label{fig:solutions_origin_flat}
	\end{figure}

\begin{figure}[H]
		\centering
		\includegraphics[width=0.47\textwidth]{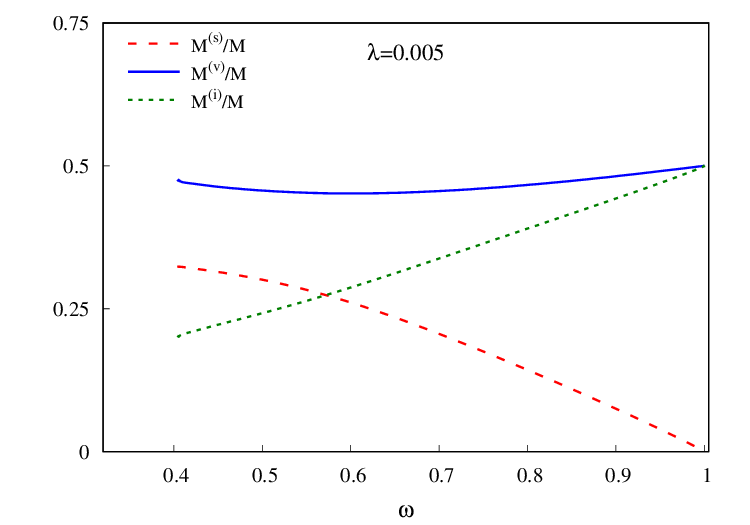}
		\caption{ The scalar field, vector field and interaction term contributions to the total mass are shown as a function
  of frequency for a typical family of spinning Proca-Higgs balls. 
  }
		\label{fig:solutions_split}
	\end{figure}

 In Fig.~\ref{fig:solutions_origin_flat} the minimum and maximum value of the function $H_1$ are displayed (top panels). We observe this function to possess nodes across all computed parameter spaces irrespective of the self-interacting parameter, $\lambda$. Notably, while the existence of nodes is independent of $\lambda$, the frequency at which these nodes occur does depend on $\lambda$. In the case of pure Proca stars, a monotonically increasing function was used to characterize the solutions, namely the maximum of $H_1$. However, as we shall observe, this is not the case for Proca-Higgs balls. Instead, we note that $\phi_0$ (the value of the scalar field at the origin), similarly to the Q-ball case, is a monotonic function that can be utilized to characterize the solutions, shown in the bottom left panel. In the bottom right panel the ratio between the mass contained in the Proca field and in the scalar field is shown. 	Notice that for all computed solutions, the Proca field carries most of the energy. This is further specified in Fig.~\ref{fig:solutions_split}.
 
\begin{figure}[H]
        \centering
		\includegraphics[width=0.47\textwidth]{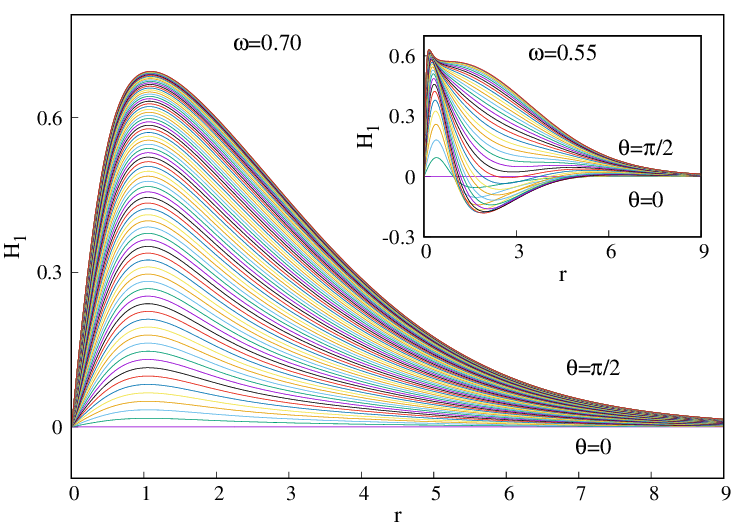} 
  		\includegraphics[width=0.47\textwidth]{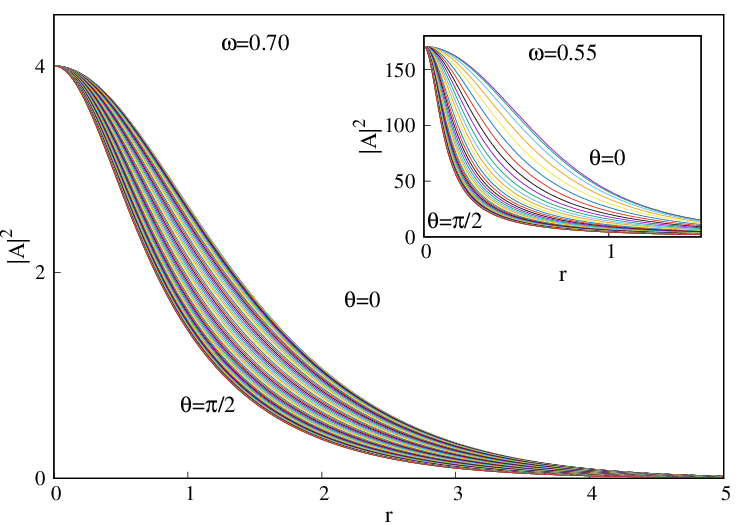}
		\includegraphics[width=0.47\textwidth]{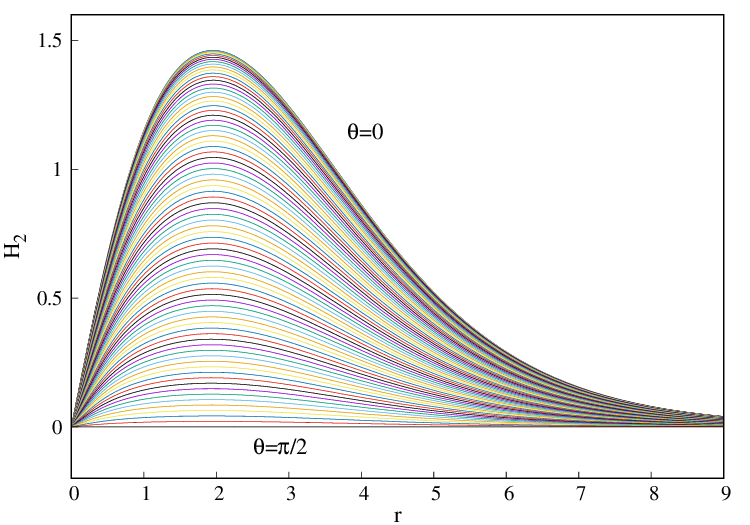}
		\includegraphics[width=0.47\textwidth]{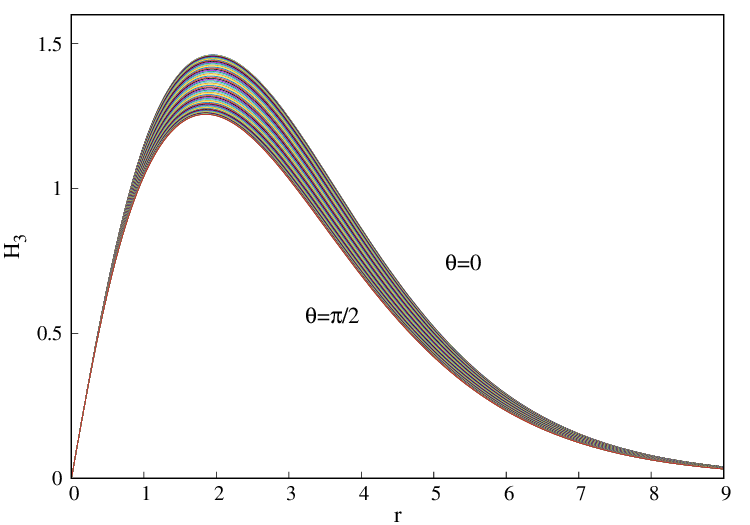}
		\includegraphics[width=0.47\textwidth]{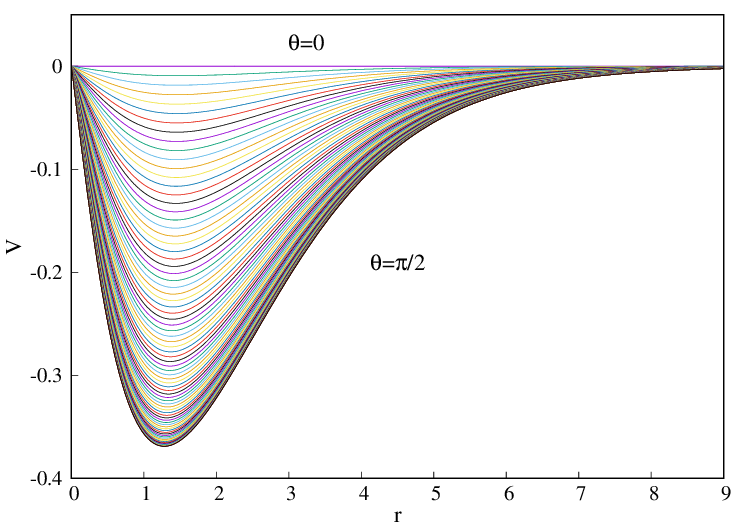}
		\includegraphics[width=0.47\textwidth]{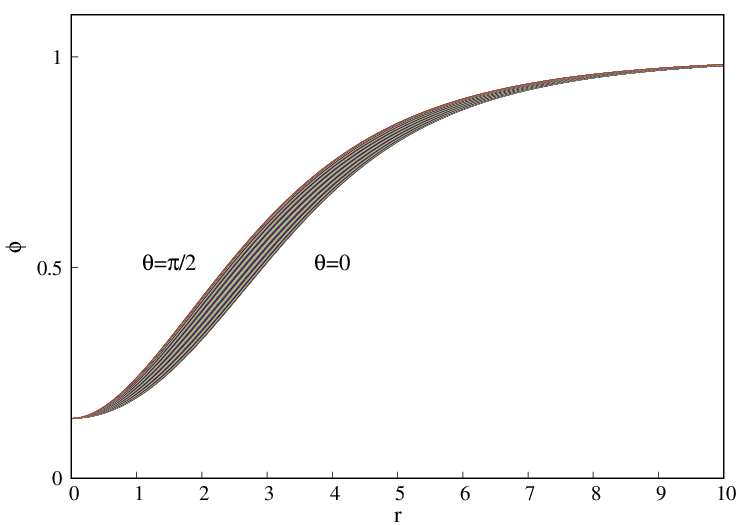}
		\caption{ The radial profile of the matter functions for a given solution with parameters $\lambda=0.05$, $\omega=0.70$.}
		\label{fig:functionslamb005}
	\end{figure}

 Finally, in Fig.~\ref{fig:functionslamb005}, the radial profile of the matter functions for a given solution with parameters $\lambda=0.05$, $\omega=0.70$ is shown. We plot the functions for each evaluated angle $\theta$ in the grid, $0\leq \theta \leq \pi/2$. In the top panels, an inset illustrates $H_1$ and $|A|^2$ for a distinct solution with parameters $\lambda=0.05$ and $\omega=0.55$. The other functions for this solution are not shown as they exhibit qualitative similarities to the solution with $\omega=0.70$. Notably, $H_1$ tends to develop nodes with decreasing frequency for the solutions under examination. Interestingly, although $H_1$ might have nodes, $|A|^2$ is nodeless. 
 
\section{Proca-Higgs Stars}
	
	Let us now present the numerical results for the rotating Proca-Higgs stars. We use the same numerical framework as in the case of balls. The solutions presented here have typical errors of order $10^{-4}-10^{-6}$. Alongside the condition specified in equation \eqref{equat_plane}, the equations are subject to the following set of boundary conditions
	
	\begin{enumerate} [label=(\roman*)]
		\item  at the origin, $r=0$,
		\begin{equation}
			\partial_r F_i=W=H_i=V=	\partial_r\phi=0 \ ;
		\end{equation}
		
		\item at infinity, $r=\infty$,
		\begin{equation}
			F_i=W=H_i=V=0\ , \qquad \phi=1 \ ;
		\end{equation}
		
		\item  on the symmetry axis, $\theta=0, \pi$,
		\begin{equation}
			\partial_\theta F_i=\partial_\theta W=H_1=\partial_\theta H_2=\partial_\theta H_3=V=0\  , \qquad \partial_\theta\phi=0 \ .
		\end{equation}
	\end{enumerate}

	We emphasize that for stars, the line element given in equation \eqref{line_BH} undergoes a minor modification, represented by the substitution $W \rightarrow r W$. In addition, the absence of conical singularities also means that, along the axis of symmetry, $\theta=0, \pi$,
	\begin{equation}
		F_1=F_2 ~,
	\end{equation}
  a condition which is satisfied at the level
 of the overall numerical accuracy.
	
We explored the Proca-Higgs stars parameter space, spanned by $(\lambda, \alpha)$ and $\omega$, studying the variation of some key physical properties, including mass and compactness (cf.~(\ref{compact}) below). The left panels of the Fig. \ref{fig:massfreqcharge} illustrate the mass $M$ as a function of the angular frequency $\omega$, highlighting the first solutions with an ergoregion and light ring (LR). The right panels show the inverse compactness as a function of $\omega$, which tends unity (thus to the compactness of a black hole), suggesting black hole foils behaviour. 	 As usual in bosonic stars, the inclusion of gravitational effects (as compared to balls), leads to a regularization of the mass of the solutions as they approach the maximal frequency $\omega\approx 1$, known as the Newtonian limit, where solutions become less compact and their mass asymptotically tending towards zero.

\begin{figure}[H]
        \centering
		\includegraphics[width=0.45\textwidth]{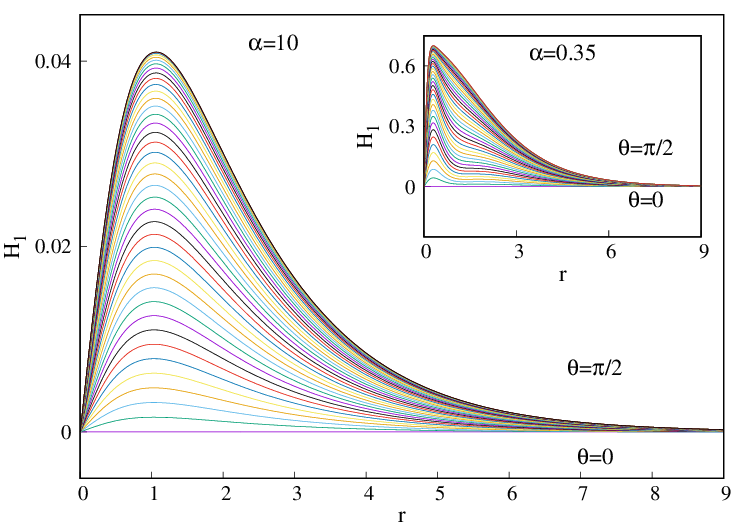} 
		\includegraphics[width=0.45\textwidth]{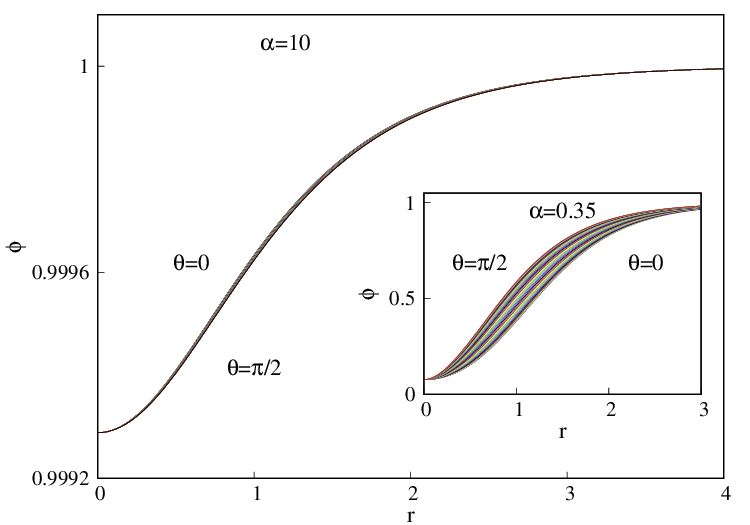}
		\includegraphics[width=0.45\textwidth]{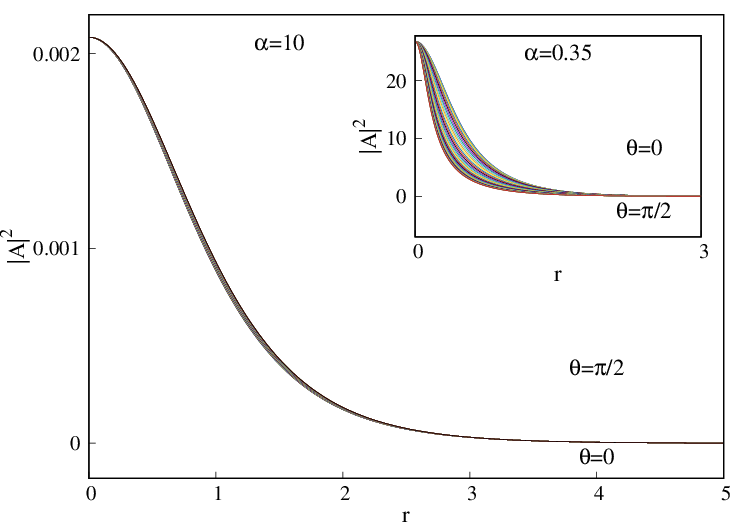}
		\caption{ The radial profile of the matter functions for a given solution with parameters $\alpha=10$, $\lambda=1$, $\omega=0.70$ and $m=1$. We plot the functions for each evaluated angle $\theta$ in the grid, $0\leq \theta \leq \pi/2$. The insets show the radial profiles for a given solution with parameters $\alpha=0.35$, $\lambda=1$, $\omega=0.70$.}
		\label{fig:functionsalpha10}
	\end{figure}

\begin{figure}[H]
		\centering
		\includegraphics[width=0.40\textwidth]{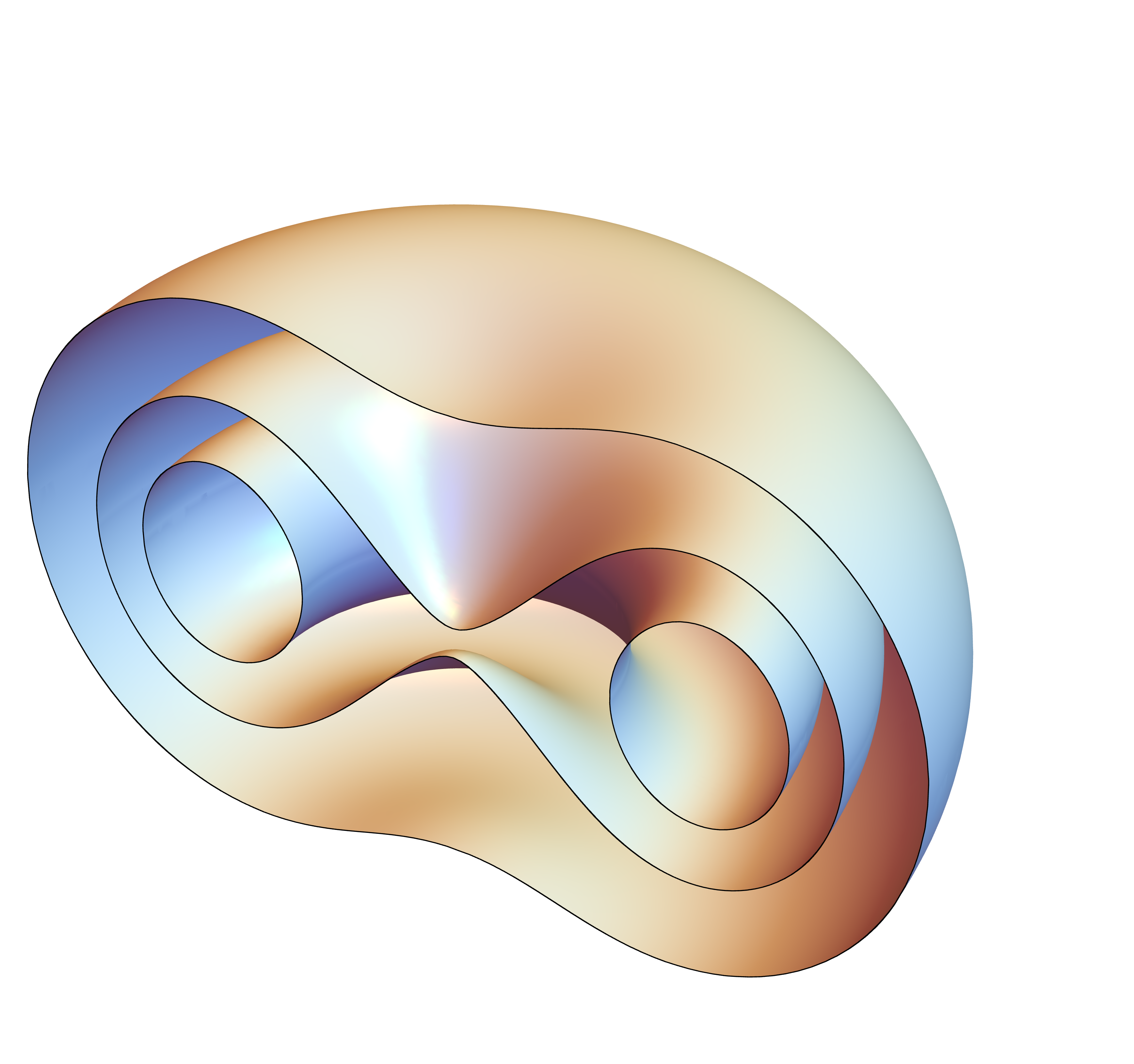} 
		\includegraphics[width=0.40\textwidth]{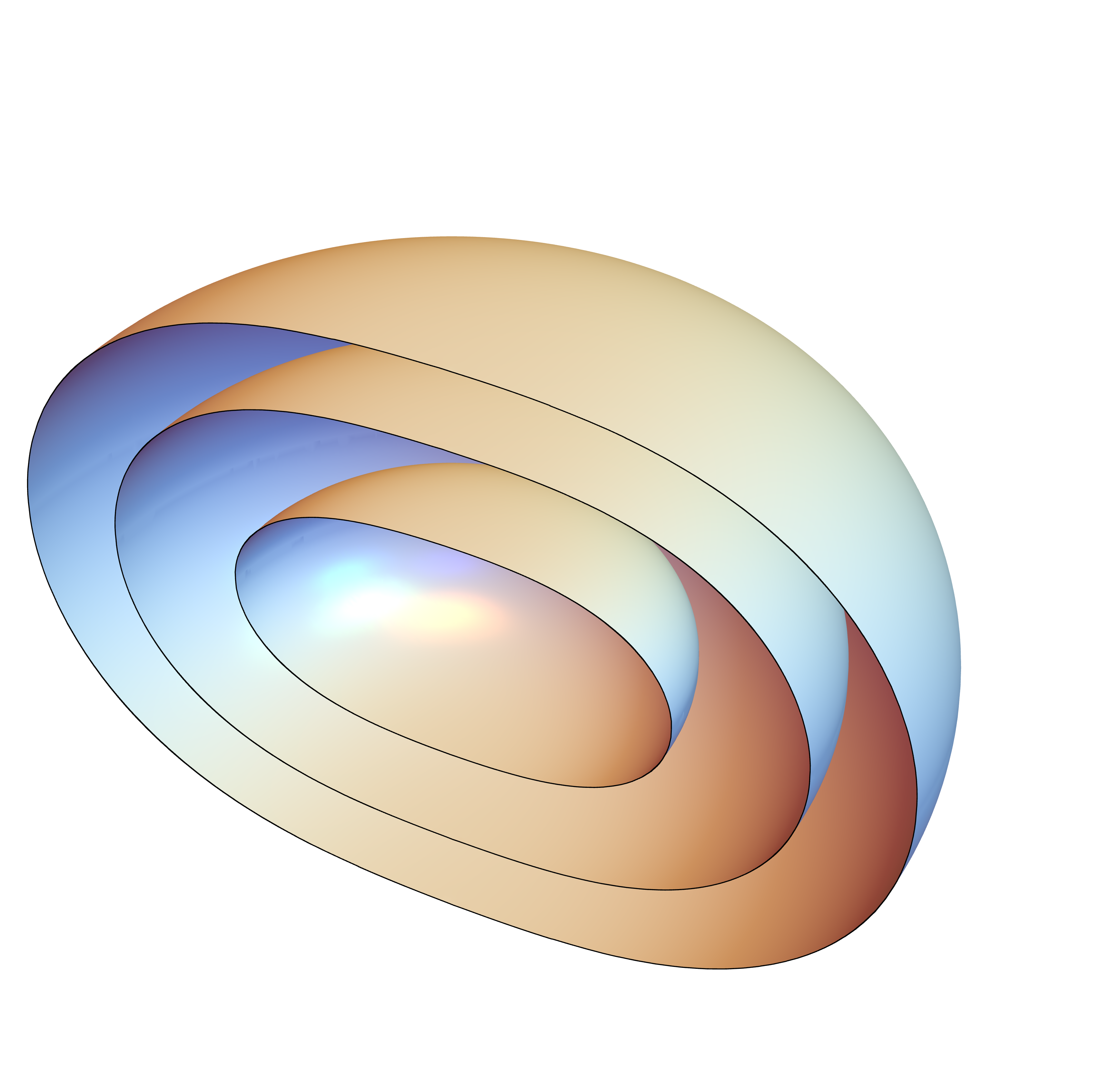}
		\caption{Morphology of the surfaces of constant Komar mass density  for two different solutions, with parameters, $\alpha=0.35,\; \lambda=1,\; m=1 \text{ and } \omega=0.70$ (left panel) and        $\alpha=10,\; \lambda=1,\; m=1 \text{ and } \omega=0.70$ (right panel).}
		\label{fig:isomass}
	\end{figure}

 In our study of static, spherically symmetric Proca-Higgs stars \cite{herdeiro2023procahiggs}, we have found that mini-Proca stars act as a maximum limit for mass and Noether charge. Essentially, we have not found any solutions where these Proca-Higgs stars are heavier than their mini-Proca counterparts. However, this upper bound does not hold in the case of rotating Proca-Higgs stars. Specially for small $\lambda$ and $\omega$. For large values of $\alpha$ and $\lambda$, the Higgs field becomes frozen to the vacuum expectation value, causing the mass of Proca-Higgs stars to converge with that of rotating mini-Proca stars. We have systematically explored the existence domain for rotating Proca-Higgs stars by varying the frequency $\omega$  across various $\alpha$ and $\lambda$ values - Fig. \ref{fig:massfreqcharge}. In Fig. \ref{fig:functionsalpha10}, one can see typical functions $H_1$, $\phi$ and $|A|^2$. Notice that $|A|^2$ is nodeless. 	Finally, to provide insight into the structure  of Proca-Higgs spinning stars in Fig.~\ref{fig:isomass} we show the morphology of the surfaces of constant Komar mass density  for two solutions. These surfaces can be either toroidal or spheroidal, depending on the specific solution and the chosen energy density value.  
 
	\begin{figure}[H]
		\includegraphics[width=0.47\textwidth]{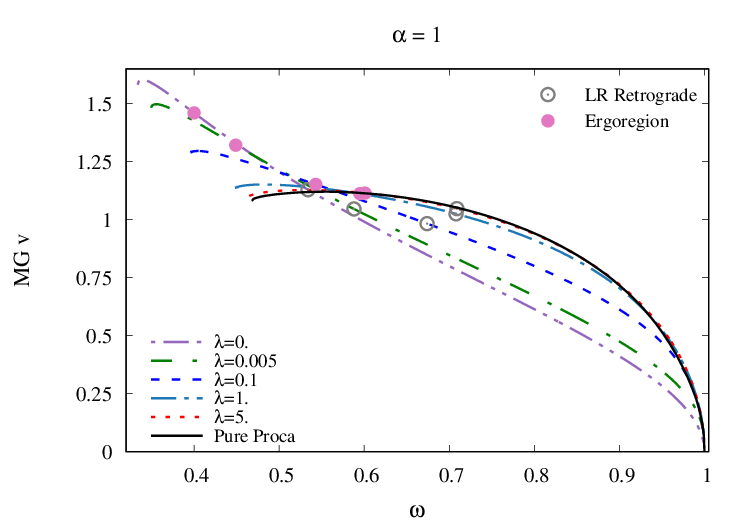}
		\includegraphics[width=0.47\textwidth]{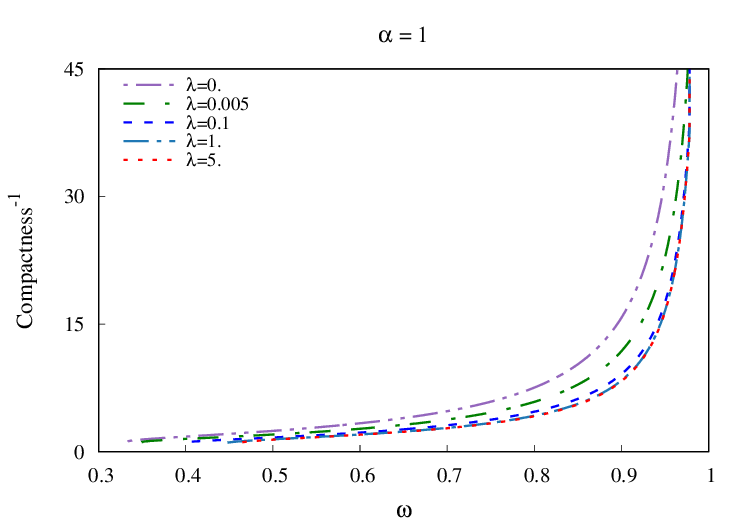}
        \includegraphics[width=0.47\textwidth]{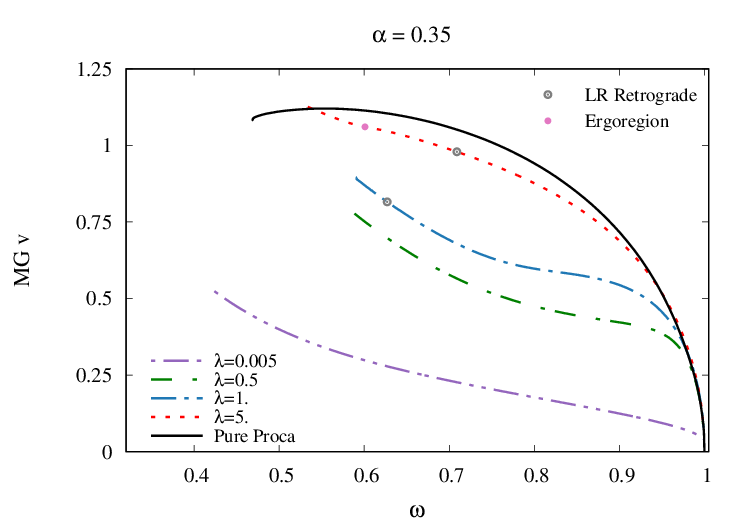}
		\includegraphics[width=0.47\textwidth]{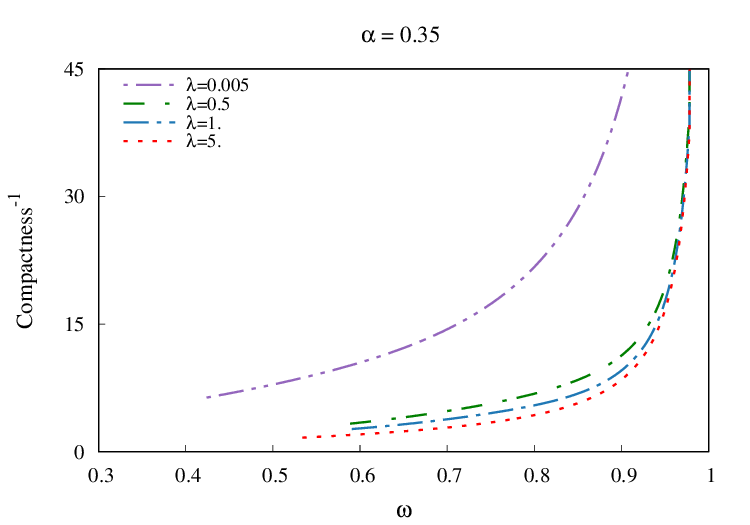}
        \includegraphics[width=0.47\textwidth]{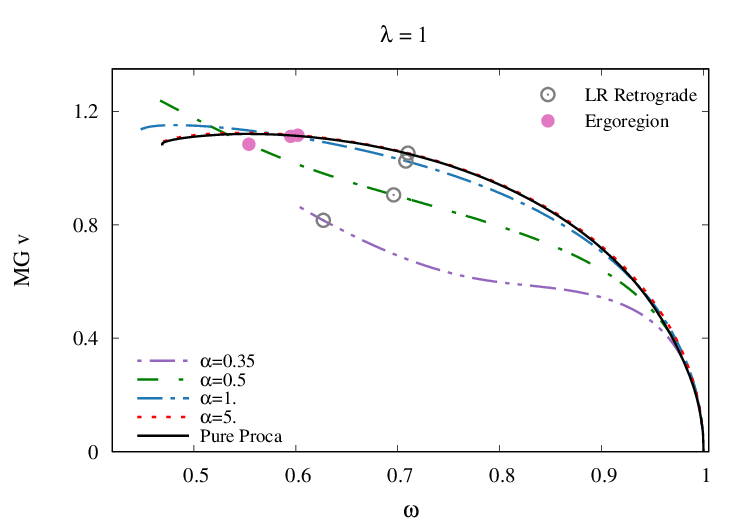}
        \hfill
		\includegraphics[width=0.47\textwidth]{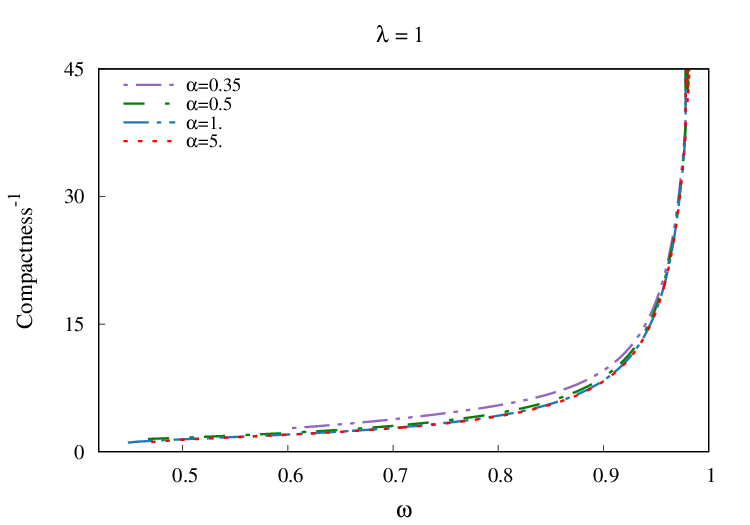}
		\caption{Left panels: mass of the Proca-Higgs stars as a function of their frequency $\omega$ for illustrative values of $\lambda$ and two different values of $\alpha$ (top and middle rows) and for different illustrative values of $\alpha$ and a fixed $\lambda$ (bottom row).  Right panel: the inverse compactness as a function of the angular frequency $\omega$.}
		\label{fig:massfreqcharge}
	\end{figure}

	\subsection{Compactness, some special geodesics and ergoregions}

 Black holes have peculiar geodesic properties. In the realm of astrophysics, black hole mimickers are expected to exhibit characteristics similar to black holes, when observed. This possible degeneracy underscores the importance of analyzing their geodesic structures. Key features such as the presence of an innermost stable circular orbit (ISCO) for timelike geodesics and light rings (LRs) for null geodesics serve as indicators of how effectively these mimickers can replicate the behavior of true black holes. Nonetheless, these characteristics do not consistently appear in bosonic stars (scalar or vector). For example, spherical mini-boson stars or mini-Proca stars (generically) lack an ISCO, but the former spinning counterparts may possess one \cite{Cao:2016zbh}, while the spinning mini-Proca stars do not have an ISCO.  Conversely, mini-boson stars, whether static or spinning, and static mini-Proca stars only exhibit LRs under sufficiently strong gravity \cite{Cunha:2015yba,Grandclement:2016eng,Herdeiro:2021lwl}, a regime in which they present instabilities. Nevertheless, even solutions without ISCO or LR, under certain observation and astrophysical conditions, may mimic the shadow of a corresponding black hole \cite{Rosa:2022tfv,Herdeiro:2021lwl,Olivares:2018abq,Sengo:2022jif,Sengo:2024pwk}.

 For all stars discussed in this paper, no solution exhibits a prograde LR neither an ISCO. However, retrograde LRs do appear in strong gravity scenarios - see also~\cite{PhysRevD.105.064026}. Along the solution line, once the first LRs appear, all further solutions will also have LRs. Moreover, LRs appear in pairs, in agreement with the general result in~\cite{Cunha:2017qtt}.

	\subsubsection{Compactness and Circular geodesics}

 	As we move along the stable branch from the Newtonian limit to the maximal mass, both scalar boson stars and Proca stars tend to become more compact. The concept of compactness is commonly defined in terms of the effective (areal, in the spherical case, or circumferential, in the axial cases) radius $R_{99}$, which encloses $99\%$ of the total mass of the star, following the literature (see, for example, \cite{Herdeiro_2015}). It is worth noting that the areal/circunferential radius R has a geometric interpretation and is related to the corresponding radius in isotropic coordinates, $r$, through $R_{99} = e^{F_2} r_{99}$. The inverse compactness can then be defined as
	\begin{equation}
		\text { Compactness }^{-1} \equiv \frac{R_{99}}{2 M_{99}} \ .
  \label{compact}
	\end{equation}
	 
	Moving to geodesics, suppose there is a test particle that orbits the star  (see \cite{PhysRevD.105.064026} for a more detailed discussion). The metric Ansatz of the metric produced by the central body can be described as follows
	\begin{equation}
		d s^2=g_{t t} d t^2+2 g_{t \varphi} d t d \varphi+g_{\varphi \varphi} d \varphi^2+g_{r r} d r^2+g_{\theta \theta} d \theta^2 \ .
	\end{equation}
	
	The motion of the test particle in a generic geometry is described by an effective Lagrangian,
	\begin{equation}
		\mathcal{L}=\frac{1}{2} g_{\mu \nu} \dot{x}^\mu \dot{x}^\nu=k \ ,
	\end{equation}
	where dots represent derivatives with respect to an affine parameter (which is equivalent to the proper time in the time-like scenario) and where $k = -1,\ 0,\ +1$ for time-like, null and spacelike geodesics, respectively. This Lagrangian density governs the particle's movement in the given geometry. Neglecting radiation reaction results in the orbit of a small object being determined by this Lagrangian. It can be interpreted as a conservation law that arises from the standard normalization condition for the object's four-velocity.
	
	Due to the spacetime symmetries, we have two conserved quantities: the energy $E$ and the angular momentum $L$ of the particle,
	\begin{equation}
		\begin{aligned}
			& E=-\xi_\mu \ \dot{x}^\mu= -g_{t t} \dot{t}-g_{t \varphi} \dot{\varphi}\ , \\
			& L=\eta_\mu \  \dot{x}^\mu =g_{t \varphi} \dot{t}+g_{\varphi \varphi} \dot{\varphi}\ .
		\end{aligned}
	\end{equation}
	
	By further assuming that the orbit is restricted to the equatorial plane, $\theta=\pi/2$, and by imposing the condition of constant orbital radius, $r=r^{\text{cir}}$ and $\dot{r}=\ddot{r}=0$, one can algebraically solve the equations of motion 
	\begin{equation}
		V_{\mathrm{k}}(r)= k+\frac{g_{\varphi \varphi} E^2+2 g_{t \varphi} E L+g_{t t} L^2}{g_{t \varphi}^2-g_{t t} g_{\varphi \varphi}},
	\end{equation}
	which simplifies to
	\begin{equation}\label{circular_orbit}
			 V_k\left(r^{c i r}\right)=0 \ ,\qquad
			 V_k^{\prime}\left(r^{c i r}\right)=0 \ .
	\end{equation}
 Furthermore, the sign of $V''(r^{\text{cir}})$ determines the radial stability of the circular orbit:
	\begin{equation}
		V_{k}^{\prime \prime}\left(r^{\mathrm{cir}}\right)>0 \Leftrightarrow \text { unstable ; } \quad V_{k}^{\prime \prime}\left(r^{\mathrm{cir}}\right)<0 \Leftrightarrow \text { stable . }
	\end{equation}
	
	Let us first consider the time-like case. For TCOs, the angular velocity of the particle along the geodesic is
	given by
	\begin{equation}
		\Omega_{\pm}=\frac{d \varphi}{d t}=\frac{-g_{t \varphi, r}\pm\sqrt{\left(g_{t \varphi, r}\right)^2-g_{t t, r} g_{\varphi \varphi, r}}}{g_{\varphi \varphi, r}}\Bigg|_{r=r^{\text{cir}}}\ .
	\end{equation}
	In this context, $\Omega_+$ denotes the value of $\Omega$ for prograde circular orbits, while $\Omega_-$ represents the value for retrograde circular orbits. Notice however that if $\left(g_{t \varphi, r}\right)^2-g_{t t, r} g_{\varphi \varphi, r}<0$ the orbits cease to exist.
	
	The ISCO is the smallest radius at which a circular orbit are still stable around a central body. The ISCO is an important concept in the study of black holes and compact objects because it determines the minimum distance that matter can orbit in a stable manner the black hole, below it, matter should plunge into the central object, emitting significant amounts of  radiation in the process. Therefore, the study of the ISCO is critical for understanding the behavior and properties of black holes and other compact objects, as well as for detecting gravitational and electromagnetic waves emitted by them. Mathematically, the ISCO is then defined as
	\begin{equation}
		\left[\left(g_{t \varphi, r}\right)^2-g_{t t, r} \ g_{\varphi \varphi, r}\right]_{r^{\text{ISCO}}}=0, \quad \text { and } \quad V_{-1}^{\prime \prime}\left(r^{\text{ISCO}}+|\delta r|\right)<0 \ .
	\end{equation}
	
	We now consider null geodesics. For LRs, $k = 0$ , the combination of the conditions \eqref{circular_orbit} yields to
	the algebraic equation
	\begin{equation}
		g_{\varphi \varphi} \sigma_{\pm}^2+2 g_{t \varphi} \sigma_{\pm}+g_{t t}=0 \ ,
	\end{equation}
	where the inverse impact parameter is defined as
	\begin{equation}
		\sigma_{\pm}=\dfrac{E_{\pm}}{L_{\pm}}=\left[\frac{-g_{t \varphi} \pm \sqrt{g_{t \varphi}^2-g_{t t} g_{\varphi \varphi}}}{g_{\varphi \varphi}}\right]_{\mathrm{LR}} \ .
	\end{equation}
	
	\subsubsection{Ergoregion}
	
	According to the definition of stationarity, the Killing vector field $\xi$ must be time-like in the vicinity of infinity. However, if $\xi$ is found to be spacelike in certain parts of the domain, designated by ergoregion. Its boundary, the ergo-surface, is the surface wherein $\xi^2$ equals zero, i.e.
	\begin{equation}
		g_{tt}=-N e^{2 F_0}+\dfrac{W^2}{r^2} e^{2 F_2} \sin ^2 \theta=0 \ .
	\end{equation}
The presence of an ergoregion can potentially mean an  instability due to the Penrose process or its field theory version, superradiance~\cite{penrose1969,PhysRevD.77.124044,Brito:2015oca,East:2017mrj,East:2017ovw}. 

In \cite{hawking_ellis_1973}, the authors prove that a stationary (non static) black hole in vacuum, inevitably possesses an ergoregion. This is not generically the case for spinning bosonic stars - see e.g.~\cite{Kleihaus:2007vk,Herdeiro:2014jaa}, including the Proca-Higgs stars reported here. However, we numerically show that some configurations give rise to an ergoregion - see Fig.~\ref{fig:massfreqcharge} and Table~\ref{table}.

For the black holes reported in the next section, on the other hand, in the vicinity of the horizon,  $g_{tt}$ is positive due to the boundary conditions. However, it will be negative at infinity, due to asymptotic flatness. Therefore, our black hole solutions will always have at least one ergoregion. Additionally, and similarly to black holes endowed with scalar hair \cite{DELGADO2019436}, in the vicinity of the star limit ($r_H$ going to zero), ergo-spheres occur in hairy black hole configurations neighbouring stars (in the parameter space) without ergoregions. On the other hand, ergo-Saturns emerge exclusively within lower frequency domains, adjacent to stars that possess an ergo-torus~\cite{Herdeiro:2014jaa}.

\begin{table}
\centering
\begin{tabular}{|cccccccc|}
\hline
\multicolumn{2}{|c|}{}     &         \multicolumn{2}{||c||}{Ergoregion}  & \multicolumn{2}{|c|}{LR}             \\ \hline
\multicolumn{1}{|c|}{$\alpha$} & \multicolumn{1}{c|}{$\lambda$} & \multicolumn{1}{||c|}{$ \omega$} & \multicolumn{1}{c||}{Mass}  & \multicolumn{1}{c|}{$ \omega$} & \multicolumn{1}{c|}{Mass} \\ \hline
\multicolumn{1}{|c|}{5}     & \multicolumn{1}{c|}{1}        & \multicolumn{1}{||c|}{0.602 }    & \multicolumn{1}{c||}{1.116} & \multicolumn{1}{c|}{0.710}         & \multicolumn{1}{c|}{1.053}                 \\
\multicolumn{1}{|c|}{1}     & \multicolumn{1}{c|}{5}       & \multicolumn{1}{||c|}{0.601}    & \multicolumn{1}{c||}{1.115} & \multicolumn{1}{c|}{0.709}         & \multicolumn{1}{c|}{1.048}   \\
\multicolumn{1}{|c|}{1}     & \multicolumn{1}{c|}{1}       & \multicolumn{1}{||c|}{0.595}    & \multicolumn{1}{c||}{1.112} & \multicolumn{1}{c|}{0.708}         & \multicolumn{1}{c|}{1.025}                \\
\multicolumn{1}{|c|}{1}     & \multicolumn{1}{c|}{0.1}       & \multicolumn{1}{||c|}{0.543}    & \multicolumn{1}{c||}{1.152} & \multicolumn{1}{c|}{0.674}         & \multicolumn{1}{c|}{0.983}                \\
\multicolumn{1}{|c|}{1}     & \multicolumn{1}{c|}{0.005}       & \multicolumn{1}{||c|}{0.449}    & \multicolumn{1}{c||}{1.321} & \multicolumn{1}{c|}{0.588}         & \multicolumn{1}{c|}{1.046}                \\
\multicolumn{1}{|c|}{1}     & \multicolumn{1}{c|}{0}       & \multicolumn{1}{||c|}{0.400}    & \multicolumn{1}{c||}{1.460} & \multicolumn{1}{c|}{0.534}         & \multicolumn{1}{c|}{1.129}                \\
\multicolumn{1}{|c|}{0.5}     & \multicolumn{1}{c|}{1}       & \multicolumn{1}{||c|}{0.554}    & \multicolumn{1}{c||}{1.084} & \multicolumn{1}{c|}{0.696}         & \multicolumn{1}{c|}{0.905}                \\
\multicolumn{1}{|c|}{0.35}     & \multicolumn{1}{c|}{5}       & \multicolumn{1}{||c|}{0.601}    & \multicolumn{1}{c||}{1.060} & \multicolumn{1}{c|}{0.709}         & \multicolumn{1}{c|}{0.979}                \\
\multicolumn{1}{|c|}{0.35}     & \multicolumn{1}{c|}{1}       & \multicolumn{1}{||c|}{-}    & \multicolumn{1}{c||}{-} & \multicolumn{1}{c|}{0.627}         & \multicolumn{1}{c|}{0.816}                \\
 \hline
\end{tabular}
 \caption{\label{table}Data of the first solution exhibiting an ergoregion or a retrograde LR, for the parameters presented in Fig. \ref{fig:massfreqcharge}.}
\end{table}
 \section{Stationary Proca clouds and black holes with Proca-Higgs hair}\label{bh_sec}

\subsection{Kerr metric}

The metric ansatz \eqref{line_BH} yields to the Kerr solution in a non-Boyer-Lindquist coordinate system. In the Kerr limit, the metric in our coordinate system takes the form~\cite{Herdeiro:2015gia}:
\begin{equation}
\begin{aligned}
& e^{2 F_1}=\left(1+\frac{b}{r}\right)^2+b\left(b+r_H\right) \frac{\cos ^2 \theta}{r^2}, \\
& e^{2 F_2}=e^{-2 F_1}\left\{\left[\left(1+\frac{b}{r}\right)^2+\frac{b\left(b+r_H\right)}{r^2}\right]^2-b\left(b+r_H\right)\left(1-\frac{r_H}{r}\right) \frac{\sin ^2 \theta}{r^2}\right\}, \\
& F_0=-F_2, \\
& W=e^{-2\left(F_1+F_2\right)} \sqrt{b\left(b+r_H\right)}\left(r_H+2 b\right) \frac{\left(1+\frac{b}{r}\right)}{r} .
\end{aligned}
\label{Knew}
\end{equation}
with $b$ a constant.
This representation provides a clear view of the metric's behavior in the presence of the parameter $b$, which introduces a spheroidal prolateness to the spacetime.

The relation between these coordinates $(t,r,\theta,\varphi)$ and the standard Boyer-Lindquist coordinates $(t,R,\theta,\varphi)$ is simply a radial shift:
\begin{equation}
r=R-\frac{a^2}{R_H} \ , 
\end{equation}
where $R_H$ is the event horizon Boyer-Lindquist radial coordinate,  $R_H\equiv M+\sqrt{M^2-a^2}$, for a Kerr BH with mass $M$ and angular momentum $J=aM$. In the new coordinate system $(t,r,\theta,\varphi)$, the Kerr solution is parameterized by $r_H$ and $b$, which relate to the Boyer-Lindquist parameters as
\begin{equation}
r_H=R_H-\frac{a^2}{R_H} \ , \qquad b=\frac{a^2}{R_H} \ .
\end{equation}
Clearly, $r_H$ fixes the event horizon radius; $b$ is a spheroidal prolateness parameter, and can be taken as a measure of non-staticity, since $b=0$ yields the Schwarzschild limit.

The ADM mass, ADM angular momentum and horizon angular velocity read, in terms of the parameters $r_H,b$ (we set $G=1=c$)
\begin{equation}
\begin{array}{c}
M=\frac{1}{2}(r_H+2b) \ , \ \ \ 
J=\frac{1}{2}\sqrt{b(b+r_H)}(r_H+2b) \ , \ \ \
\label{Kerr2}
\displaystyle{\Omega_H=\frac{1}{r_H+2b}\sqrt{\frac{b}{r_H+b}}} \ ,
\end{array}
\end{equation}
\begin{equation}
    A_H=4\pi (r_H+b)(r_H+2b)\ , \ \ \
T_H=\frac{r_H}{4\pi (r_H+b)(r_H+2b)}\ . \ \ \
\end{equation}
The choice $r_H=-2b\neq 0$
yields Minkowski spacetime expressed in spheroidal prolate coordinates. Extremality occurs when $r_H=0$.

\subsection{Boundary Conditions}

The event horizon is located at a surface with constant radial variable $r = r_H > 0$. The boundary conditions there and also the numerical treatment of the problem are
simplified by introducing a new radial coordinate
\begin{equation}\label{newr}
    x=\sqrt{r^{2}-r_{H}^{2}}.
\end{equation}

Using the new coordinates $(t,x,\theta,\varphi)$, the Ansatz for the matter fields is given, once more, by:
\begin{equation}
\mathcal{A}=\left(i V d t+\frac{H_1}{x} d x+H_2 d \theta+i H_3 \sin \theta d \varphi\right) e^{i(m \varphi-w t)}\,,\qquad \phi=\phi(x,\theta)\,.
\label{matteransatzBH}
\end{equation}
At the horizon, 
 \begin{equation}\partial_x F_i=H_1=\partial_x H_2=\partial_x H_3 =\partial_x\phi=0, \qquad W= r_H^2 \Omega_H,\qquad\qquad \left(V+\frac{w}{m} H_3 \sin \theta\right)=0 .
\end{equation}
At infinity, $x=\infty$:
\begin{equation}
F_i=W=H_i=V=0\ , \qquad \phi=1 \ .
\end{equation}

An additional constraint from the boundary conditions at the horizon is $\omega= m \Omega_H$. This requirement - the synchronization condition - might be regarded as due to horizon properties, cf. Section~\ref{komar_quantities}.

For the study of Proca-Higgs clouds around  a Kerr black hole, we solve the matter fields equations
using the  ansatz
(\ref{matteransatzBH}), for a fixed
Kerr
background as given by (\ref{Knew}). 
The corresponding equations 
are found by setting $\alpha$ to zero in the general 
expressions displayed in the Appendix \ref{eomap}.

 \subsection{Non-linear Q-clouds on the Kerr black hole background}

The key finding of this section is the generalizability of all flat space ball solutions to (nonlinear) cloud configurations around a Kerr black hole, as a test field. The parameters of the black hole are not arbitrary, constrained by the synchronization condition \eqref{sync}.  Hence, without back-reaction ($i.e.$ the self-gravity of the Proca-Higgs field is turned off), inserting a tiny black hole inside the Proca-Higgs balls is possible, resulting in a stationary configuration. Furthermore, we may increase the black hole size for a given $\omega$ by varying the parameter $r_H$. Then, stationary nonlinear clouds continue to exist as the black hole area increases, but only up to a certain point. 
At this specific point, they exist only in a linear form,
while beyond this limit no cloud solutions are found.
The set of these special points admitting only \textit{linear} clouds is known as {\t the existence line}. Consequently, the domain of existence of Proca-Higgs clouds on a Kerr background, similarly to scalar Q-clouds with self-interactions \cite{HERDEIRO2014302}, exhibits two distinct boundaries: (i) flat spacetime balls, and (ii) linear clouds around Kerr black holes.

Here, we do not directly solve for Kerr linear clouds; these solutions are extrapolated, defined as the boundary for nonlinear clouds and hairy black holes. Our numerical analyses suggest that the Proca-Higgs existence line coincides with the Proca model's, which was analyzed in \cite{Santos:2020pmh}, particularly when $\lambda$ is not too small. This alignment is observed consistently across the solutions we computed. As the existence line is approached, the amplitude of the matter fields decreases to zero. Like flat spacetime cases, decreasing $\omega$ implies an increase in Proca field components, complicating the numerical analysis.

\begin{figure}[h!]
         \centering
        \includegraphics[width=0.47\textwidth]{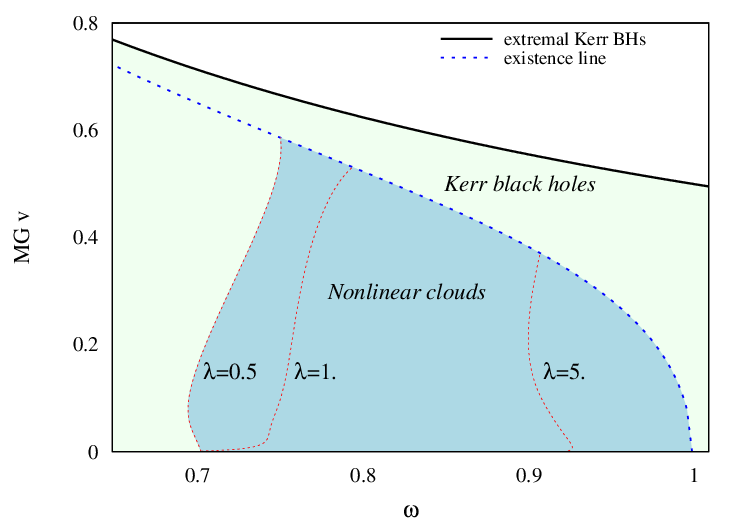}
  		\includegraphics[width=0.47\textwidth]{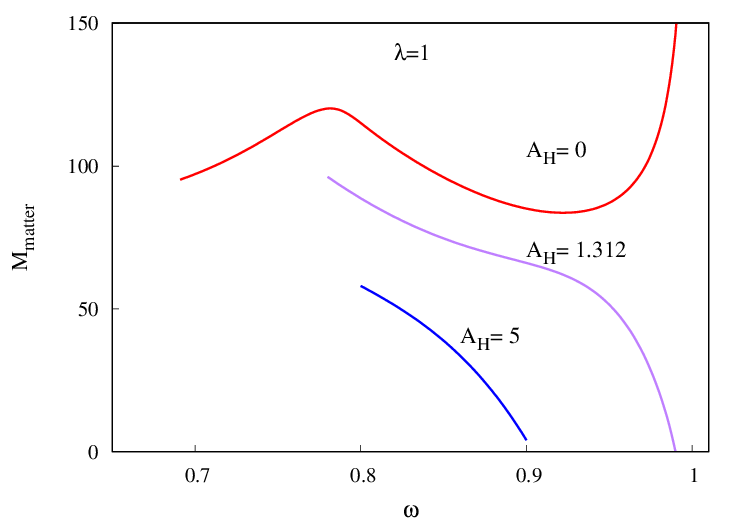}
      	\includegraphics[width=0.47\textwidth]{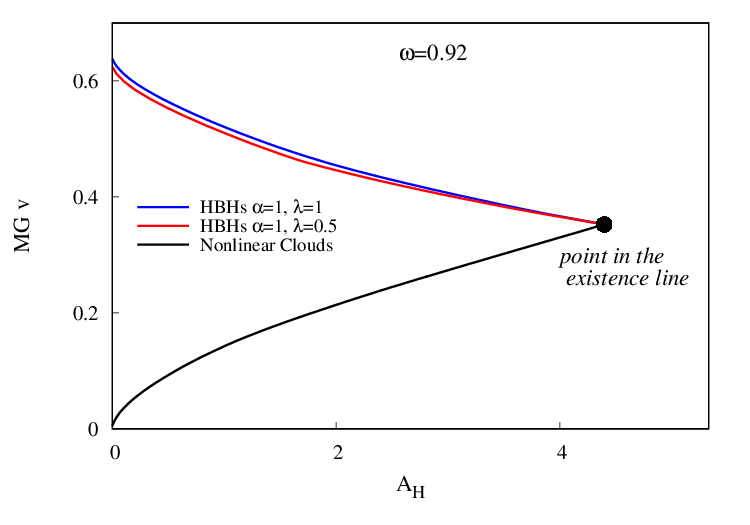}
		\caption{ (Top Left Panel) 
  Existence line (blue dotted)  for linear clouds on the Kerr background. Kerr BHs exist below the
solid black line, which corresponds to extremal Kerr solutions. The dashed red curves are numerical curves of the last obtained solution and do not impose a physical boundary. Nonlinear clouds, for different values of $\lambda$, exist from the corresponding red dashed curve to the existence line. (Top Right Panel) The spectrum of clouds for different fixed values of the event horizon area $A_H$. The solutions with $A_H=0$ are the flat spacetime  balls. (Bottom Panel)   The ADM mass is shown versus the event horizon area for two different configurations of hairy black holes and for the nonlinear clouds (Kerr metric). As we increase the black hole size, 
the solutions  tend to the existence line. Beyond this point, nonlinear clouds cease to exist, and hairy black holes bifurcate from there.}
		\label{figbh:clouds}
	\end{figure}

 In Fig.~\ref{figbh:clouds} (top panels) we plot the 2-dimensional subspace of the full parameter space of Kerr BHs allowing for nonlinear clouds.  The matter functions profiles look rather similar to the ones in the flat spacetime limit (see Fig.~\ref{fig:functionslamb005}) and $|\mathcal{A}|^2$ is non-negative.

  \subsection{Hairy Kerr black holes}

\begin{figure}[h!]
  \centering
		\includegraphics[width=0.47\textwidth]{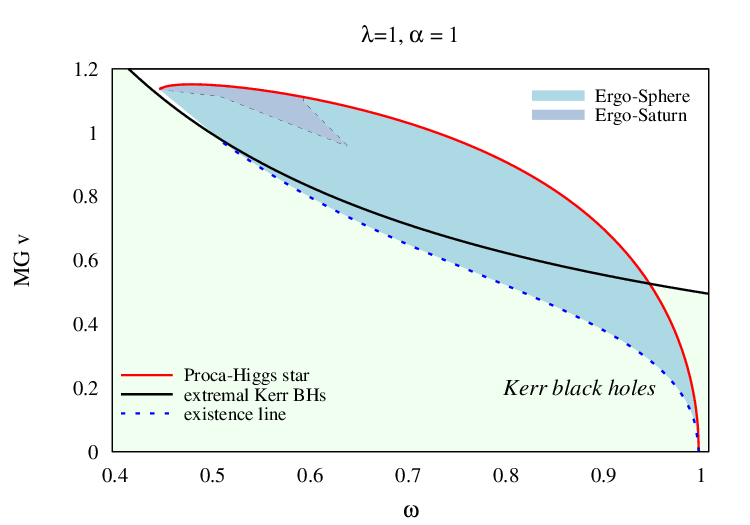}
        \includegraphics[width=0.47\textwidth]{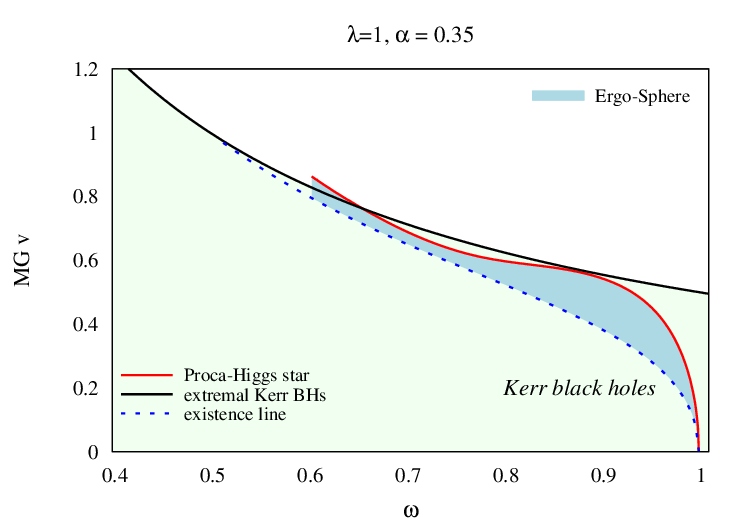}
        \includegraphics[width=0.47\textwidth]{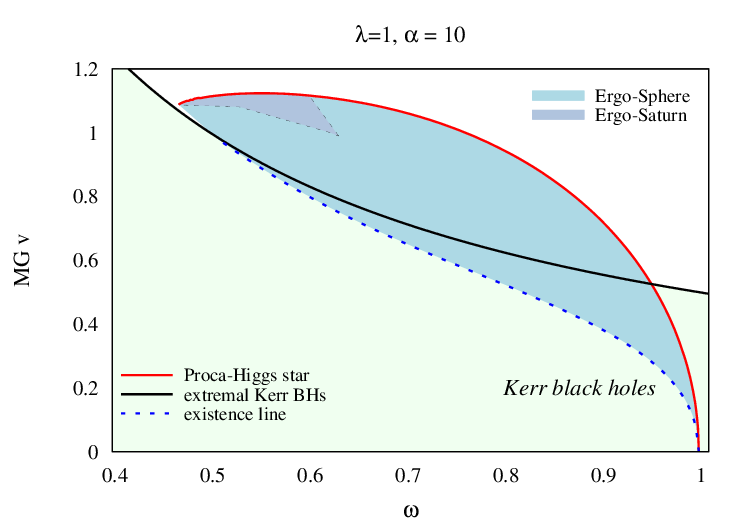}
        \includegraphics[width=0.47\textwidth]{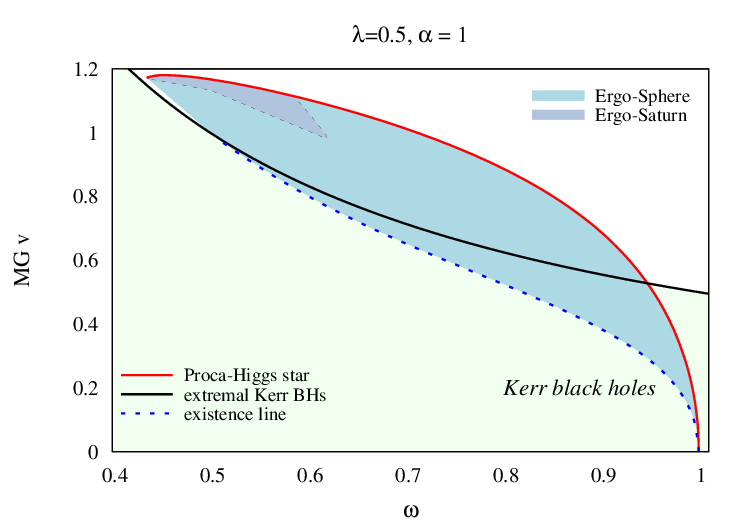}
		\caption{Diagrams of ADM mass versus frequency $\omega$ for $m = 1$. The solitonic limit is represented by the red solid line. Additionally, the ergo-region typologies are displayed: the light (steel) blue shaded area indicates the region of the parameter space where hairy black holes have an ergo-sphere (ergo-Saturn). }
  		\label{figbh:MW}
	\end{figure}

We have systematically explored the black hole existence domain by varying the horizon radius, $r_H$, for constant angular frequencies, $\omega$, and vice versa, across various $\alpha$ and $\lambda$ values. Specifically, we present results for $(\alpha, \lambda)$ pairs: $(0.35, 1)$, $(1, 1)$, $(1, 0.5)$, and $(10, 1)$, illustrated in Figure \ref{figbh:MW}. The overview of the domain of existence generically agrees with the previous literature on scalar and Proca fields, e.g.~\cite{Herdeiro:2015gia}, which has established that three boundaries confine the hairy black holes' existence domain: the corresponding clouds, stars, and extremal black holes lines.

\begin{figure}[h!]
		\centering
		\includegraphics[width=0.47\textwidth]{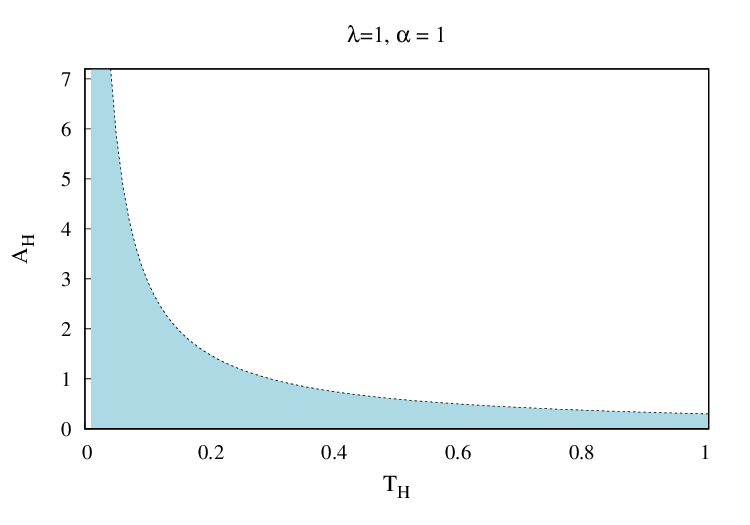} 
		\caption{Event horizon area, $A_H$, in relation to the Hawking temperature, $T_H$. The observed trend remains consistent across varying values of $\alpha$ and $\lambda$.}
		\label{figbh:A}
	\end{figure}

\begin{figure}[h!]
		\centering
		\includegraphics[width=0.47\textwidth]{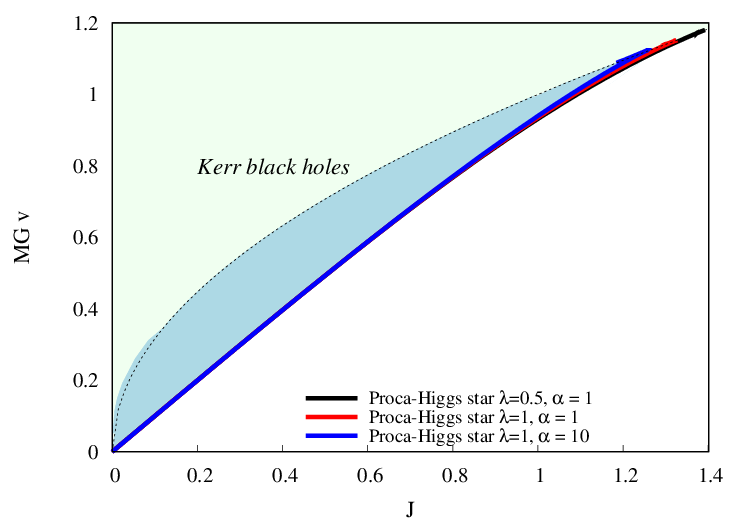}
  		\includegraphics[width=0.47\textwidth]{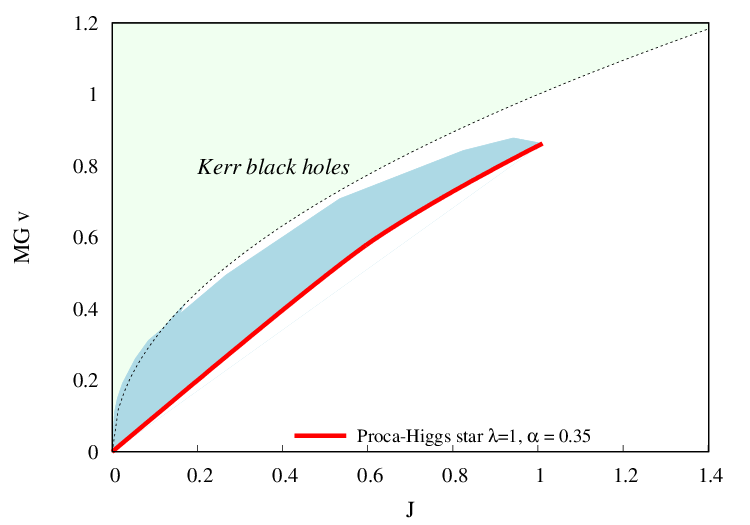}
		\caption{ADM against total angular momentum. The black dashed line represents extremal Kerr solutions, with non-extremal BHs located above it. The Proca-Higgs stars are represented by the red solid line. }
		\label{figbh:MJ}
	\end{figure}

Fig. \ref{figbh:A} displays how the area of the horizon changes in relation to the horizon temperature. In Fig.~\ref{figbh:MJ}, we can see that a great part of the existence domain lies in the overspinning regime, i.e., violating the  Kerr bound, the solutions obeying $a^2= \dfrac{J^2}{M^2}>M^2$. This is not surprising since the stars themselves violate such limit and the parameter spaces of black holes and stars are connected.
It should be emphasized that contrary to vacuum Kerr black holes, the Proca-Higgs hairy black holes do not have a static limit (with non-trivial hair). The minimum frequency $\omega$ of the corresponding stars establishes a minimum bound for the angular velocity of the horizon due to the synchronization condition.


\begin{figure}[h!]
         \centering
        \includegraphics[width=0.32\textwidth]{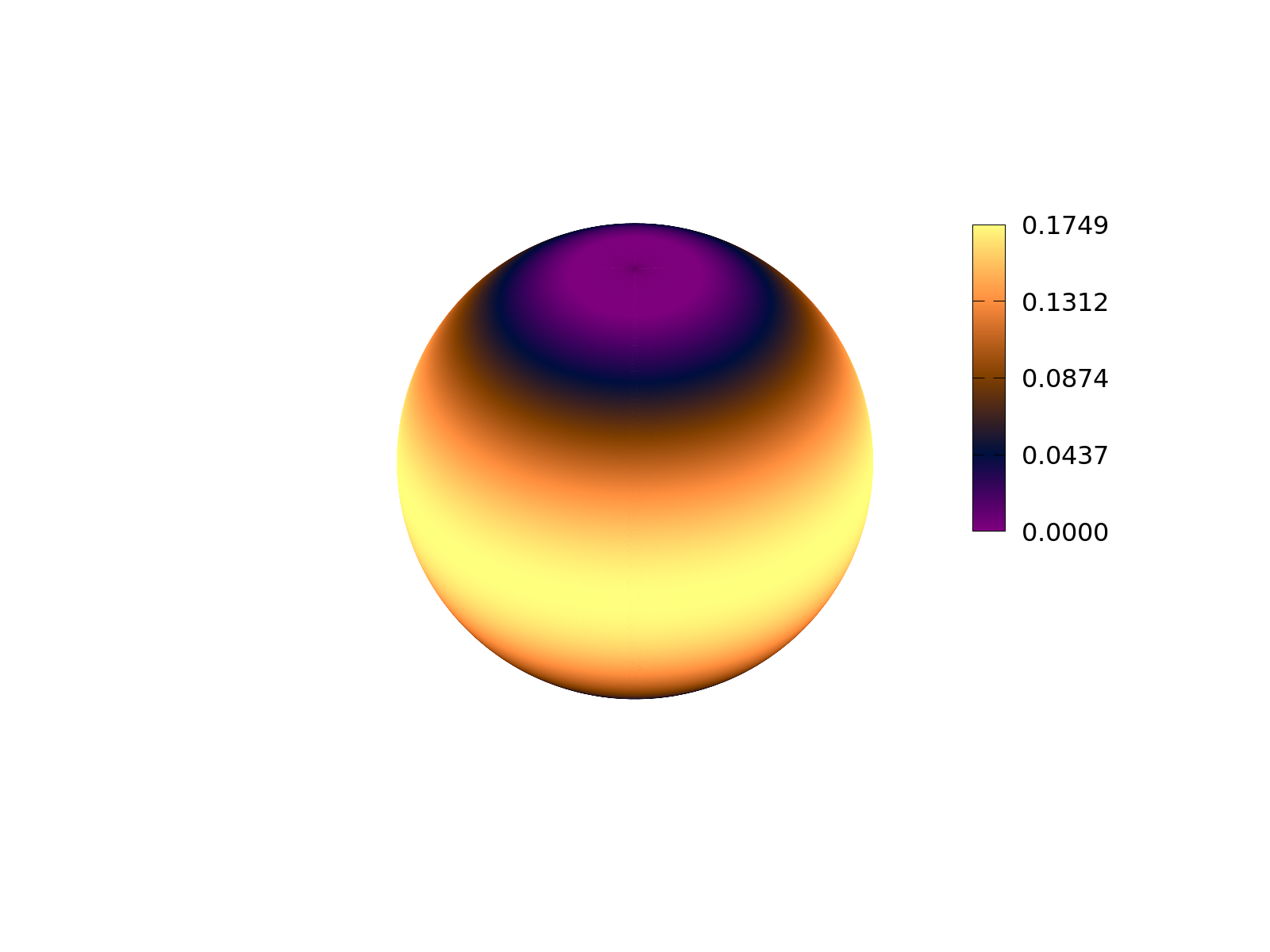}
  		\includegraphics[width=0.32\textwidth]{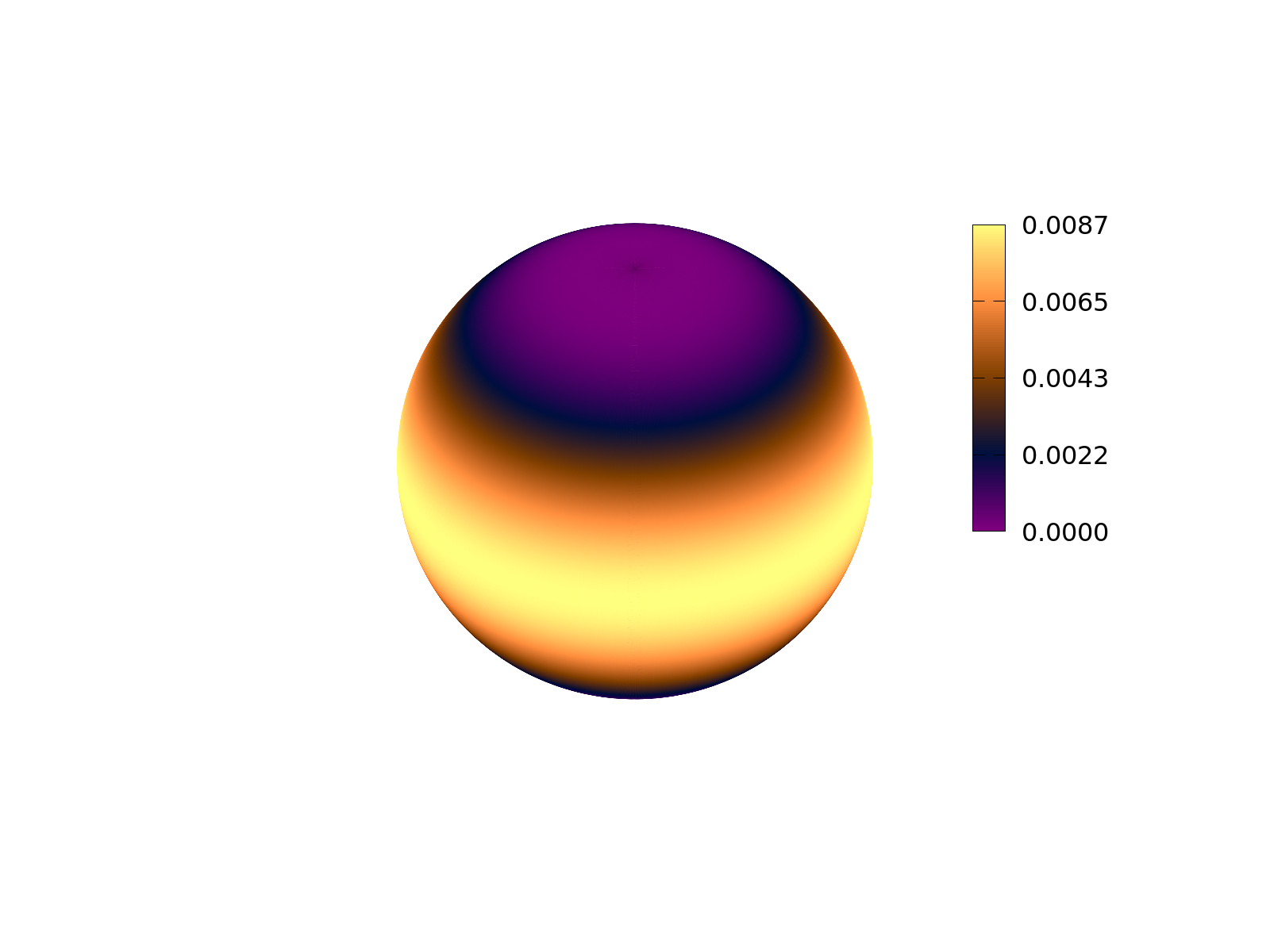}
      	\includegraphics[width=0.32\textwidth]{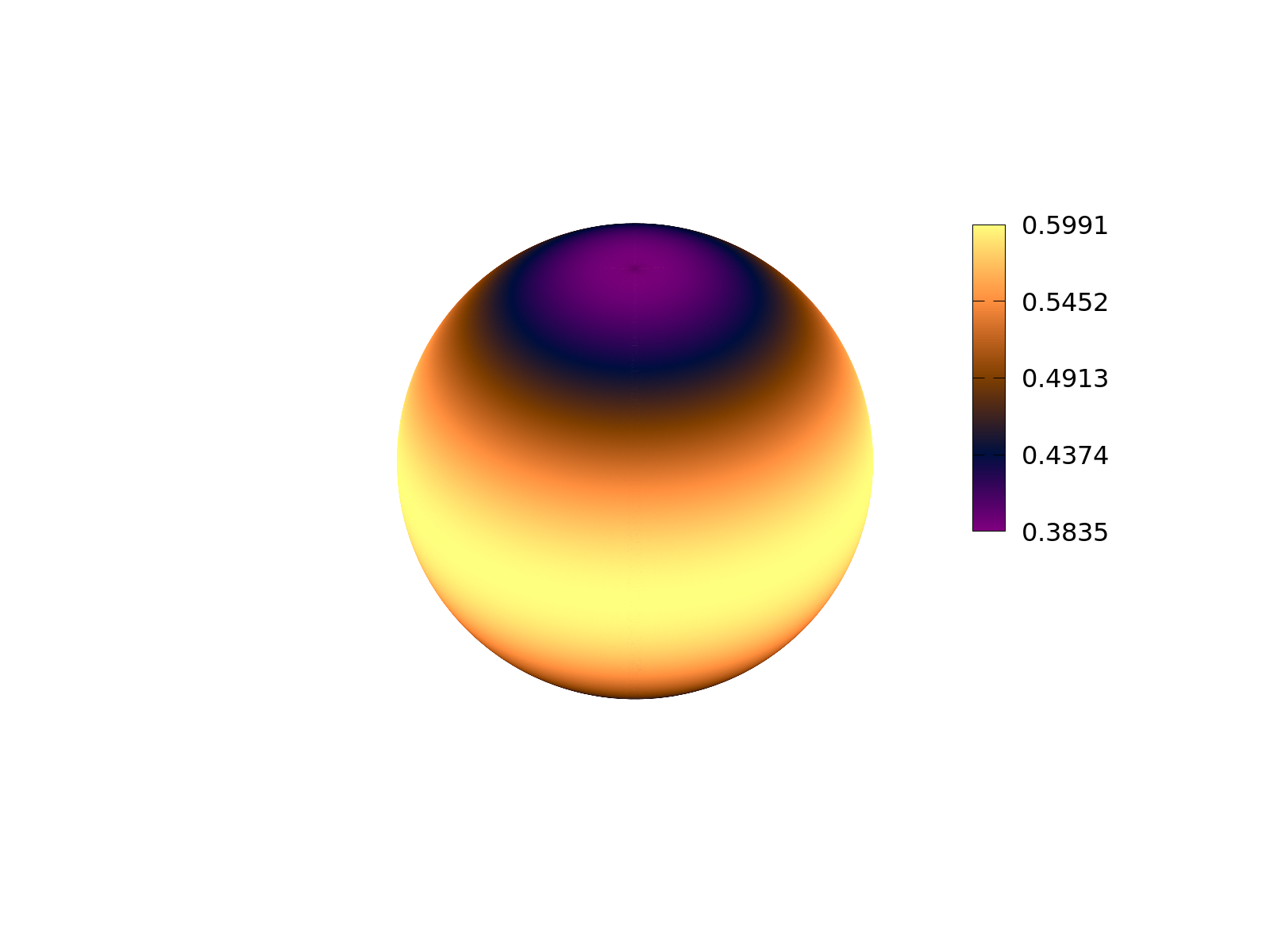}
		\caption{   Komar mass density (Left Panel), angular momentum density  (Central Panel) and Noether charge density (Right Panel) of the matter fields on the horizon for the solution $\alpha=1.0$, $\lambda=0.5$, $r_H=0.05$, $\omega=0.70$, $M=0.9698$ and $J=1.0130$.} 
		\label{figbh:sol}
	\end{figure}

   For a typical solution, the behavior of the Komar mass density, angular momentum density, and Noether charge density at the horizon is illustrated in Fig \ref{figbh:sol}.

\section{Conclusions}

In this paper we have reported the construction of spinning Proca-Higgs balls, stars and hairy black holes, furthering the analysis in~\cite{herdeiro2023procahiggs} which introduced the Proca-Higgs model and computed the corresponding balls and stars in spherical symmetry.

Overall, our results confirm that the Proca-Higgs models allow spinning generalizations of the basic solitonic solutions found in~\cite{herdeiro2023procahiggs}, and, furthermore, allow for the construction of hairy black holes anchored on the synchronization mechanism.  Moreover, we have found numerical evidence for
 the existence of  non-linear Proca-Higgs clouds on the background of Kerr black holes.
 There are qualitative differences with respect to, say, the pure Proca model - namely the existence of the aformentioned balls and non-linear clouds on Kerr - but overall one recognizes familiar features from other bosonic field models (scalar and vector) leading to $Q$-balls, boson stars and black holes with synchronized hair.

Apart from constructing the solutions, we have conducted a preliminary study of the model's general properties. This includes developing a mass formula for Proca-Higgs hairy black holes and examining various energy conditions within the model (Appendix~\ref{sec_ec}). Furthermore, we have started to explore astrophysical aspects of the gravitating solutions, particularly focusing on how they can be differentiated from a Kerr solution, such as through their distinct multipolar structures (Appendix~\ref{sec_multipole}). 
We have also discussed briefly the appearance of LRs and ergo-regions in the model. Clearly, our constructed solutions can now be analysed in greater depth concerning their phenomenology. 

Another important avenue for further research is to explore the dynamics (and stability) of these configurations, in particular since some surprising results were found recently in the case of the spherical models~\cite{Brito:2024biy}. An additional motivation is that the spinning Proca stars were seen to be more dynamically robust than their scalar cousins~\cite{Sanchis-Gual:2019ljs}, leading to recent interesting applications concerning the LR instability~\cite{Cunha:2022gde}. Proca-Higgs spinning stars may provide another fertile ground for such explorations.  

 Finally, 
the recent results in Ref. \cite{Herdeiro:2023wqf}  
indicate the existence  of
 new families of (static) solutions in Einstein-Proca model,
with a variety of interesting features
($e.g.$ the ground state of the model is not spherically symmetric, bound state configurations with more than one  component existing  as well).
We expect these solutions to possess generalizations in 
(Einstein-)Proca-Higgs model;
in particular spinning multi-component stars should exist, together with their extensions with a  (single) black hole horizon.

\bigskip

\section*{Acknowledgement}
E.S.C.F. would like to thanks Prof. Hideki Maeda for his invaluable contributions and discussions regarding energy conditions. This work is supported by the Center for Research and Development in Mathematics and Applications (CIDMA) through the Portuguese Foundation for Science and Technology (Funda\c c\~ao para a Ci\^encia e a Tecnologia), UIDB/04106/2020, 
https://doi.org/10.54499/UIDB/04106/2020 
and UIDP/04106/2020, https://doi.org/10.54499/UIDP/04106/2020.
The authors acknowledge support from projects
http://doi.org/10.54499/PTDC/FISAST/3041/2020,
http://doi.org/10.54499/CERN/FIS-PAR/0024/2021 and https://doi.org/10.54499/2022.04560.PTDC.  This work has further been supported by the European Horizon Europe staff exchange (SE) programme HORIZON-MSCA2021-SE-01 Grant No. NewFunFiCO-101086251. E.S.C.F. is supported by the FCT grant PRT/BD/153349/2021 under the IDPASC Doctoral Program. Computations have been performed at the Argus and Blafis cluster at the U. Aveiro.
\appendix
\section{Quadrupole Moment}\label{sec_multipole}

 A Kerr black hole is uniquely defined by its first two multipole moments \cite{PhysRevLett.11.237,Hansen74}: mass and angular momentum, fulfilling the no-hair conjecture \cite{10.1063/1.3022513} (discussions on no-hair theorems can be found in \cite{doi:10.1142/S0218271815420146,Herdeiro:2016tmi} for the scalar case and \cite{Ayon-Beato:2002kxy}). Conversely, due to potential deformities in their mass distribution, neutron stars require consideration of higher-order multipole moments, like the quadrupole moment, alongside their mass and radius for a full characterization \cite{PhysRevD.91.103003,Laarakkers_1999,PhysRevLett.112.121101,PhysRevD.88.023009}. The discovery of gravitational waves from binary black hole mergers \cite{LIGOScientific:2016aoc} emphasizes the importance of these multipolar moments. Specifically, the spin-induced quadrupole moment in such mergers can help distinguish between binary black hole systems and other exotic compact objects \cite{PhysRevLett.119.091101,Sanchis-Gual:2018oui,CalderonBustillo:2020fyi,CalderonBustillo:2022cja}. Furthermore, these moments can play an important role in astrophysical experiments, allowing researchers to study black holes and their no-hair theorems, test General Relativity itself\cite{Cardoso_2016,Will_2008}, or other theories of gravity such as scalar-tensor theories \cite{PhysRevD.91.044011,10.1093/mnras/stv1819,PhysRevD.99.104014}, all of it using the multipole data from gravitational waves and/or geodesic observations. Given the profound significance of multipole moments in characterizing the spacetime, we are motivated to investigate the quadrupole moment within the Proca-Higgs model in this appendix.

In the context of Newtonian gravitational systems, the multipolar expansion provides a holistic representation of the gravitational field for static mass distributions. Geroch's seminal work \cite{Geroch:1970cc} on asymptotically flat static spacetimes in vacuum introduced a tensorial characterization of multipole moments, with these spacetimes inherently governed by the Laplace equation, $\nabla^2 \Phi=0$. This equation echoes its Newtonian counterpart, yielding multipoles as a series of symmetric, trace-free tensors. Geroch's moments, coordinate-independent by design, converge to Newtonian multipoles in weak gravitational fields\cite{Geroch70}. Subsequent advancements by Hansen \cite{Hansen74}, incorporating stationary, and Simon \cite{Simon84} expanded this framework to electrovacuum scenarios.

Within a four-dimensional, stationary, and asymptotically flat spacetime characterized by metric $g_{\mu\nu}$, the time-like Killing vector, $\xi$, plays a pivotal role. Its norm as $\lambda=\xi^{2}$, defines a scalar field $\lambda$  and the positive definite, 3-dimensional, metric $h_{\mu \nu}=\lambda g_{\mu \nu}+\xi_\mu \xi_\nu$. In vacuum spacetimes, the twist vector $\omega_{\mu}=\epsilon_{\mu\nu\rho\sigma}\xi^{\nu}\nabla^{\rho}\xi^{\sigma}$ can be expressed as a total derivative, introducing a new scalar function $\omega$. Hansen's approach further defines scalar potentials for multipole moments linked to mass and angular momentum as $\Phi_{M}=\frac{1}{4\lambda}(\lambda^{2}+\omega^{2}-1)$ and $\Phi_{J}=\frac{1}{2\lambda}\omega$. The challenge in generalizing Geroch-Hansen's multipole moments beyond vacuum spacetimes lies in constructing an appropriate scalar function $\omega$. While Simon achieved this for electrovacuum spacetimes in the 70s, a generic construction was only recently proposed \cite{Mayerson:2022ekj}. It is crucial to note that while the Geroch-Hansen formalism has been applied to more general theories, such as scalar fields minimally coupled to General Relativity, these generalization have a limited scope.

Further advancements addressing axially symmetric spacetimes  can be found in~\cite{Fodor89,Backdahl05,Hoenselaers90,Fodor:2020fnq,Filho:2021exr}. Although the previous studies set the ground for the study of multipole moments in axistationary spacetimes, their results were difficult to use in numerically constructed metric. Under the assumption of analyticity, Ryan and Pappas' studies \cite{ryan1995gravitational,PhysRevLett.108.231104} defined the quadrupole moment as a coefficient extractable from metric functions. While valid for vacuum spacetimes, its applicability to other frameworks was ambiguous. Fortunately, this technique was primarily applied to exponentially decaying fields, ensuring the validity of their results due to the rapid decay of these fields \cite{Adam:2022nlq,Adam:2023qxj,Vaglio:2022flq}.

Here, for the computation of the quadrupole moment, we adopt the methodology delineated in \cite{ryan1995gravitational,PhysRevLett.108.231104}, which extracts the moment from the asymptotics of a stationary and axially symmetric spacetime. The references mentioned previously utilize a line element expressed in quasi-isotropic coordinates. This can be represented as:

\begin{equation}
\label{isotropic}
ds^2 =- e^{2\nu}dt^2+ e^{2(\zeta-\nu)}(dR^2 + R^2 d\theta^2) + R^2\sin^2\theta B^2 e^{-2\nu}(d\varphi-W dt)^2 ,
\end{equation}
where the variables $\zeta, \nu, B,$ and $W$ are functions dependent on $R$ and $\theta$. The behavior of the metric functions $\nu$ and $B$ as $R$ approaches infinity determines the spacetime's quadrupole moment. To a leading order approximation, these functions are:

\begin{equation}
\label{as1}
\nu = -\frac{M}{R} + \left(\frac{B_0M}{3} + \nu_2 P_2(\cos\theta)\right)\frac{1}{R^3} + \dots, \quad B = 1 + \frac{B_0}{R^3} + \dots,
\end{equation}

where $P_n$ denotes the Legendre polynomials. Consequently, the quadrupole moment, $M_2$, is:

\begin{equation}
\label{quadrupole}
M_2 = -\nu_2 - \frac{4}{3}\left(\frac{1}{4} + \frac{B_0}{M^2}\right)M^3.
\end{equation}

To evaluate the quadrupole moment for black hole solutions using equation (\ref{quadrupole}), we first need to express the line element (\ref{line_BH}) in the form of (\ref{isotropic}) (for the stars, the result is equivalent by setting $r_H=0$). This is achieved through the coordinate transformation given by $r = R(1 + \frac{r_H}{4R})^2$. After this transformation, the coefficients $B_0$ and $\nu_2$ from the asymptotic behavior (\ref{as1}) can be derived from the far-field form of the solution (expressed in terms of the quasi-isotropic $R$):
\begin{equation}
\label{quadrupole1}
B_0 = a_5 - \frac{1}{16}(2c_t - r_H)^2, \quad \nu_2 = \frac{1}{12}\left(-24 b_1 - 8 a_5c_t + 2c_t^3 + 4 a_5 r_H - 3c_t^2r_H - 3c_t r_H^2\right) \ ,
\end{equation}
where $a_5$, $b_1$ and $c_t$ can be read from the leading order terms the asymptotic expansion 
\begin{equation}
\begin{aligned}
& F_0(r, \theta)=\frac{c_t}{r}+\frac{c_t r_H}{2 r^2}+\frac{f_{03}(\theta)}{r^3}+\ldots, \\
& F_1(r, \theta)=-\frac{c_t}{r}+\frac{f_{12}(\theta)}{ r^2}+\frac{f_{13}(\theta)}{r^3}+\ldots, \\
& F_2(r, \theta)=-\frac{c_t}{r}+\frac{a_5-\frac{1}{4} c_t\left(c_t+r_H\right)}{r^2}+\frac{f_{23}(\theta)}{r^3}+\ldots,
\end{aligned}
\end{equation}
and the expressions for the $\theta$-dependent coefficients are:
\begin{eqnarray}
\label{r-infty-s1}
\nonumber
&&
f_{03}(\theta)=b_1+\frac{c_t r_H^2}{2}-\frac{1}{8}\left(24 b_1+8 a_5 c_t-2c_t^3-4a_5 r_H+3 c_t^2 r_H+3c_t r_H^2 \right)\cos^2 \theta,
\\
&&
\nonumber
f_{12}(\theta)= -\frac{c_t}{4}(c_t+r_H)+a_5\cos 2\theta,
\\
&&
\nonumber
f_{13}(\theta)=  \frac{1}{16}\bigg(
8 b_1+a_5(8c_t-4 r_H)
-c_t(2c_t^2+c_tr_H+r_H^2)
\\
\nonumber
&&
{~~~~~~~~~~~~~~~~~~~~~}+
(
24 b_1+8 a_5c_t-2c_t^3+12 a_5r_H+3c_t^2r_H+3c_tr_H^2
)\cos 2\theta
\bigg) ,
\\
&&
\label{inf-2}
f_{23}(\theta)=  
\frac{1}{16}\bigg(
8 b_1+4a_5(2c_t+3 r_H)
-c_t(2c_t^2+c_tr_H+r_H^2)
\\
\nonumber
&&
{~~~~~~~~~~~~~~~~~~~~~}+
(
24 b_1+8 a_5c_t-2c_t^3-4 a_5r_H+3c_t^2r_H+3c_tr_H^2
)\cos 2\theta
\bigg) .
\end{eqnarray} 

In this convention, the Kerr quadrupole moment is

\begin{equation}
    M_2=- \dfrac{J^2}{M}\,.
\end{equation}
and
\begin{equation}
 b_1=\frac{1}{6} c_t^2\left(2 c_t-3 r_H\right),\qquad\qquad a_5=\frac{1}{4} c_t\left(c_t-r_H\right)\,.
\end{equation}

Even though analiticity has been established solely for stationary vacuum equations \cite{Hagen_1970,10.1063/1.525148}, given the exponential decay observed in the matter fields, this assumption appears to remain valid for our case. To estimate the error associated with the quadrupole moment, we derive the coefficient $b_1 $ using two distinct metric functions, $F_0$ and  $F_2$. We then compute the ratio between these coefficients. Typically, the discrepancies we observe are in the range $10^{-3} - 10^{-5}$.

\begin{figure}[h!]
		\includegraphics[width=0.47\textwidth]{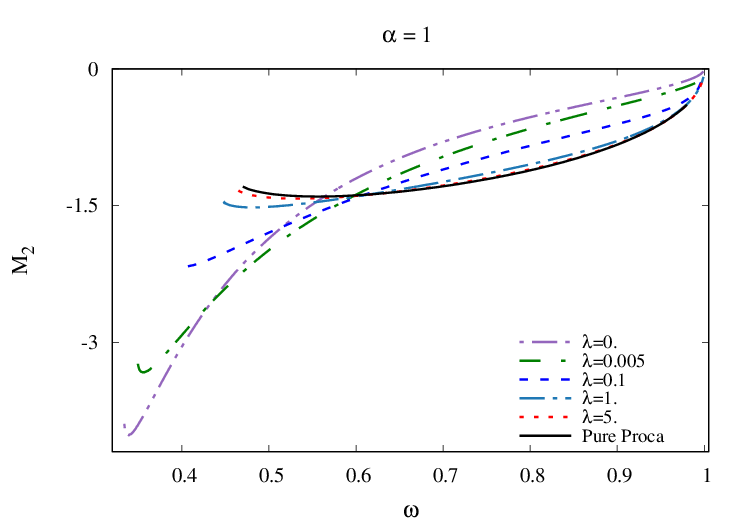} 
		\includegraphics[width=0.47\textwidth]{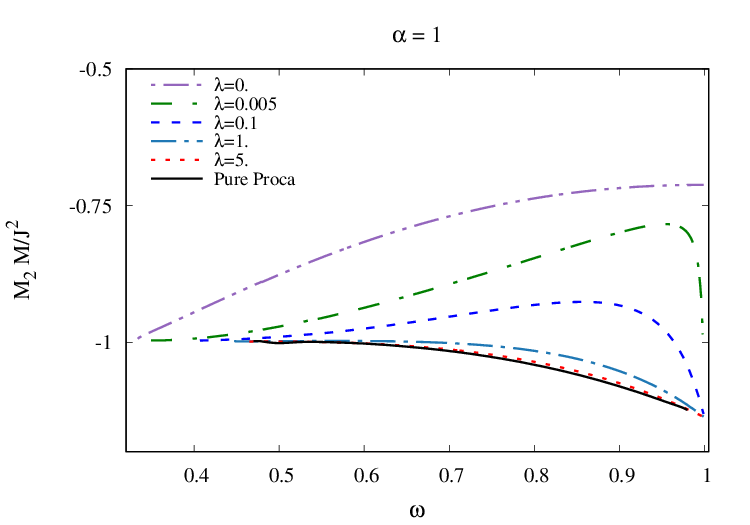}
  \includegraphics[width=0.47\textwidth]{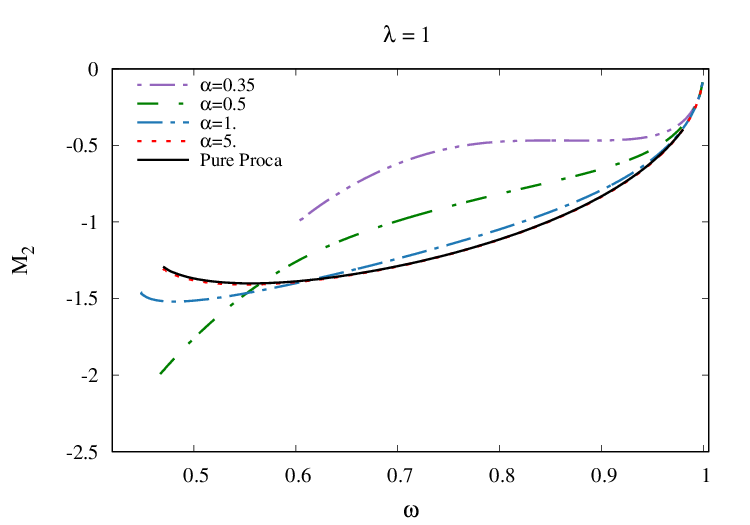} 
		\includegraphics[width=0.47\textwidth]{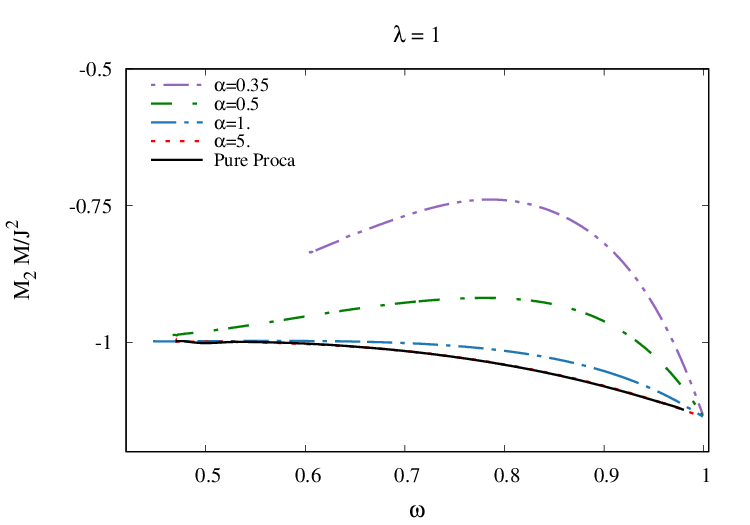}
  \includegraphics[width=0.47\textwidth]{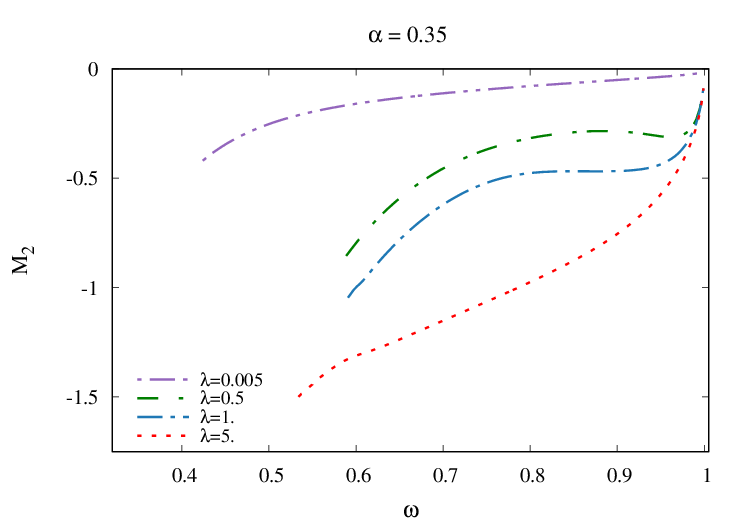} 
  \hfill
		\includegraphics[width=0.47\textwidth]{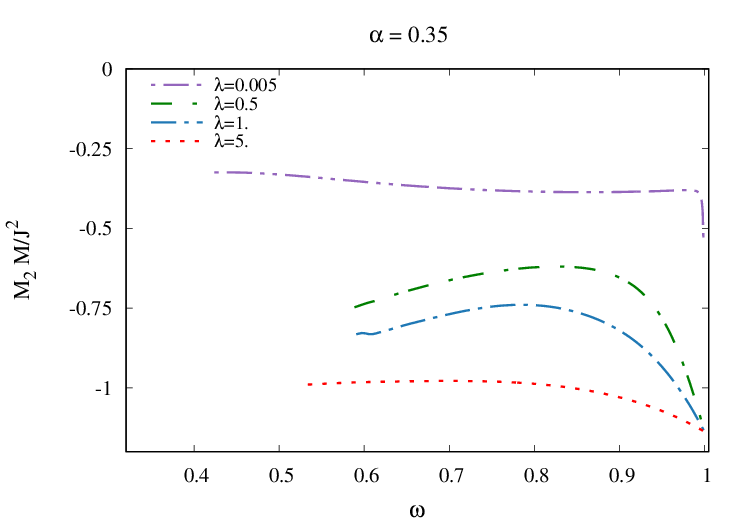}
		\caption{(Left) Quadrupole moment for Proca-Higgs Stars $vs.$ their frequency, $\omega$, across various values of: $\lambda$ for fixed $\alpha=1$ (top),  $\alpha$ and fixed $\lambda=1$ (middle) and $\lambda$ for fixed $\alpha=0.35$ (bottom). (Right) Comparison with Kerr black hole. 
  Here, the quadrupole moment is normalized by dividing it by the corresponding value for Kerr. 
  Notably, as the Proca-Higgs solutions become increasingly compact, this normalized quadrupole converges to 1.}
		\label{figmult:phaselamb}
	\end{figure}

\begin{figure}[h!]
 \centering
		\includegraphics[width=0.45\textwidth]{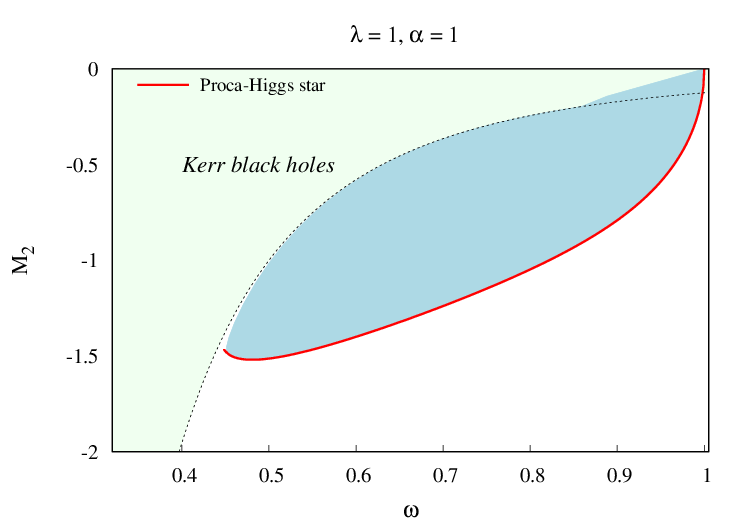} 
		\includegraphics[width=0.45\textwidth]{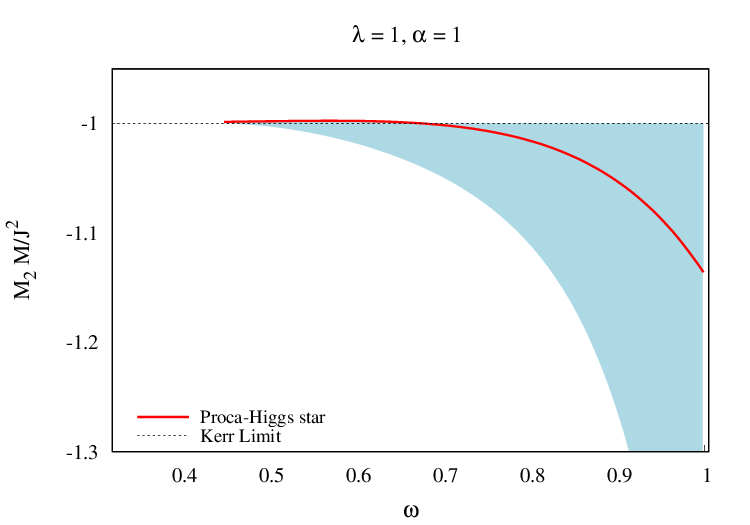}
  		\includegraphics[width=0.45\textwidth]{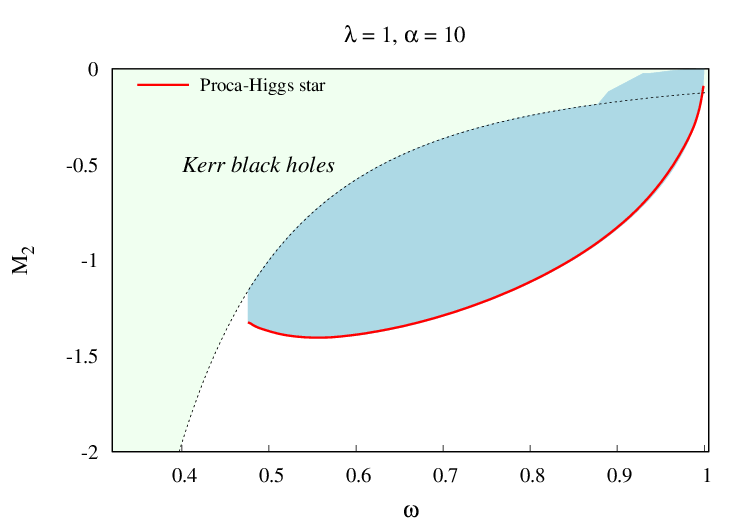} 
		\includegraphics[width=0.45\textwidth]{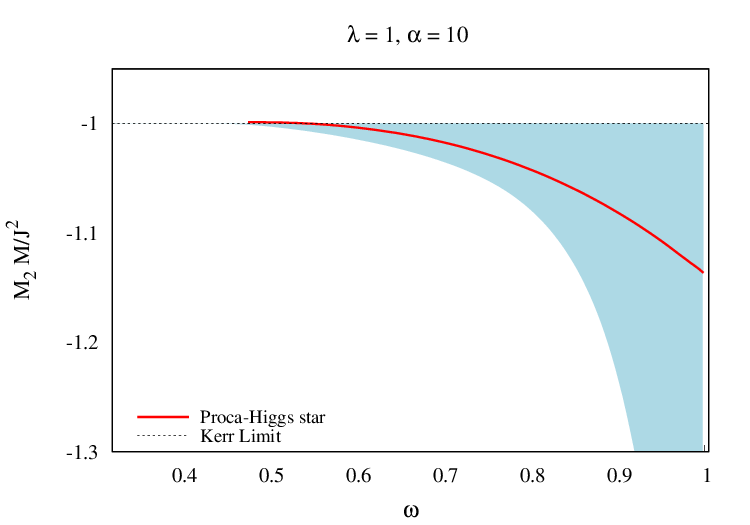}
    		\includegraphics[width=0.45\textwidth]{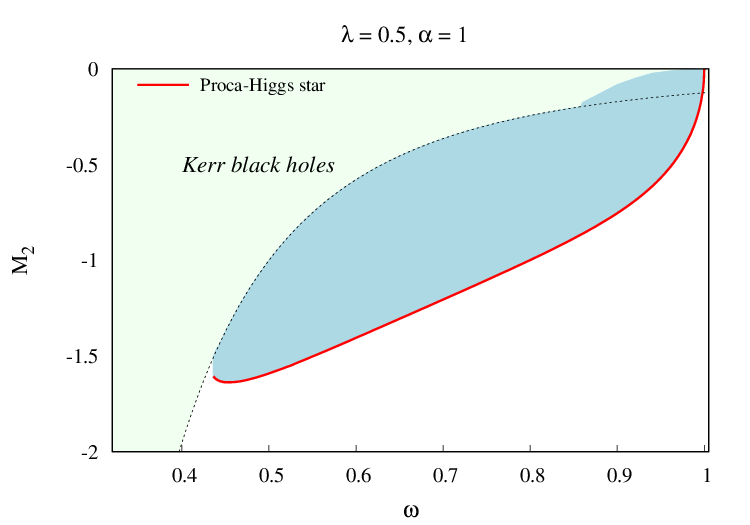} 
		\includegraphics[width=0.45\textwidth]{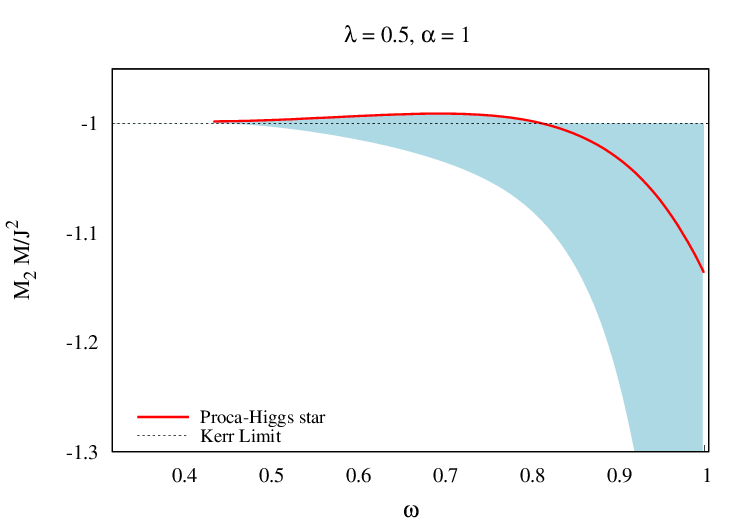}
    		\includegraphics[width=0.45\textwidth]{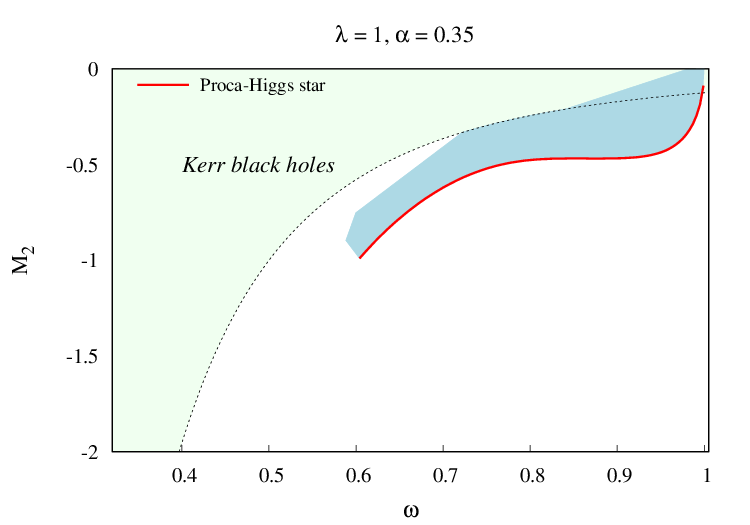} 
		\includegraphics[width=0.45\textwidth]{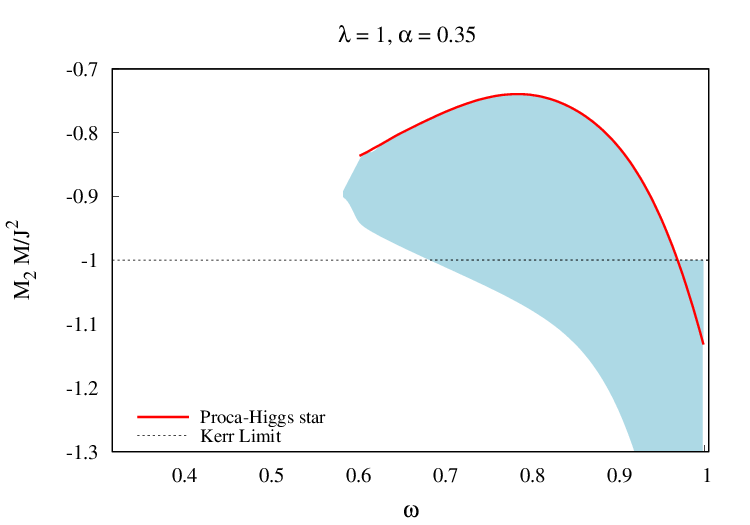}
		\caption{
Same as Figure
\ref{figmult:phaselamb}
  for Proca-Higgs 
  black holes.  }
  \label{figmultbh:alpha1}
	\end{figure}

Unlike other horizonless objects like Fuzzballs, which generally have multipole moments larger than their corresponding Kerr black holes~\cite{PhysRevLett.125.221601}, the multipolar structure of the Proca-Higgs star is profoundly influenced by the parameters $\alpha$, $\lambda$, and the frequency $\omega$ and can either be greater of smaller the the Kerr correspondent. A selection of illustrations is provided in Figs.~\ref{figmult:phaselamb}-\ref{figmultbh:alpha1}.

	\section{Energy conditions}\label{sec_ec}
	There are four primary pointwise energy conditions: Weak, Strong, Dominant, and Null, and they can be interpreted in various ways (for a good review on ECs, see \cite{Lahoz}). These conditions are used to eliminate exotic solutions that violate causality, such as wormholes and time travel, by restricting what is considered ``physically plausible" sources of matter and energy. Originally introduced ad hoc to establish key theorems in GR, ECs are now essential to the theory.
	
	The standard energy conditions for $T_{\mu\nu}$ are as follows:
	\begin{itemize}
		\item {\it Null} energy condition (NEC): $T_{\mu\nu} k^\mu k^\nu\ge 0$ for any null vector $k^\mu$.
		\item {\it Weak} energy condition (WEC): $T_{\mu\nu} v^\mu v^\nu\ge 0$ for any timelike vector $v^\mu$.
		\item {\it Strong} energy condition (SEC): $\left(T_{\mu\nu}-\frac{1}{n-2}Tg_{\mu\nu}\right) v^\mu v^\nu\ge 0$ for any timelike vector $v^\mu$.
		\item {\it Flux} energy condition (FEC): $J_\mu J^\mu\le 0$ holds for $J^\mu:=-T^\mu_{\phantom{\mu}\nu}v^\nu$; namely, $-T^\mu_{\phantom{\mu}\nu}v^\nu$ is a causal vector or a zero vector, for any time-like vector $v^\mu$.
		\item {\it Dominant} energy condition (DEC): $T_{\mu\nu} v^\mu v^\nu\ge 0$ and $J_\mu J^\mu\le 0$ hold for any timelike vector $v^\mu$. 
	\end{itemize}
	
	It is important to highlight that the NEC is the least stringent of the energy conditions, as it is satisfied by all the other conditions. As a result, in our discussion of the pointwise energy conditions, we will concentrate on the SEC and the WEC, which are more restrictive than the NEC.
	
	A reference frame called ZAMO (Zero Angular Momentum Observers) is introduced to verify the Energy Condition for the numerical solutions obtained. ZAMO is a time-like observer with a four-velocity $u_\mu$ at each point, where $u_\mu$ is chosen to be $(u_t,0,0,0)$. Hence, the WEC and SEC become
	
	\begin{itemize}
		\item WEC: $-T_t^t-\frac{g^{\varphi t}}{g^{t t}} T_{\varphi}^t \geqslant 0$\ .
		\item SEC: $-\left(T_t^t-\frac{1}{2} T\right)-\frac{g^{\varphi t}}{g^{t t}}T_{\varphi}^t  \geqslant 0$ .
	\end{itemize}

	\subsection{Energy conditions in arbitrary dimensions}

	In the papers \cite{Maeda_2022} and \cite{Maeda:2018hqu}, the authors explore the standard energy conditions for matter fields in arbitrary dimensions, while also presenting particular criteria for the combined sum of multiple distinct stress-energy tensors. Additionally, they examine the energy conditions for several physics-motivated matter fields, such as the Maxwell equations, real Proca field, and Maxwell(Proca)-dilaton field. Building upon these results, we aim to prove the energy conditions for the Proca-Higgs model and, as a preliminary step, proving the energy conditions for a complex Proca field.
	
	In order to demonstrate that an $n$-dimensional Proca-Higgs model, without assuming any symmetry and under reasonable assumptions, satisfies the common ECs, let us split the proof into parts. Specifically, we will show that each component of the energy-momentum tensor $T_{\alpha \beta} = T_{\alpha \beta}^{(v)} + T_{\alpha \beta}^{(s)}$, defined in equations \eqref{Einstein-eqs}, \eqref{vector_EMT} and \eqref{scalar_EMT}, satisfies the standard energy conditions.
	
	The energy-momentum tensor for a Proca field, $T_{\alpha \beta}^{(v)}$, can be written as
	\begin{align}
		& T_{\mu \nu}^{(v)}=M_{\mu \nu}+\phi^{2} \tau_{\mu \nu}, \label{cproca}\\
		& M_{\mu \nu}:=\mathcal{F}_{(\mu}{ }^{\rho} \bar{\mathcal{F}}_{\nu) \rho}-\frac{1}{4} g_{\mu \nu} \mathcal{F}_{\rho \sigma} \bar{\mathcal{F}}^{\rho \sigma}, \label{maxwell}\\
		& \tau_{\mu \nu}:=A_{(\mu} \bar{A}_{\nu)}-\frac{1}{2} g_{\mu \nu} A^{\rho} \bar{A}_{\rho} \label{massterm} .
	\end{align}
	
	In this demonstration, we adopt the definitions from \cite{Maeda:2018hqu} and employ an orthonormal basis, which consists of a set of $n$ vectors denoted as ${E}^\mu_{(a)}=({E}^\mu_{(0)},{E}^\mu_{(1)},\cdots,{E}^\mu_{(n-1)})$ satisfying ${E}_{(a)}^{\mu}{E}_{(b) \mu}=\eta_{(a)(b)}=\mbox{diag}(-1,1,\cdots,1)$, forming an orthonormal basis in the local Lorentz frame of a specific spacetime. The spacetime metric, $g_{\mu\nu}$, written in this local frame is given by $g_{\mu\nu}=\eta_{(a)(b)}E^{(a)}{\mu}E^{(b)}{\nu}$. An orthonormal basis ${E}^\mu_{(a)}$ has a degree of freedom provided by the local Lorentz transformation ${E}^\mu_{(a)}\to{\tilde E}^\mu_{(a)}:=L_{(a)}^{(b)}{E}^\mu_{(b)}$, satisfying $L_{(a)}^{(c)}L_{(b)}^{~~(d)}\eta_{(c)(d)}=\eta_{(a)(b)}$. For a given vector field $v^\mu$, its corresponding local Lorentz vector can be defined as $v^{(a)}:=v^\mu {E}_\mu^{(a)}$, which transforms as a vector under local Lorentz transformations but as a scalar under coordinate transformations. $\eta_{(a)(b)}$ and its inverse $\eta^{(a)(b)}$ are respectively utilized to lower and raise the indices $(a)$, and $v_\mu v^\mu=v_{(a)}v^{(a)}$ is satisfied. The orthonormal components of the energy-momentum tensor can be expressed as $T_{(a)(b)}=T_{\mu\nu}E_{(a)}^{\mu}E_{(b)}^{\nu}$.
	
	\begin{Prop}
		A complex Maxwell field \eqref{maxwell} satisfies all the standard energy conditions.
	\end{Prop}
	
	{\it Proof}. We write $F_{\mu \nu}$ in the orthonormal frame such as
	\begin{equation}
		\begin{aligned}
			\mathcal{F}_{\mu \nu}= & 2 f_{(0)(1)} E_{[\mu}^{(0)} E_{\nu]}^{(1)}+2 f_{(0)(2)} E_{[\mu}^{(0)} E_{\nu]}^{(2)}+\cdots+2 f_{(0)(n-1)} E_{[\mu}^{(0)} E_{\nu]}^{(n-1)} \\
			& +2 f_{(1)(2)} E_{[\mu}^{(1)} E_{\nu]}^{(2)}+2 f_{(1)(3)} E_{[\mu}^{(1)} E_{\nu]}^{(3)}+\cdots+2 f_{(1)(n-1)} E_{[\mu}^{(1)} E_{\nu]}^{(n-1)} \\
			& +2 f_{(2)(3)} E_{[\mu}^{(2)} E_{\nu]}^{(3)}+2 f_{(2)(4)} E_{[\mu}^{(2)} E_{\nu]}^{(4)}+\cdots+2 f_{(2)(n-1)} E_{[\mu}^{(2)} E_{\nu]}^{(n-1)} \\
			& +2 f_{(3)(4)} E_{[\mu}^{(3)} E_{\nu]}^{(4)}+\cdots+2 f_{(n-2)(n-1)}^{(n-2)} E_{[\mu]}^{(n-1)} \\
			= & 2 \sum_{i=1}^{n-1} f_{(0)(i)} E_{\mu}^{[(0)} E_{\nu}^{(i)]}+2 \sum_{i=1}^{n-1} \sum_{j>i}^{n-1} f_{(i)(j)} E_{\mu}^{[(i)} E_{\nu}^{(j)]}
		\end{aligned}
	\end{equation}
	where $f_{(a)(b)}$ are complex functions. For any given time-like vector $v^{\mu}$, we set the frame such that $v^{(i)}=0$ for all $i$ by a local Lorentz transformation without loss of generality. In this frame, we have
	\begin{equation}
		v^{\mu} \mathcal{F}_{\mu \nu}=v^{(0)} E_{(0)}^{\mu} \sum_{i=1}^{n-1} f_{(0)(i)} E_{\mu}^{(0)} E_{\nu}^{(i)}=v^{(0)} \sum_{i=1}^{n-1} f_{(0)(i)} E_{\nu}^{(i)} .
	\end{equation}
	
	This is a spacelike vector. We can still use a freedom of the Lorentz transformation in the spacelike section, namely spacelike rotation, such that $v^{\mu} F_{\mu \nu}$ is pointing the direction of $E_{\nu}^{(1)}$, in which frame we have $f_{(0)(i)}=0$ for $i=2,3, \cdots, n-1$ and hence $v^{\mu} \mathcal{F}_{\mu \nu}=$ $v^{(0)} f_{(0)(1)} E_{\nu}^{(1)}$. In this frame, we have
	\begin{equation}
		\mathcal{F}_{\mu \nu} \bar{\mathcal{F}}^{\mu \nu}=-2\left|f_{(0)(1)}\right|^{2}+2 \sum_{i=1}^{n-1} \sum_{j>i}^{n-1}\left|f_{(i)(j)}\right|^{2}
	\end{equation}
	
	and	
 \begin{equation}
	M_{\mu \nu} v^{\mu} v^{\nu}=\frac{1}{2}\left(v_{(0)}\right)^{2}\left(\left|f_{(0)(1)}\right|^{2}+\sum_{i=1}^{n-1} \sum_{j>i}^{n-1}\left|f_{(i)(j)}\right|^{2}\right)
\end{equation}
	which is non-negative; therefore, the WEC is respected (and hence also NEC). Now, let us prove that the FEC is obeyed. To do so, consider
	\begin{align}
		J_\mu & =-T_{\mu \nu} v^\nu=-\left(\frac{1}{2}
		( \mathcal{F}_{\mu \sigma }v^{(0)}  \bar{f}_{(0)(1)} E_{\gamma}^{(1)}
		+\bar{\mathcal{F}}_{\mu \sigma } v^{(0)}  f_{(0)(1)} E_{\gamma}^{(1)}
		)g^{\sigma \gamma}-\frac{1}{4} v_\mu \mathcal{F}_{\rho \sigma} \bar{\mathcal{F}}^{\rho \sigma}\right)\nonumber \\
		& = -v^{(0)}\left\{\frac{1}{2} E_\mu^{(0)}\left(|f_{(0)(1)}|^2+\sum_{i=1}^{n-1} \sum_{j>i}^{n-1}|f_{(i)(j)}|^2\right)-\frac{1}{2} \sum_{i=2}^{n-1} \left(\bar{f}_{(0)(1)} f_{(1)(i)}+f_{(0)(1)} \bar{f}_{(1)(i)} \right)E_\mu^{(i)}\right\} \label{j_maxwell}.
	\end{align}
	
	To compute the FEC condition, we take
	\begin{equation}
		J_\mu J^\mu=\left(v^{(0)}\right)^2\left\{\dfrac{1}{4}\sum_{i=2}^{n-1}\left(\bar{f}_{(0)(1)} f_{(1)(i)}+f_{(0)(1)} \bar{f}_{(1)(i)} \right)^2-\frac{1}{4}\left(|f_{(0)(1)}|^2+\sum_{i=1}^{n-1} \sum_{j>i}^{n-1}|f_{(i)(j)}|^2\right)^2\right\}
	\end{equation}
	
	But we can write $\left(\bar{f}_{(0)(1)} f_{(1)(i)}+f_{(0)(1)} \bar{f}_{(1)(i)} \right)^2=4 |f_{(0)(1)}|^2| f_{(1)(i)}|^2+\left(\bar{f}_{(0)(1)} f_{(1)(i)}-f_{(0)(1)} \bar{f}_{(1)(i)} \right)^2$. Notice that the term in parenthesis is zero or negative. Hence, we can write
	\begin{align}
		J_\mu J^\mu=&-(v^{(0)})^2\biggl\{\frac14\biggl(|f_{(0)(1)}|^2-\sum_{j=2}^{n-1}|f_{(1)(j)}|^2-\sum_{j=3}^{n-1}|f_{(2)(j)}|^2-\cdots-\sum_{j=n-1}^{n-1}|f_{(n-2)(j)}|^2\biggl)^2 \nonumber \\
		&+|f_{(0)(1)}|^2\biggl(\sum_{j=3}^{n-1}|f_{(2)(j)}|^2+\cdots+\sum_{j=n-1}^{n-1}|f_{(n-2)(j)}|^2\biggl)-\dfrac{1}{4}\sum_{i=2}^{n-1}\left(\bar{f}_{(0)(1)} f_{(1)(i)}-f_{(0)(1)} \bar{f}_{(1)(i)} \right)^2\biggl\}\ .
	\end{align}

	Therefore,  a complex Maxwell field satisfies the FEC. Since both WEC and FEC hold for a complex Maxwell field, it follows that the NEC and DEC are also respected.
	
	Now, we need to prove the SEC. Hence, we take
	\begin{equation}
		T=- \frac{n-4}{4} F_{\rho \sigma} \bar{F}^{\rho \sigma},
	\end{equation}
	then,
	\begin{equation}
		\left(T_{\mu \nu}-\frac{1}{n-2} T g_{\mu \nu}\right) v^\mu v^\nu=\frac{\left(v^{(0)}\right)^2}{n-2}\left\{(n-3)|f_{(0)(1)}|^2+\sum_{i=1}^{n-1} \sum_{j>i}^{n-1}|f_{(i)(j)}|^2\right\},
	\end{equation}
concluding that, in fact, the SEC is satisfied.

	\begin{Prop}
		All the standard energy conditions are respected for a complex Proca field \eqref{cproca} with $\phi^2>0$.
	\end{Prop}
	
	{\it Proof}. As in the proof of Proposition 1, we consider the frame such that $v^{(i)}=0$ for all $i$. In this frame, $v^{\mu}$ and $A_{\mu}$ are expressed as
	
	$$
	v^{\mu}=v^{(0)} E_{(0)}^{\mu}, \quad A_{\mu}=A_{(a)} E_{\mu}^{(a)}
	$$
	
	and hence we have
	\begin{equation}
		\tau_{\mu \nu} v^{\mu} v^{\nu}=\frac{1}{2}\left(v^{(0)}\right)^{2} \sum_{a=0}^{n-1}\left|A_{(a)}\right|^{2} \geq 0
	\end{equation}
	\begin{equation}
		\left(\tau_{\mu \nu}-\frac{1}{n-2} \tau g_{\mu \nu}\right) v^\mu v^\nu=\left(v^{(0)}\right)^2|A_{(0)}|^2 \geq 0 .
	\end{equation}
	
	The above equation shows that $\tau_{\mu \nu}$ satisfies the WEC. In order to prove the FEC and DEC, we take
	\begin{equation}
		\begin{aligned}
			\hat{J}_\mu & =-\frac{1}{2}
			(
			\mathcal{A}_{\mu}\bar{\mathcal{A}}_{\nu}
			+\bar{\mathcal{A}}_{\mu}\mathcal{A}_{\nu}
			)v^\nu
			+\frac{1}{2}v_{\mu}\mathcal{A}_\sigma\bar{\mathcal{A}}^\sigma \\
			& =-\frac{1}{2} v^{(0)} E_\mu^{(0)} \sum_{a=0}^{n-1}|A_{(a)}|^2-\frac{1}{2} v^{(0)}\left( \bar{A}_{(0)} A_{(i)} + A_{(0)} \bar{A}_{(i)} \right)E_\mu^{(i)},\label{j_proca}
		\end{aligned}
	\end{equation}
	\begin{equation}
		\hat{J}_\mu\hat{J}^\mu=-\dfrac{1}{4}\left(v_{(0)}\right)^2\left(\bar{A}^\sigma A_\sigma\right)^2+ \dfrac{1}{4}\sum\limits_{i=1}^{n-1}\left(A_{(0)}\bar{A}_{(i)}-\bar{A}_{(0)}{A}_{(i)}\right)^2 \leq 0
	\end{equation}
	
	Consequently, we have shown that $\tau_{\mu\nu}$ satisfies FEC and hence DEC as well. Finally, in order to show that the full complex-Proca model satisfies the FEC and DEC, following Lemma 3 and Proposition 22 of \cite{Maeda:2018hqu}, we introduce ${\bar J}_\mu:=-{\bar T}_{\mu\nu}v^\nu={\bar j}_{(a)}E_\mu^{(a)}$ and ${\hat J}_\mu:=-{\tau}_{\mu\nu}v^\nu={\hat j}_{(a)}E_\mu^{(a)}$. Hence, we need to verify that ${\bar j}_{(0)}{\hat j}_{(0)}\ge 0$. From equations \eqref{j_maxwell} and \eqref{j_proca}
	\begin{equation}
		{\bar j}_{(0)}  =-\frac{1}{2}v^{(0)}\left(|f_{(0)(1)}|^2+\sum_{i=1}^{n-1} \sum_{j>i}^{n-1}|f_{(i)(j)}|^2\right) ,
	\end{equation}
	\begin{equation}
		{\hat j}_{(0)}= -\frac{1}{2} v^{(0)} \sum_{a=0}^{n-1}|A_{(a)}|^2,
	\end{equation}
	we verify ${\bar j}_{(0)}{\hat j}_{(0)}\ge 0$.
	
	Based on Propositions above and Lemma 3 presented in \cite{Maeda:2018hqu}, we have demonstrated that a complex Proca field with a non-negative mass $\phi^2$ satisfies all of the standard energy conditions.  
	
\begin{figure}[h!]
		\includegraphics[width=0.47\textwidth]{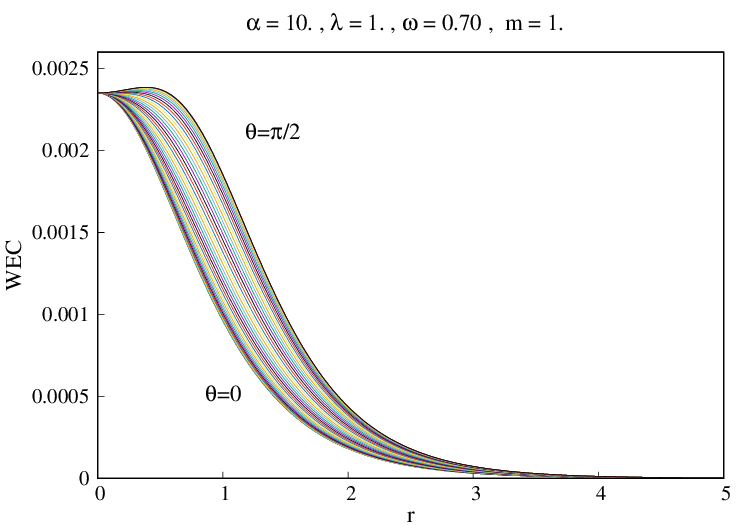} 
		\includegraphics[width=0.47\textwidth]{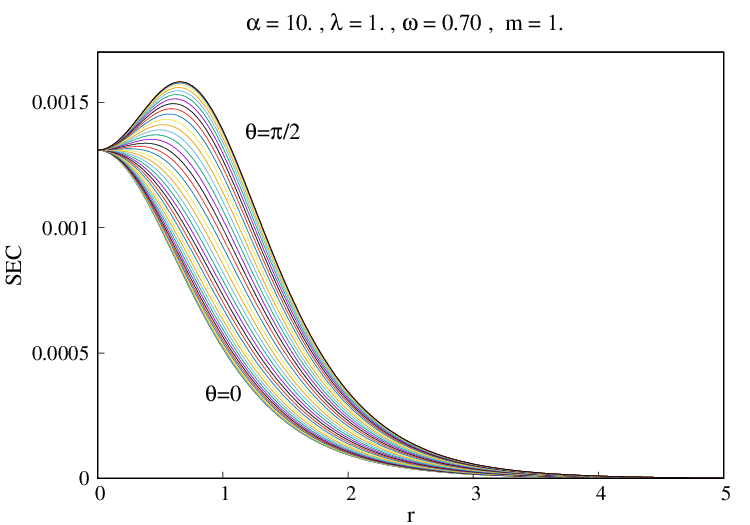}
  \includegraphics[width=0.47\textwidth]{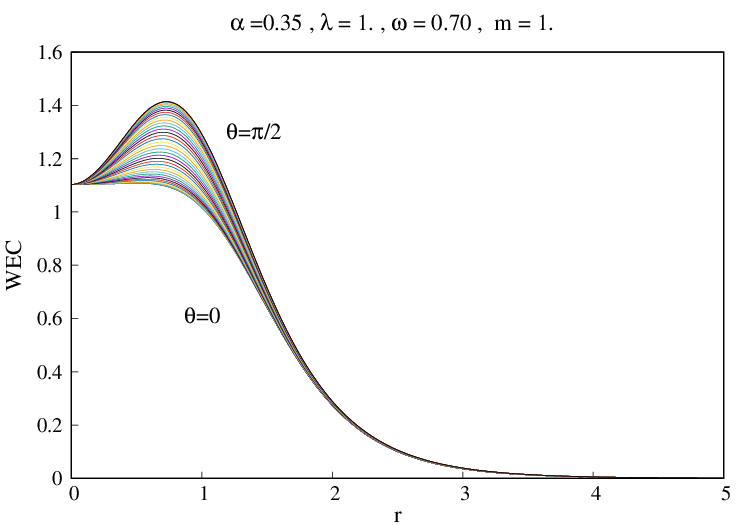} 
  \hfill
		\includegraphics[width=0.47\textwidth]{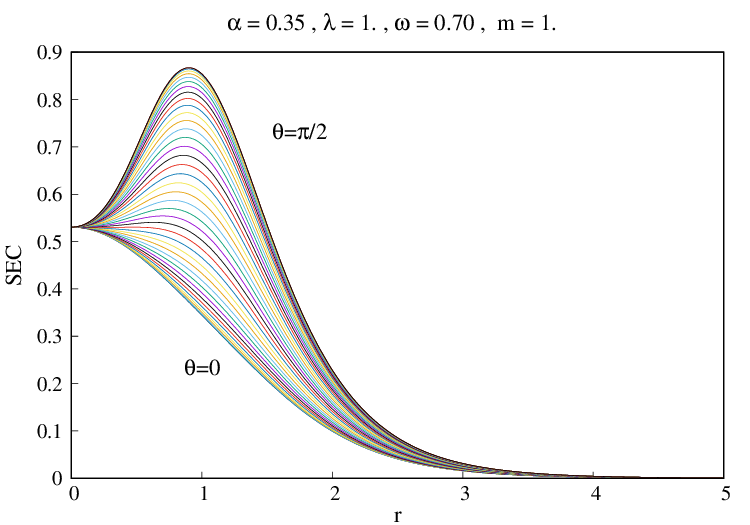}
		\caption{Density profiles of the WEC (left panel) and SEC (right panel) as a function of the radial coordinate, evaluated for a ZAMO for a particular solution with parameters: (top) $\alpha=10.$, $\lambda=1.$, $\omega=0.70$ and (bottom) $\alpha=0.35$, $\lambda=1.$, $\omega=0.70$. We plot the densities for each evaluated angle $\theta$ in the grid, $0\leq \theta \leq \pi/2$. It can be seen that for a ZAMO (and in fact in general), the solutions never violate the WEC or the SEC.}
		\label{fig:ecalpha10}
	\end{figure}

	To conclude, it was shown in \cite{Maeda_2022} and \cite{Maeda:2018hqu} that the scalar field with a positive potential satisfies the NEC, WEC, FEC, and DEC, but it is well known that the SEC can potentially be violated. Hence, the same argument as before leads to the conclusion that the full Proca-Higgs model also satisfies all the standard energy conditions, including the SEC, as long as the scalar part obeys it. In Fig.~\ref{fig:ecalpha10} we illustrate the density profiles of the WEC and SEC for some Proca-Higgs solutions.

\section{Equations of motion}
\label{eomap}

Turning our attention to the system derived from the given ansatz, we explicitly give the coupled partial differential equations (PDEs) at hand. The Klein-Gordon equation is used as it is in Eq. \eqref{scalarfe}. In order to derive the Proca field equations, we made use of the gauge condition inside the equations of motion due to better numerical convergence.

\begingroup
\footnotesize

\begin{multline}
	\frac{r \sin ^2\theta \phi_{,r} \left(N \left(r \left(F_{0,r}+F_{2,r}\right)+2\right)+r N'\right)}{N}+\frac{\sin ^2\theta \phi_{,\theta} \left(F_{0,\theta}+F_{2,\theta}+\cot \theta\right)}{N}+\frac{\sin ^2\theta  \phi_{,\theta\theta}}{N}+r^2 \sin ^2\theta \phi_{,rr}-\\-\frac{\lambda  r^2 \sin ^2\theta e^{2 F_{1}} \phi \left(\phi^2-1\right)}{N}+ \frac{\sin ^2\theta \phi e^{-2 (F_{0}+F_{2})}}{r^2 N^2} \Biggl(r^2 N H_{3}^2 \left(-e^{2 (F_{0}+F_{1})}\right)-r^2 N e^{2 (F_{0}+F_{2})} \left(N H_{1}^2+H_{2}^2\right)+\\
 +e^{2 (F_{1}+F_{2})} \left(\sin \theta H_{3} W+r^2 V\right)^2\Biggl)=0\,,
\end{multline}

\begin{multline}
\frac{r^2 \sin ^2\theta  H_{1,r} \left(N \left(F_{0,r}-2 F_{1,r}+F_{2,r}\right)+2 N'\right)}{N}+\frac{\sin \theta  H_{1,\theta} \left(\sin \theta  F_{0,\theta}-2 \sin \theta  F_{1,\theta}+\sin \theta  F_{2,\theta}+\cos \theta \right)}{N}+\frac{\sin ^2\theta  H_{1,\theta\theta}}{N}+r^2 \sin ^2\theta  H_{1,rr}+\\
+\frac{2 r H_{2} \phi_{,r\theta} \sin ^2\theta }{N \phi}+\frac{2 r^2 H_{1} \phi_{,rr} \sin ^2\theta }{\phi}-\frac{6 r^2 H_{1} \phi_{,r}^2 \sin ^2\theta }{\phi^2}+\left(\frac{2 \sin ^2\theta  \left(r H_{2,r}-2 H_{2} \left(r F_{1,r}+1\right)\right)}{N \phi}-\frac{6 r H_{2} \sin ^2\theta  \phi_{,r}}{N \phi^2}\right) \phi_{,\theta}+ \\
+\frac{2 \sin \theta }{r N^2 \phi} \Biggl\{N \left(e^{2 F_{1}-2 F_{2}} H_{3}-H_{2} \left(\cos \theta +\sin \theta  \left(F_{0,\theta}+F_{2,\theta}\right)\right)-\sin \theta  H_{2,\theta}\right) r^2-N^2 H_{1} \sin \theta  \left(r \left(F_{0,r}+2 F_{1,r}+F_{2,r}\right)+2\right) r^2+\\
+ e^{2 F_{1}-2 F_{0}} \sin \theta  \left(r^2 \omega -W\right) \left(V r^2+H_{3} \sin \theta  W\right)\Biggl\} \phi_{,r}+\frac{e^{-2 (F_{0}+F_{2})}}{2 r^2 N^3} \Biggl\{-2 e^{2 (F_{0}+F_{2})} r^2 H_{1} \sin ^2\theta \Biggl[r \Biggl(F_{0,r} \left(2 r F_{1,r}+1\right)+F_{2,r}+\\
+ 2 F_{1,r} \left(r F_{2,r}+1\right)-r \left(F_{0,rr}+F_{2,rr}\right)\Biggl)+2\Biggl] N^3-2 e^{2 F_{0}} r^2 \Biggl[e^{2 F_{1}} \left(H_{1}-2 H_{3} \sin \theta  \left(r F_{2,r}+1\right)\right)+e^{2 F_{2}} \sin \theta  \Biggl(2 \cos \theta  H_{2} \left(r F_{1,r}+1\right)+\\
+\sin \theta \biggl(2 \left(H_{2} \left(F_{0,\theta}+F_{2,\theta}\right)+H_{2,\theta}\right)+r \biggl(e^{2 F_{1}} r H_{1} \phi^2+2 H_{2,\theta} F_{1,r}-r H_{1} \left(N''+N' \left(F_{0,r}-2 F_{1,r}+F_{2,r}\right)\right)-\\
-2 F_{1,\theta} H_{2,r}-H_{2} \left(-2 \left(F_{0,\theta}+F_{2,\theta}\right) F_{1,r}+F_{0,r,\theta}+F_{2,r,\theta}\right)\biggl)\biggl)\Biggl)\Biggl] N^2+2 e^{2 (F_{1}+F_{2})} \sin ^2\theta  \Biggl[V \left(2 \omega  F_{0,r} r^3+W_{,r} r-2 W \left(r F_{0,r}+1\right)\right) r^2+\\
+ H_{1} \left(W-r^2 \omega \right)^2+H_{3} \sin \theta  \left(-\omega  W_{,r} r^3+2 W \left(r \omega  \left(r F_{0,r}+1\right)+W_{,r}\right) r-2 W^2 \left(r F_{0,r}+2\right)\right)\Biggl] N+\\
+2 e^{2 (F_{1}+F_{2})} r \sin ^2\theta  \left(r^2 \omega -W\right) \left(V r^2+H_{3} \sin \theta  W\right) N'\Biggl\}=0\,,
\end{multline}

\begin{multline}
\frac{r^2 \sin ^2\theta  H_{2,r} \left(N \left(F_{0,r}-2 F_{1,r}+F_{2,r}\right)+N'\right)}{N}+\frac{\sin \theta  H_{2,\theta} \left(\sin \theta  F_{0,\theta}-2 \sin \theta  F_{1,\theta}+\sin \theta  F_{2,\theta}+\cos \theta \right)}{N}+\frac{\sin ^2\theta  H_{2,\theta\theta}}{N}+r^2 \sin ^2\theta  H_{2,rr}+\\+
\phi_{,\theta} \Biggl\{\frac{2 \sin \theta }{N^2 \phi} \Biggl[-N H_{2} \left(\sin \theta  \left(F_{0,\theta}+2 F_{1,\theta}+F_{2,\theta}\right)+\cos \theta \right)-\sin \theta  N \left(H_{1} \left(r N \left(F_{0,r}+F_{2,r}\right)+r N'+N\right)+r N H_{1,r}\right)+ \\
+\frac{\sin \theta  e^{2 F_{1}-2 F_{0}} }{r^2}\left(r^2 \omega -W\right) \left(\sin \theta  H_{3} W+r^2 V\right)+N H_{3} e^{2 F_{1}-2 F_{2}}\Biggl]-\frac{6 r \sin ^2\theta  H_{1} \phi_{,r}}{\phi^2}\Biggl\} + \\
+\frac{1}{N}\Biggl\{-H_{2} \left(\sin \theta  \left(2 F_{1,\theta} \left(\sin \theta  \left(F_{0,\theta}+F_{2,\theta}\right)+\cos \theta \right)-\sin \theta  \left(F_{0,\theta\theta}+F_{2,\theta\theta}\right)+r^2 \sin \theta  e^{2 F_{1}} \phi^2\right)+1\right)-2 r \sin ^2\theta  N' F_{1,\theta} H_{1}+  \\
+ e^{2 F_{1}-2 F_{2}} \left(2 H_{3} \left(\sin \theta  F_{2,\theta}+\cos \theta \right)-H_{2}\right)\Biggl\}-\sin ^2\theta  \Biggl(H_{1} \left(2 F_{1,\theta} \left(r \left(F_{0,r}+F_{2,r}\right)+1\right)-r \left(F_{0,r,\theta}+F_{2,r,\theta}\right)\right)-2 \left(r F_{1,r}+1\right) H_{1,\theta}+ \\
+2 r F_{1,\theta} H_{1,r}\Biggl)+\frac{\sin ^2\theta  e^{2 F_{1}-2 F_{0}} }{r^2 N^2}\left(2 F_{0,\theta} \left(r^2 \omega -W\right) \left(\sin \theta  H_{3} W+r^2 V\right)+H_{2} \left(W-r^2 \omega \right)^2+W_{,\theta} \left(\sin \theta  H_{3} \left(2 W-r^2 \omega \right)+r^2 V\right)\right)+\\
+\frac{2 r \sin ^2\theta  \phi_{,r} \left(H_{1,\theta}-2 F_{1,\theta} H_{1}\right)}{\phi}+\frac{2 \sin ^2\theta  H_{2} \phi_{,\theta\theta}}{N \phi}-\frac{6 \sin ^2\theta  H_{2} \phi_{,\theta}^2}{N \phi^2}+\frac{2 r \sin ^2\theta  H_{1} \phi_{,r\theta}}{\phi}=0\,,
\end{multline}

\begin{multline}
 \frac{\sin \theta  H_{3,\theta} \left(N \left(\sin \theta  \left(F_{0,\theta}-F_{2,\theta}\right)+\cos \theta \right)-\frac{\sin ^3\theta W W_{,\theta} e^{2 F_{2}-2 F_{0}}}{r^2}\right)}{N^2}+\frac{\sin ^2\theta  H_{3,\theta\theta}}{N}+r^2 \sin ^2\theta  H_{3,rr}+\\
 +\frac{\sin ^2\theta  H_{3,r} \left(r^3 \left(N \left(F_{0,r}-F_{2,r}\right)+N'\right)+\sin ^2\theta  W \left(2 W-r W_{,r}\right) e^{2 F_{2}-2 F_{0}}\right)}{r N}+\\
 +\frac{e^{-2 (F_{0}+F_{2})}}{2 r^2 N^2 \phi} \Biggl\{N \Biggl[\phi \Biggl(r^2 e^{2 F_{0}} H_{3} \left(-e^{2 F_{2}} \left(\sin 2 \theta  \left(F_{2,\theta}-F_{0,\theta}\right)+2 r^2 \sin ^2\theta  e^{2 F_{1}} \phi^2+2\right)-2 e^{2 F_{1}}\right)-\\
 - 2 \sin ^3\theta  e^{4 F_{2}} \left(r W_{,r}-2 W\right) \left(H_{1} \left(r^2 \omega -W\right)+r^3 V_{,r}\right)\Biggl)+4 r^2 H_{2} e^{2 (F_{0}+F_{2})} \left(\phi \left(\sin \theta  F_{2,\theta}+\cos \theta \right)+\sin \theta  \phi_{,\theta}\right)\Biggl]+\\
 + 4 r^2 \sin \theta  N^2 H_{1} e^{2 (F_{0}+F_{2})} \left(\left(r F_{2,r}+1\right) \phi+r \phi_{,r}\right)+2 \sin ^2\theta  e^{2 F_{2}} \phi \Biggl[e^{2 F_{1}} H_{3} \left(W-r^2 \omega \right)^2-\sin \theta  e^{2 F_{2}} W_{,\theta} \biggl(H_{2} \left(r^2 \omega -W\right)+  \\
 +\cos \theta  H_{3} W+r^2 V_{,\theta}\biggl)\Biggl]\Biggl\}
=0\,,
\end{multline}

\begin{multline}
	\frac{\sin \theta  V_{,\theta} \left(N \left(\sin \theta  \left(F_{2,\theta}-F_{0,\theta}\right)+\cos \theta \right)+\frac{\sin ^3\theta W W_{,\theta} e^{2 F_{2}-2 F_{0}}}{r^2}\right)}{N^2}+\frac{\sin ^2\theta  V_{,\theta\theta}}{N}+ r^2 \sin ^2\theta  V_{,rr}+\\
 +V_{,r} \left(r \sin ^2\theta  \left(-r F_{0,r}+r F_{2,r}+2\right)+\frac{\sin ^4\theta W \left(r W_{,r}-2 W\right) e^{2 F_{2}-2 F_{0}}}{r N}\right)+ \\
 +\frac{e^{-2 (F_{0}+F_{2})} }{2 r^4 N^2 \phi} \Biggl\{2 N \Biggl[-r^2 \sin \theta  e^{2 (F_{0}+F_{2})} \Biggl(-\phi \biggl(\sin \theta  \biggl(H_{3,\theta} \left(2 W \left(\sin \theta  \left(F_{2,\theta}-F_{0,\theta}\right)+\cos \theta \right)+\sin \theta  W_{,\theta}\right)+r N' H_{1} \left(W-r^2 \omega \right)+\\
 + r^2 (-\sin \theta ) N' H_{3,r} W\biggl)+\cos \theta  H_{3} \left(2 W \left(\sin \theta  \left(F_{2,\theta}-F_{0,\theta}\right)+\cos \theta \right)+\sin \theta  W_{,\theta}\right)\biggl)+H_{2} \Biggl(\phi \biggl(2 W \biggl(\sin \theta  \left(F_{2,\theta}-F_{0,\theta}\right)+\cos \theta \biggl)+\\
 + \sin \theta  \left(2 r^2 \omega  F_{0,\theta}+W_{,\theta}\right)\biggl)+2 r^2 \omega  \sin \theta  \phi_{,\theta}\Biggl)+r^4 \sin \theta  e^{2 F_{1}} V \phi^3\Biggl)-r^4 V \phi e^{2 (F_{0}+F_{1})}+\sin ^4\theta  e^{4 F_{2}} W \left(r W_{,r}-2 W\right) \phi \Biggl(H_{1} \left(r^2 \omega -W\right)+\\
 +r \sin \theta  H_{3,r} W\Biggl)\Biggl]-2 r^3 \sin ^2\theta  N^2 e^{2 (F_{0}+F_{2})} \Biggl[H_{1} \left(\phi \left(2 F_{0,r} \left(r^2 \omega -W\right)+2 F_{2,r} W+W_{,r}\right)+2 r^2 \omega  \phi_{,r}\right)+ \\
 +r \sin \theta  H_{3,r} \phi \left(2 W \left(F_{0,r}-F_{2,r}\right)-W_{,r}\right)\Biggl]+2 \sin ^2\theta  e^{2 F_{2}} \phi \Biggl(e^{2 F_{1}} V \left(r^3 \omega -r W\right)^2+\sin ^2\theta  e^{2 F_{2}} W W_{,\theta} \biggl(H_{2} \left(r^2 \omega -W\right)+\\
 +W \left(\sin \theta  H_{3,\theta}+\cos \theta  H_{3}\right)\biggl)\Biggl)\Biggl\}
	=0\,,
\end{multline}

\endgroup

We address each of the four Einstein equations, given by 
$ E_{a b} \equiv R_{a b}-\frac{1}{2} g_{a b} R-2\alpha^2 T_{a b}=0$.
Each equation is expected to have second derivatives of a unique metric function. To realize this, we utilize particular combinations of the Einstein equations

\begin{equation}
\begin{aligned}
& E_r^r+E_\theta^\theta-E_{\varphi}^{\varphi}-E_t^t=0, \\
& E_r^r+E_\theta^\theta-E_{\varphi}^{\varphi}+E_t^t+2 \dfrac{W}{r} E_{\varphi}^t=0, \\
& E_r^r+E_\theta^\theta+E_{\varphi}^{\varphi}-E_t^t-2 \dfrac{W}{r}  E_{\varphi}^t=0, \\
& E_{\varphi}^t=0 .
\end{aligned}
\end{equation}

\begingroup
\footnotesize
\begin{multline}
	\frac{r \sin ^2 \theta\,  F_{1,r} \left(2N +r N'\right)}{2 N}+\frac{r^2 \sin ^2 \theta\,  F_{1,\theta\theta}}{N}+r^2 \sin ^2 \theta\,  N F_{1,rr}+\\+
	\frac{\sin  \theta\,}{4 H^2}  \Bigg\{-2 N \left[r \sin \theta  \left(2 N F_{0,r} \left(r F_{2,r}+1\right)+r N' F_{2,r}\right)+2 F_{0,\theta} \left(\sin \theta  F_{2,\theta}+\cos \theta \right)\right] - \frac{\sin ^3 \theta\,  e^{2 F_{2}-2 F_{0}} \left(N \left(2 W-r W_{,r}\right)^2+ W_{,\theta}^2\right)}{r^2}\Bigg\}+\\
	+	\alpha ^2  \Bigg\{\frac{\sin ^2\theta  \phi_{,\theta}^2}{N} +r^2 \sin ^2\theta  \phi_{,r}^2+\sin ^2\theta  \phi^2 \left(\frac{e^{2 F_{1}-2 F_{0}} \left(\sin \theta  H_{3} W+r^2 V\right)^2}{r^2 N^2}+\frac{H_{2}^2-H_{3}^2 e^{2 F_{1}-2 F_{2}}}{N}+H_{1}^2\right)+\\+\frac{2 e^{-2 (F_{0}-F_{1}+F_{2})} (\omega \sin \theta  H_{3}+V)^2}{N^2}+\frac{2 \sin ^2\theta  e^{-2 F_{1}} \left(H_{1,\theta}-r H_{1,r}\right)^2}{r^2}\Bigg\}=0\,,
\end{multline}

\begin{multline}
   \frac{r^2 \sin ^2 \theta\,  F_{2,\theta\theta}}{H}+r^2 \sin^2 \theta\,  F_{2,r}^2+r^2 \sin^2 \theta\,  N F_{2,rr}+  \frac{r \sin ^2 \theta\,  F_{2,r} \left(H \left(r  F_{0,r}+3\right)+r N'\right)}{H}+\frac{r^2 \sin ^2 \theta\,  F_{2,\theta} \left(F_{0,\theta}+2 \cot  \theta\, \right)}{H}+\\
   +\frac{r^2 \sin ^2 \theta\,  F_{2,\theta}^2}{H}+ \frac{\sin \theta  }{2 N^2}\left(2 N \left(\sin \theta  \left(r N F_{0,r}+r N'+N-1\right)+\cos \theta  F_{0,\theta}\right)+\frac{\sin ^3\theta e^{2 F_{2}-2 F_{0}} \left(N \left(r W_{,r}-2 W\right)^2+W_{,\theta}^2\right)}{r^2}\right)+\\+
    	\frac{\alpha ^2 e^{-2 (F_{0}+F_{1}+F_{2})}}{8 r^4 N^2}\Bigg\{8 r^2 N^2 e^{2 F_{0}} \left(e^{2 F_{1}} \left(H_{1}-r \sin \theta  H_{3,r}\right)^2-\sin ^2\theta  e^{2 F_{2}} \left(H_{1,\theta}-r H_{2,r}\right)^2\right)+\\
     +4 N e^{2 F_{1}} \Bigg(\sin ^2\theta  e^{2 F_{2}} \left(\lambda  r^6 \left(\phi^2-1\right)^2 e^{2 (F_{0}+F_{1})}+2 \left(H_{1} \left(r^2 \omega -W\right)+r \sin \theta  H_{3,r} W+r^3 V_{,r}\right)^2\right)+\\
     +2 r^2 e^{2 F_{0}} \left(H_{3}^2 \left(2 r^2 \sin ^2\theta  e^{2 F_{1}} \phi^2+\cos ^2\theta\right)-2 H_{2} \left(\sin \theta  H_{3,\theta}+\cos \theta  H_{3}\right)+H_{2}^2+\sin ^2\theta  H_{3,\theta}^2+\sin (2 \theta ) H_{3} H_{3,\theta}\right)\Bigg)+\\
     +8 \sin ^2\theta  e^{2 (F_{1}+F_{2})} \left(H_{2} \left(r^2 \omega -W\right)+W \left(\sin \theta  H_{3,\theta}+\cos \theta  H_{3}\right)+r^2 V_{,\theta}\right)^2-8 r^4 e^{4 F_{1}} (\omega  \sin \theta  H_{3}+V)^2\Bigg\}=0\,,
\end{multline}

\begin{multline}
\frac{r \sin ^2\theta  F_{0,r} \left(2 N \left(r F_{2,r}+2\right)+3 r N'\right)}{2 N}+\frac{\sin \theta  F_{0,\theta} \left(\sin \theta  F_{2,\theta}+\cos \theta \right)}{N}+\frac{\sin ^2\theta  F_{0,\theta\theta}}{N}+r^2 \sin ^2\theta  F_{0,r}^2+r^2 \sin ^2\theta  F_{0,rr}+\\+\frac{\sin ^2\theta  F_{0,\theta}^2}{N}+ 8 r^2 \sin ^4\theta e^{2 F_{2}-2 F_{0}} \left(N \left(r W_{,r}-2 W\right)^2+W_{,\theta}^2\right)-8 r^5 \sin ^2\theta  N \left(N' \left(r F_{2,r}+2\right)+r N''\right)+\\+\frac{\alpha ^2 e^{-2 (F_{0}+F_{1}+F_{2})}}{8 r^4 N^2}\Biggl\{8 r^2 N^2 e^{2 F_{0}} \left(-e^{2 F_{2}}\sin ^2\theta  \left(H_{1,\theta}-r H_{2,r}\right)^2-e^{2 F_{1}} \left(H_{1}-r \sin \theta  H_{3,r}\right)^2\right)+ \\+
4 N e^{2 F_{1}} \Biggl[\sin ^2\theta  e^{2 F_{2}} \left(\lambda  r^6 \left(\phi^2-1\right)^2 e^{2 (F_{0}+F_{1})}-2 \left(H_{1} \left(r^2 \omega -W\right)+r \sin \theta  H_{3,r} W+r^3 V_{,r}\right)^2\right)- \\
-2 r^2 e^{2 F_{0}} \left(-H_{2}+\sin \theta  H_{3,\theta}+\cos \theta  H_{3}\right)^2\Biggl]-8 \sin ^2\theta  e^{2 (F_{1}+F_{2})} \Biggl[2 r^2 e^{2 F_{1}} \phi^2 \left(\sin \theta  H_{3} W+r^2 V\right)^2+\big(H_{2} \left(r^2 \omega -W\right)+\\+
W \left(\sin \theta  H_{3,\theta}+\cos \theta  H_{3}\right)+r^2 V_{,\theta}\big)^2\Biggl]-8 r^4 e^{4 F_{1}} (\omega  \sin \theta  H_{3}+V)^2\Biggl\}=0\,,
\end{multline}

\begin{multline}
-\frac{\sin \theta  W_{,\theta} \left(\sin \theta  F_{0,\theta}-3 \left(\sin \theta  F_{2,\theta}+\cos \theta \right)\right)}{r N}-r \sin ^2\theta  W_{,r} \left(F_{0,r}-3 F_{2,r}\right)+\frac{\sin ^2\theta  W_{,\theta\theta}}{r N}+r \sin ^2\theta  W_{,rr}+\\+\frac{2 \sin ^2\theta  W \left(r F_{0,r}-3 r F_{2,r}-1\right)}{r}- 
\frac{4 \alpha ^2 e^{-2 F_{2}}}{r^3 N}\Biggl\{r^2 \sin \theta  e^{2 F_{1}} H_{3} \phi^2 \left(\sin \theta  H_{3} W+r^2 V\right)+N \left(r \sin \theta  H_{3,r}-H_{1}\right) \Biggl[H_{1} \left(r^2 \omega -W\right)+ \\
+r \sin \theta  H_{3,r} W+r^3 V_{,r}\Biggl]+  H_{2} \left(\left(r^2 \omega -2 W\right) \left(\sin \theta  H_{3,\theta}+\cos \theta  H_{3}\right)-r^2 V_{,\theta}\right)+H_{2}^2 \left(W-r^2 \omega \right)+r^2 \sin \theta  H_{3,\theta} V_{,\theta}+\sin ^2\theta  H_{3,\theta}^2 W+\\
+\sin 2 \theta  H_{3} H_{3,\theta} W+r^2 \cos \theta  H_{3} V_{,\theta}+\cos ^2\theta H_{3}^2 W\Biggl\}
	=0\,,
\end{multline}
\endgroup

	\bibliographystyle{hhieeetr}
	\bibliography{biblio}

\end{document}